\documentclass[english,twocolumn,aps, pra, showpacs, amsmath]{revtex4}
\usepackage[OT1]{fontenc}
\usepackage[latin1]{inputenc}
\usepackage{float}
\usepackage{graphicx}
\usepackage{amssymb}

\makeatletter

\providecommand{\boldsymbol}[1]{\mbox{\boldmath $#1$}}

\providecommand{\tabularnewline}{\\}

\usepackage{float}

\makeatletter

\usepackage{float}

\makeatletter

\makeatletter
\usepackage{bm}

\usepackage{latexsym}

\makeatother

\usepackage{hyperref}

\hypersetup{%
pdftitle={A theory of intense-field dynamic alignment and high harmonic generation
from coherently rotating molecules and interpretation of intense-field
ultrafast pump-probe experiments},
pdfauthor={A. Abdurrouf, F.H.M. Faisal},
pdfkeywords={alignment, HHG, dynamic signal, Fourier spectrum, polarization angle, intense field},
bookmarksnumbered,
pdfstartview={FitH},
urlcolor=blue
}%

\makeatother

\makeatother

\makeatother

\makeatother

\makeatother

\usepackage{babel}
\makeatother
\begin{document}

\title{A theory of intense-field dynamic alignment and high harmonic generation
from coherently rotating molecules and interpretation of intense-field
ultrafast pump-probe experiments}

\author{A. Abdurrouf$^{1,2}$ }

\email{a.abdurrouf@googlemail.com}

\author{and F.H.M. Faisal$^{1}$}

\email{ffaisal@physik.uni-bielefeld.de}

\address{$^{1}$Fakultät für Physik, Universität Bielefeld, Postfach 100131,
D-33501 Bielefeld, Germany}

\address{\emph{$^{2}$}Department of Physics, University of Brawijaya, Malang,
Indonesia 65145}

\begin{abstract}
A theory of ultra-fast pump-probe experiments proposed by us earlier
{[}F.H.M. Faisal \emph{et al}., Phys. Rev. Lett. \textbf{98}, 143001
(2007) and F.H.M. Faisal and A. Abdurrouf, Phys. Rev. Lett. \textbf{100},
123005 (2008)] is developed here fully and applied to investigate
the phenomena of dynamic alignment and high harmonic generation (HHG)
from coherently rotating linear molecules. The theory provides essentially
analytical results for the signals that allow us to investigate the
simultaneous dependence of the HHG signals on the two externally available
control parameters, namely, the relative angle between the polarizations,
and the delay-time between the two pulses. It is applied to investigate
the characteristics of high harmonic emission from nitrogen and oxygen
molecules that have been observed experimentally in a number of laboratories.
The results obtained both in the time-domain and in the frequency-domain
are compared diwith the observed characteristics as well as directly
with the data and are found to agree remarkably well. In addition
we have predicted the existence of a ``magic'' polarization angle
at which all modulations of the harmonic emission from nitrogen molecule
changes to a steady emission at the harmonic frequency. Among other
things we have also shown a correlation between the existence of the
\char`\"{}magic'' or critical polarization angles and the symmetry
of the active molecular orbitals, that is deemed to be useful in connection
with the ``inverse problem'' of molecular imaging from the HHG data.
\end{abstract}


\maketitle

\section{Introduction}

In recent years there has been much interest and progress in understanding
the interaction of atoms and molecules with intense laser fields (e.g.
reviews \cite{pos-04,bec-05}). Among the phenomena observed, the
high-order harmonic generation (HHG) is of particular interest, no
less because of its potential applications as a source of coherent
ultraviolet light and/or for generation of ultrashort attosecond laser
pulses. In contrast to atoms, molecules have extra degrees of freedom
such as vibration, and rotation of the molecular frame, and have additional
symmetry properties, that give rise to richer physical phenomena when
they interact with intense laser pulses. Among them is the phenomenon
of alignment of linear molecules by strong and long laser pulses which
has been investigated in the past \cite{sei-99,ort-99,cai-01,sta-03}.
Much interest has recently been generated by the observation of recurrent
dynamic alignments of linear molecules like $\mathrm{N_{2}}$ and
$\mathrm{O_{2}}$ \cite{lit-03,doo-03}, interacting with intense
ultrashort laser pulses. They are monitored, for example, by non-destructive
high harmonic generation signals from intense-field pump-probe experiments
with delayed pairs of intense ultrashort pulses \cite{kak-04,zei-04,ita-05,kan-05,miy-05}.
The dynamic HHG signals have been used also to {}``reconstruct''
the molecular orbitals \cite{ita-04,lev-06,pat-06}, to investigate
proton motions \cite{bak-06} and molecular dynamics \cite{wag-06}.

In this paper we derive fully a recently proposed \cite{fai-07,fai-08}
quantum theory of intense-field dynamic alignment and high harmonic
generation from linear molecules and apply it to analyze the observed
dynamical HHG signals for $\mathrm{N_{2}}$ and $\mathrm{O_{2}}$
molecules. Theoretical expressions for the signals are given analytically
as a simultaneous function of the two external operational parameters
-- the delay time, $t_{d}$, and the relative polarization angle,
$\alpha$, between the pump and the probe pulse \cite{kak-04,zei-04,ita-05,kan-05,miy-05}.

Before proceeding further, we briefly discuss the main experimental
characteristics of dynamic alignment and the HHG signals as observed
for $\mathrm{N_{2}}$ and $\mathrm{O_{2}}$. We recall at the out
set that the quantum measure of dynamical alignment of a rotating
molecule is the quantum expectation value (with respect to the rotational
wave-packet states induced by the pump pulse) of the {}``alignment
operator'' $\cos^{2}\theta$, that is averaged over the Boltzmann
distribution of the initially occupied rotational states: $A\left(t_{d}\right)\equiv\left\langle \left\langle \cos^{2}\theta\right\rangle \right\rangle \left(t_{d}\right)$,
where $\theta$ is the angle between the molecular axis and the probe
polarization direction; 
 the double angular brackets stand for the expectation value with
respect to the rotational wave-packets (inner brackets) and the statistical
average with respect to the Boltzmann distribution (outer brackets)
of the initially occupied rotational states. 
It was observed experimentally \cite{kak-04,zei-04,ita-05,kan-05,miy-05}
that the dynamic (or delay-time dependent) HHG signal for $\mathrm{N_{2}}$
mimicked the {}``alignment measure'' $A\left(t_{d}\right)$. It
exhibited the phenomenon of rotational revivals \cite{sei-99,ros-02,sta-03}
including the {}``full-revival'' with a period $T_{r}=\frac{1}{2Bc}$,
where $B$ is the rotational constant \cite{her-50}, as well as a
$\frac{1}{2}$-revival, and a $\frac{1}{4}$-revival. They are consistent
with the time dependence of $A\left(t_{d}\right)$ defined above,
since the operator $\cos^{2}{\theta}$ can couple the rotational states
with $\Delta J=\pm2$ (Raman allowed transitions) among the rotational
states of the induced wavepackets, and thus can give rise to fractional
revival periods associated with the corresponding beat frequencies.
In the case of $\mathrm{O_{2}}$, unexpectedly, an additional $\frac{1}{8}$-revival
appeared in the HHG signal \cite{ita-05,kan-05,miy-05}. The latter
is impossible for the alignment measure $A\left(t_{d}\right)$ to
account for, since it can not couple the rotational states with $\Delta J=\pm4$,
that could give rise to a beat period $\frac{1}{8}T_{r}$. Thus, to
fit their data of $\mathrm{O_{2}}$, Itatani \emph{et al}. \cite{ita-05}
proposed, empirically, to consider the expectation value of the operator
$B\left(t_{d}\right)\equiv\left\langle \left\langle \sin^{2}2\theta\right\rangle \right\rangle \left(t_{d}\right)$.
Subsequently, some of the early theoretical models of the HHG signal
(e.g. \cite{zho-05a,zho-05b,mad-06}) gave a similar result for $\mathrm{O_{2}}$
and thus appeared to justify the empirical fit. Such a model also
suggests that the maximum HHG signal for $\mathrm{N}_{2}$ can occur
when the field polarization and the molecular axis were parallel,
whereas the maximum signal of $\mathrm{O_{2}}$ would occur when they
are {}``diagonal'' (i.e. make an angle $\theta=45^{0}$). Unlike
the time dependent signals themselves, their Fourier transform ($F.T.$),
with sharply defined individual spectral lines and series, provide
an alternative (and rather more precise) means of studying the dynamic
alignment phenomenon. More recent experimental observations of the
dynamic HHG signals for $\mathrm{N_{2}}$ and $\mathrm{O_{2}}$, and
their $F.T.$ have revealed surprising characteristics that can not
be fully understood in terms of the earlier considerations. Thus:\\
 (a) Kanai \emph{et al.} \cite{kan-05} found that their experimental
HHG signals for $\mathrm{N_{2}}$ and $\mathrm{O_{2}}$ could not
be well fitted, respectively, by the expectation values of the operators
$\cos^{2}\theta$ and $\sin^{2}2\theta$, alone. They considered empirically
additional operators involving \textit{higher} powers of $\cos^{2}{\theta}$,
or Legendre polynomials, to fit their data.\\
 (b) Miyazaki \emph{et al.} \cite{miy-05} measured the dynamical
HHG signals of $\mathrm{N_{2}}$ and $\mathrm{O_{2}}$ and Fourier
transformed their signals and found not only spectral series containing
strong Raman allowed but also weak Raman forbidden and anomalous lines,
for both $\mathrm{N_{2}}$ and $\mathrm{O_{2}}$. \\
 (c) Itatani \emph{et al.} \cite{ita-04} observed that the HHG signal
from dynamically aligned $\mathrm{N_{2}}$ was enhanced when the pump
polarization was taken parallel to the probe polarization, and were
suppressed when the polarizations were taken to be perpendicular.\\
 (d) Kanai \emph{et al.} \cite{kan-05} and Miyazaki \emph{et al.}
\cite{miy-05,kak-05} measured the HHG signal for the diatomic $\mathrm{N_{2}}$,
$\mathrm{O_{2}}$, and the triatomic $\mathrm{CO_{2}}$, for different
relative angles $\alpha$ between the pump and probe polarizations,
and observed that the HHG signal modulations are not only smaller
in the perpendicular case, compared to the parallel case, but also
are of \textit{opposite phase} in the two geometries.\\
 (e) Kanai \emph{et al}. \cite{kan-05} proposed a planar emission
model of HHG which produced an opposite phase relation, as observed,
but it did not yield the \textit{unequal} modulation amplitudes, observed
in the two geometries.\\
 (f) The present theory predicted (cf. \cite{fai-08} and below) a
{}``magic'' polarization angle, $\alpha_{c}\approx55^{0}$, at which
the harmonic emission from coherently rotating molecules with $\sigma_{g}$
orbital symmetry (e.g. $\mathrm{N_{2}}$) becomes equal for {\textit{all}}
delay times $t_{d}$. Most recent observations by Yoshii \emph{et
al.} \cite{yos-07,miy-06} appear to confirm the same.

In this paper we present an \textit{ab initio} development of the
above mentioned theory \cite{fai-07,fai-08} that is shown to provide
a unified theoretical account of all the phenomena noted above and
other related characteristics of dynamic alignments and the HHG signals
from the aligning molecules, as well as their Fourier spectra, that
have been observed experimentally. To this end, below we begin with
a short schematic description of a typical intense-field pump-probe
experiment on dynamic alignments and the molecular HHG signals as
a function of (a) the time delay, $t_{d}$, and (b) the relative polarization
angle, $\alpha$, between the pump and the probe pulse. In Sec. II,
III and IV, we systematically derive the S-matrix theory of molecular
alignment and dynamic HHG signal from an ensemble of freely rotating
linear molecule, discuss the connection between the {}``one'' -
and the {}``many''-molecule signals, and the relation between the
quantum amplitude for the emission of the HHG photons and the expectation
value of the dipole transition moment. In Sec. V, we apply the theory
to $\mathrm{N_{2}}$ and $\mathrm{O_{2}}$ molecules and obtain analytic
expressions for the {}``HHG operators'' and the HHG {\textit{signals}}
for an arbitrary, $\alpha$, and delay times, $t_{d}$. In Sec. VI.
we use the theoretical expressions to explicitly calculate the HHG
signals for specific experimental parameters, for the linear molecules
$\mathrm{N_{2}}$ and $\mathrm{O_{2}}$, both in the time domain and
in the frequency domain, and discuss the results with reference to
the corresponding experimental observations. In addition, we investigate
in section VII a number of related problems of general interest including
the influence of the probe pulse on the dynamic alignment, the effect
of the initial temperature on the HHG signal, and the {\textit{mean
energy}} of the molecule after interaction with the pump pulse. We
also discuss two other definitions used earlier for the HHG signal,
as well as investigate the case of {}``adiabatic alignment'' of
a linear molecule, within the present theory, in the limit of long
pulse durations. We end with a concluding summary in Sec. VIII.

\begin{figure}[h]
\begin{centering}
\includegraphics[scale=0.5]{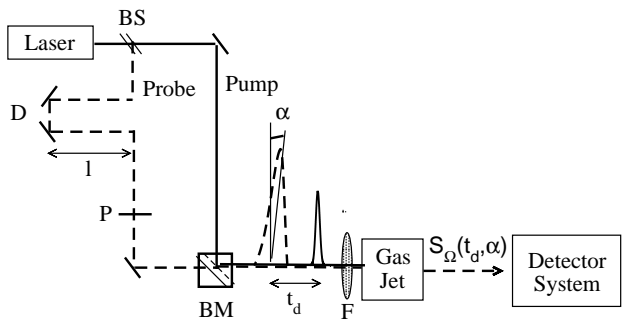} 
\par\end{centering}

\caption{\label{fig:PumpProbe} A scheme of a typical pump-probe experiment.
See text for further explanation. }
\end{figure}

In Fig. \ref{fig:PumpProbe} we show a schematic of a typical intense-field
pump-probe experiment. A laser beam is first split into two parts,
$L_{1}$ and $L_{2}$, by a beam splitter ($BS$) with a desired ratio
of the beam intensities. The probe-pulse $L_{2}$ is delayed by passing
through a delay line system ($D$), by a finite amount $t_{d}$, with
respect to the pump-pulse $L_{1}$ and both are sent through a beam
mixer ($BM$) to the target gas molecules from a gas jet. The high
harmonic signal produced by the probe pulse is recorded by the detector
system for each selected values of $t_{d}$. In addition, a polarizer
$P$ can be inserted to rotate the angle of polarization of the probe
pulse with respect to the polarization direction of the pump pulse
at any desired angle $\alpha$. The pulses are generally assumed to
be effectively non-overlapping ($t_{d}\ne0$) and that they are shorter
than the period of the rotational degrees of freedom of interest.
Note that both $t_{d}$ and $\alpha$ provide controllable parameters
on the high harmonic emission process from the outside.

\section{A Quantum Theory of Intense-Field Pump-Probe Experiments and Molecular
High Harmonic Generation Signals}

\subsection{Total Hamiltonian and Equations of Motions of the Dynamical System}

Within the adiabatic Born-Oppenheimer approximation of the target
molecule, the total Hamiltonian of the system can be written \cite{fai-07}:
(in a.u.: $e=\hbar=m=\alpha c=1$) as, \begin{equation}
H_{tot}(t)=H_{N}+V_{N-L_{1}}(t)+H_{e}+V_{e-L_{2}}\left(t-t_{d}\right)\label{TotHam}\end{equation}
 where $H_{N}$ is the nuclear Hamiltonian, $V_{N-L_{1}}(t)$ is the
interaction due to the pump pulse with the nuclear motion at time
$t$, $H_{e}$ is the electronic Hamiltonian, and $V_{e-L_{2}}\left(t-t_{d}\right)$
is the interaction of the probe pulse with the active electron at
a delay $t_{d}$. We describe the two laser pulses (in the long-wavelength
dipole-approximation) of the laser fields, $\mathrm{F}(\phi_{j}(t))\equiv\mathrm{f}(t)\cos\left(\phi_{j}(t)\right)$
and the corresponding the vector potentials by $\mathrm{A}(\phi_{j}(t))=-\frac{c}{\omega}\mathrm{f}(t)\sin\left(\phi_{j}(t)\right)$,
where $\mathrm{f}(t)$ is the slowly varying envelope of the electric
field (compared to the period of the high harmonics, or the electron
motion). The phase $\phi_{j}(t)$ of the field at the position of
the active electron of the molecule is given by $\phi_{j}(t)=\left(\omega t-\mathrm{k}_{\omega}.\mathrm{X}_{j}\right)$,
where the $C.M.$ of the molecule is assumed to be located at a position
{}``$\mathrm{X}_{j}$''; $\omega$ and $\mathrm{k}_{\omega}$ are
the laser of frequency and the wavenumber, respectively. For the sake
of simplicity of writing, we may suppress the notation the full $\mathrm{X}_{j}$
and $t$ dependence of the phase factor $\phi_{j}(t)$ unless otherwise
needed explicitly, e.g. while summing coherently the {}`` many-molecule''
emission amplitudes from different locations $\left\{ \mathrm{\mathrm{X}_{j}}\right\} $
to obtain the total amplitude associated with the signal macroscopically
coherent signal. It will be found that the coherent signal appears
significantly for the {}``elastic'' scattering (the final state
of the molecule is the same as the initial state) with respect to
the target molecule, and (for an ideal gas medium) along the forward
direction of the incident field (cf. \cite[§4]{bec-05}, and references
cited in that section).

Thus, we may write the laser-molecule interaction Hamiltonians appearing
above as 
given by \begin{equation}
V_{N-L_{1}}(t)=-\mu\cdot\mathrm{F}_{1}(t)-\frac{1}{2}\mathrm{F}_{1}\left(\phi_{j}(t)\right)\bm{:}\alpha\bm{:}\mathrm{F}_{1}\left(\phi_{j}(t)\right)\label{NucHam}\end{equation}
 where $\bm{\mu}$ is the permanent dipole moment (if non-zero) and
$\bm{\alpha}$ with Cartesian components $\alpha_{ii'};\,(i,i')=(1,2,3)$
is the polarizability \textit{tensor} of the molecule (always non-zero);
and\begin{equation}
V_{e-L_{2}}\left(\phi_{j}(t-t_{d})\right)=-\hat{\bm{d}}_{e}\cdot\bm{F}\left(\phi_{j}(t-t_{d})\right)\label{Eleham}\end{equation}
 where $\hat{\bm{d}}_{e}$ stands for the electronic dipole \textit{operator}.

\subsection{Total Wavefunction in Intense-field S-Matrix Theory}

We first consider a systematic solution of the time-dependent Schrödinger
equation of the system \begin{equation}
i\frac{\partial}{\partial t}\Psi(t)=H_{tot}(t)\Psi(t)\label{TotSchEqu}\end{equation}
 using the general technique of intense-field many-body $S$-matrix
theory (IM$S$T) \cite{bec-05}. In this approach the total wavefunction
of the system satisfying a given initial (final) condition can be
written as a series expansion in such a way that the dominant virtual
states, when present, can appear already in the leading terms of the
series. To this end we introduce three partitions of the same total
Hamiltonian, referring to the initial, {}``$i$'', the final, {}``$f$'',
and the (deemed to be relevant) intermediate virtual state, {}``$0$'',
interactions plus the corresponding {}``reference'' Hamiltonians:
\begin{eqnarray}
H_{tot}(t) & = & H_{i}+V_{i}(t)\nonumber \\
 & = & H_{f}(t)+V_{f}(t)\nonumber \\
 & = & H_{0}(t)+V_{0}(t).\label{ParHam}\end{eqnarray}
 It is also useful to define the reference Green's functions associated
with the reference Hamiltonians, $H_{s}(t);\, s\equiv i,f,0$ : \begin{equation}
\left(i\frac{\partial}{\partial t}-H_{s}(t)\right)G_{s}(t,t')=\delta(t-t').\label{RefGreEqu}\end{equation}
 In general, the Green functions can be obtained from the complete
set of the fundamental solutions, $\left|\psi_{j}^{(s)}(t)\right\rangle $
of the Schrödinger equations governed by the reference Hamiltonians
$H_{s}(t);\, s=i,f,0$: \begin{equation}
G_{s}(t,t')=-i\theta(t-t')\sum_{all\,\ j}\left|\psi_{j}^{(s)}(t)\right\rangle \left\langle \psi_{j}^{(s)}(t')\right|\label{RefGreFun}\end{equation}
 The validity of the solutions Eq. (\ref{RefGreEqu}) can be readily
established by operating on the left hand side of Eq. (\ref{RefGreFun})
with $\left(i\frac{\partial}{\partial t}-H_{0}(t)\right)$, using
Eq. (\ref{RefGreEqu}) and the completeness of the fundamental solutions,
$\sum_{j}\left|\psi_{j}^{(s)}(t)\right\rangle \left\langle \psi_{j}^{(s)}(t)\right|=\bm{1}$
and the relation $\frac{\partial}{\partial t}\theta(t-t')=\delta(t-t')$,
to obtain a delta-function integration on the right hand side, followed
by the obvious simplification. Thus, we can express the total wavefunction
of the interacting system, evolving from an arbitrary initial state,
$\left|\chi_{i}(t)\right\rangle $, as a series: \begin{equation}
\left|\Psi(t)\right\rangle =\sum_{j=0}^{\infty}\left|\Psi_{i}^{(j)}(t)\right\rangle \label{KFRWavFun}\end{equation}
 with\begin{equation}
\left|\Psi_{i}^{(0)}(t)\right\rangle =\left|\chi_{i}(t)\right\rangle \label{KFRWavFun0}\end{equation}
 \begin{equation}
\left|\Psi_{i}^{(1)}(t)\right\rangle =\int_{t_{i}}^{t_{f}}dt_{1}G_{f}^{0}\left(t,t_{1}\right)V_{i}\left(t_{1}\right)\left|\chi_{i}(t_{1})\right\rangle \label{KFRWavFun1}\end{equation}
 \begin{eqnarray}
\left|\Psi_{i}^{(2)}(t)\right\rangle  & = & \int_{t_{i}}^{t_{f}}\int_{t_{i}}^{t_{f}}dt_{2}dt_{1}G_{f}^{0}\left(t,t_{2}\right)V_{f}\left(t_{2}\right)G_{0}\left(t_{2},t_{1}\right)\nonumber \\
 &  & \times V_{i}\left(t_{1}\right)\left|\chi_{i}(t_{1})\right\rangle \nonumber \\
 &  & \cdots\cdots\label{KFRWavFun2}\end{eqnarray}
 and

\begin{eqnarray}
\left|\Psi_{i}^{(n)}(t)\right\rangle  & = & \int_{t_{i}}^{t_{f}}.......\int_{t_{i}}^{t_{f}}\int_{t_{i}}^{t_{f}}dt_{n}......dt_{2}dt_{1}G_{f}^{0}\left(t,t_{n}\right)\nonumber \\
 &  & \times V_{f}\left(t_{n}\right).....G_{0}\left(t_{3},t_{2}\right)V_{f}\left(t_{2}\right)G_{0}\left(t_{2},t_{1}\right)\nonumber \\
 &  & \times V_{i}\left(t_{1}\right)\left|\chi_{i}(t_{1})\right\rangle .\label{KFRWavFunn}\end{eqnarray}

\section{Many-Molecule vs. One-Molecule Signals}

\subsection{Transition Amplitudes for High Harmonic Generation}

Emission of a harmonic photon of frequency $\Omega=n\omega$ and wavevector
$\bm{K}_{\Omega}$, from its vacuum state $\left|0_{\Omega}\right\rangle $
(zero occupation number in Fock-space), into a singly occupied number
state, $\left|1_{\Omega}\right\rangle $, is fundamentally a quantum
electrodynamical process i.e. due to the interaction of the active
electron with the vacuum-field albeit in the presence of the intense
external laser field. Its theoretical formulation therefore clearly
requires one to consider at least the combined state of the interacting
{}``laser field (semiclassical) + molecule + vacuum-field''- system
in the extended space consisting of the direct product of the ordinary
space of {}``laser field (semiclassical) + molecule'' and the occupation
number space of the vacuum and the emitted photon (cf. e.g. \cite[§4.5]{bec-05}).
Nevertheless, exactly the same result for the single photon HHG emission
amplitude can also be obtained using the ordinary quantum mechanics,
simply by taking the quantum electrodynamically normalized interaction
$V^{*}(t)$ for the spontaneous emission of a photon of frequency
$\Omega$ and wavevector $\bm{K_{\Omega}}$ (cf. \cite[Lecture 2]{fey-ed,sak-qm}):
\begin{equation}
V^{*}(t)=N_{\Omega}e^{i\Phi_{j}(t)}\bm{\epsilon}_{\Omega}\cdot\hat{\bm{d}}_{e}\label{SakInt}\end{equation}
In the above, $N_{\Omega}\equiv\sqrt{\frac{2\pi\hbar\Omega}{L^{3}}}$,
$L^{3}$ is the quantization volume, $\bm{\epsilon_{\Omega}}$ is
the polarization vector of the emitted photon, and $\bm{d}_{e}$ is
the usual electronic transition dipole operator; the phase $\Phi_{j}(t)=\left(\Omega t-\bm{k}_{\omega}\cdot\bm{X}_{j}\right)$.
As usual in the present dipole approximation, we have neglected the
retardation factor, $e^{-i\bm{K}_{\Omega}\cdot{\bm{r}}}\approx1$;
we may note explicitly that the exact position of the electron with
respect to an arbitrary coordinate origin is given by $\bm{X}_{j}+\bm{r}$,
where as before $\bm{X}_{j}$ is the C.M. of the $j$th molecule and
$\bm{r}$ is the position of the electron with respect to the C.M.
of the molecule. %
\begin{figure}
\begin{centering}
\includegraphics[scale=0.3]{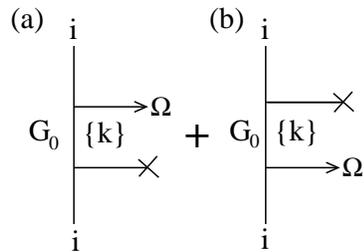} 
\par\end{centering}

\caption{\label{fig:FeyDia} Quantum amplitude for coherent emission of a
high harmonic photon (frequency $\Omega$) is the sum of a direct
(a) and a time-reversed (b) diagram.}
\end{figure}

The HHG amplitude for the emission of a harmonic frequency $\Omega$
from the $j$th molecule is given by (cf. \cite[§4]{bec-05}) by the
sum of two {}``Feynman-like'' diagrams, (a) and (b), shown in Fig.
\ref{fig:FeyDia}. The diagram (a) corresponds to the so-called {}``direct''
amplitude (associated with the retarded Green's function), whereas
the diagram (b) corresponds with the {}``time-reversed'' amplitude
(associated with the advanced Green's function). The amplitude for
the harmonic emission process can be written down analytically from
diagram (a) by reading in the forward (upward) direction of time:
First, the molecule is prepared by the {}``pump'' laser pulse in
the state $i\equiv\left|\Phi_{i}(t)\right\rangle $. Next, the molecule
interacts with the probe laser field (short horizontal line with a
cross), then it propagates through $G_{0}=G_{0}^{(+)}$ (vertical
line). Next it interacts with the vacuum-field by $V^{(*)}$ (horizontal
line ending in $\Omega$), and emits the harmonic photon of frequency
$\Omega$, and finally returns to the ${\textit{same}}$ state $i\equiv\left|\Phi_{i}(t)\right\rangle $
as before. A similar interpretation holds for the time reversed diagram
(b) except that here the system evolves backward in time thorough
$G_{0}=G^{(-)}$. The quantum amplitude $A_{j}(\Omega)$, for the
emission of the HHG photon of frequency $\Omega$, and wavenumber
$\bm{K_{\Omega}}$, from the molecule located at $\bm{X}_{j}$, is
given by the {\textit{sum}} of the two diagrams: \begin{equation}
A_{j}(\Omega)=diag.(a)+diag.(b)\label{SumHhgAmp0}\end{equation}
 Mathematically, we have \begin{eqnarray}
diag.(a) & = & -i\int_{-\infty}^{\infty}dt\int_{-\infty}^{\infty}dt'\left\langle \phi_{i}(t)\right|V^{*}(t)\nonumber \\
 &  & \times G_{0}^{(+)}(t,t')V_{e-L_{2}}(t'-t_{d})\left|\chi_{i}(t')\right\rangle \label{DirHhgAmp}\end{eqnarray}
 and, \begin{eqnarray}
diag.(b) & = & -i\int_{-\infty}^{\infty}dt\int_{-\infty}^{\infty}dt'\left\langle \phi_{i}(t')\right|V_{e-L_{2}}\left(t'-t_{d}\right)\nonumber \\
 &  & \times G_{0}^{(-)}(t',t)V^{*}(t)\left|\chi_{i}(t)\right\rangle \label{TimRevHhgAmp}\end{eqnarray}

\subsection{A Relation between HHG Amplitude and $F.T.$ of Dipole Expectation
Value}

\subsubsection{Recasting the quantum HHG amplitude}

Before proceeding further, we consider the relation between the quantum
HHG amplitude, Eq. (\ref{SumHhgAmp0}), and the expectation value
of the electric dipole operator, $\hat{\bm{d}}_{e}$, that is popularly
used for calculations of HHG signals. To this end we first rewrite
the quantum amplitude Eq. (\ref{DirHhgAmp}) by introducing the first
order wavefunction $\Psi^{(1)}$ (cf. Eq. (\ref{KFRWavFun1})) that
arises from the initial state $\left|\chi_{i}(t)\right\rangle $ due
to the interaction with the probe pulse: \begin{eqnarray}
\left|\Psi^{(1)}(t)\right\rangle  & = & \int_{-\infty}^{\infty}dt'G_{0}^{(+)}(t,t')\nonumber \\
 &  & \times V_{e-L_{2}}(t'-t_{d})\left|\chi_{i}(t')\right\rangle \label{KFRWavFun1-a}\end{eqnarray}
 We may first rewrite Eq. (\ref{DirHhgAmp}) as: \begin{eqnarray}
diag.(a) & = & -i\int_{-\infty}^{\infty}dt\left\langle \chi_{i}(t)\left|V^{*}(t)\right|\Psi^{(1)}(t)\right\rangle \nonumber \\
 & = & -i\int_{-\infty}^{\infty}dte^{i(\Omega t-\bm{K}_{\Omega}\cdot\bm{X}_{j})}\nonumber \\
 &  & \times\left\langle \chi_{i}(t)\left|N_{\Omega}\bm{\epsilon}_{\Omega}\cdot\hat{\bm{d}}_{e}\right|\Psi^{(1)}(t)\right\rangle \label{DirHhgAmpPri}\end{eqnarray}
 where we have used the explicit form of the interaction $V^{*}(t)$.
Next we rewrite Eq. (\ref{TimRevHhgAmp}) using a standard relation
satisfied by the advanced and the retarded Green functions (e.g. \cite{sak-qm}):
\begin{equation}
G_{0}^{(-)}(t',t)=\left[G_{0}^{(+)}(t,t')\right]^{*}\label{RelGreFun}\end{equation}
 We also note that the laser-molecule interaction is real (Hermitian),
i.e. \begin{equation}
V_{e-L_{2}}(t)=\left[V_{e-L_{2}}(t)\right]^{*}\label{HerInt}\end{equation}
 Thus, the integral over $dt'$ in Eq. (\ref{TimRevHhgAmp}) can be
rewritten as, \begin{eqnarray}
diag.(b) & = & -i\int_{-\infty}^{\infty}dt\int_{-\infty}^{\infty}dt'\left\langle \chi_{i}(t')\right|V_{e-L_{2}}(t'-t_{d})\nonumber \\
 &  & \times G_{0}^{(-)}(t',t))V^{*}(t)\left|\chi_{i}(t)\right\rangle \nonumber \\
 & = & -i\int_{-\infty}^{\infty}dt\int_{-\infty}^{\infty}dt'\nonumber \\
 &  & \times\left[G_{0}^{(+)}(t,t')V_{e-L_{2}}(t'-t_{d})\left|\chi_{i}(t')\right\rangle \right]^{*}\nonumber \\
 &  & \times V^{*}(t)\left|\chi_{i}(t)\right\rangle \nonumber \\
 & = & -i\int_{-\infty}^{\infty}dt\left\langle \Psi^{(1)}(t)\left|V^{*}(t)\right|\chi_{i}(t)\right\rangle \nonumber \\
 & = & \int_{-\infty}^{\infty}dte^{i(\Omega t-\bm{K}_{\Omega}\cdot\bm{X}_{j})}\nonumber \\
 &  & \times\left\langle \Psi^{(1)}(t)\left|N_{\Omega}\bm{\epsilon}_{\Omega}\cdot\hat{\bm{d}}_{e}\right|\chi_{i}(t)\right\rangle \label{TimRevHhgAmpPri}\end{eqnarray}
 Hence, adding Eqs. (\ref{DirHhgAmpPri}) and (\ref{TimRevHhgAmpPri}),
we get the quantum HHG amplitude in the suggestive form: \begin{eqnarray}
A_{j}(\Omega) & = & -iN_{\Omega}\bm{\epsilon}_{\Omega}\cdot\int_{-\infty}^{\infty}dte^{i(\Omega t-\bm{K}_{\Omega}\cdot\bm{X}_{j})}\nonumber \\
 &  & \times\left\{ \left\langle \chi_{i}(t)\left|\hat{\bm{d}}_{e}\right|\Psi^{(1)}(t)\right\rangle +\left\langle \Psi^{(1)}(t)\left|\hat{\bm{d}}_{e}\right|\chi_{i}(t)\right\rangle \right\} \nonumber \\
 & = & -iN_{\Omega}\bm{\epsilon}_{\Omega}\cdot\int_{-\infty}^{\infty}dte^{i(\Omega t-\bm{K}_{\Omega}\cdot\bm{X}_{j})}\nonumber \\
 &  & \times\left\{ \left\langle \chi_{i}(t)\left|\hat{\bm{d}}_{e}\right|\Psi^{(1)}(t)\right\rangle +c.c.\right\} \label{HhgAmpPri}\end{eqnarray}
 where {}``c.c.'' stands for the complex conjugate.

\subsubsection{Dipole expectation value}

The expectation value of the dipole operator, $\bm{D}_{i,i}(t)$,
of the transition dipole operator can be calculated within the lowest
order KFR approximation \cite{kel-64,fai-73,rei-80} of the wavefunction
of the system as follows: \begin{eqnarray}
\bm{D}_{i,i}(t) & = & \left\langle \Psi(t)\left|\hat{\bm{d}}_{e}\right|\Psi(t)\right\rangle \nonumber \\
 & = & \left\langle \chi_{i}(t)\left|\hat{\bm{d}}_{e}\right|\chi_{i}(t)\right\rangle \nonumber \\
 &  & +\left\{ \left\langle \chi_{i}(t)\left|\hat{\bm{d}}_{e}\right|\Psi^{(1)}(t)\right\rangle \right.\nonumber \\
 &  & \left.+\left\langle \Psi^{(1)}(t)\left|\hat{\bm{d}}_{e}\right|\chi_{i}(t)\right\rangle \right\} +\cdots\nonumber \\
 & = & \left\{ \left\langle \chi_{i}(t)\left|\hat{\bm{d}}_{e}\right|\Psi^{(1)}(t)\right\rangle +c.c.\right\} +\cdots\label{DipExpVal}\end{eqnarray}
 Note that the zeroth order term in the first line above vanishes
for centrosymmetric systems; we have also neglected the quadratic
powers of the first order KFR-correction (and the higher order terms).

Combining Eq. (\ref{DipExpVal}) with Eq. (\ref{HhgAmpPri}), the
quantum HHG amplitude can be expressed in the form \begin{equation}
A_{j}(\Omega)=-i\, e^{-i\bm{K}_{\Omega}\cdot\bm{X}_{j}}N_{\Omega}\bm{\epsilon}_{\Omega}\cdot\int_{-\infty}^{\infty}dt\, e^{i\Omega t}\bm{D}_{i,i}(t)\label{HhgAmp-1}\end{equation}
 Thus, the quantum HHG amplitude is clearly {\textit{proportional}}
to the Fourier transform ($F.T.$) of the expectation value of the
transition dipole operator, Eq. (\ref{DipExpVal}). We may recall
that the proportionality constant $N_{\Omega}$ above is of quantum
electrodynamical origin and can {\textit{not}} be derived from the
classical electrodynamics alone \cite{jac-62}.

In practice, the $F.T.$ of interest can be conveniently obtained
by Fast Fourier transform (or $FFT$) numerically \cite{pre-nr}.
Alternatively, for {}``slowly varying'' pulse envelopes (compared
to the high harmonic frequency) one may express the \emph{F.T.} of
$\bm{D}_{i,i}(t)$ as a Fourier series \cite{foot-10}: \begin{equation}
\bm{D}_{i,i}(t)=\sum_{n}e^{-in(\omega t-\bm{k}_{\omega}\cdot\bm{X}_{j})}\tilde{\bm{D}}(n\omega)\label{FouTraDipExpVal}\end{equation}
 where $\tilde{\bm{D}}(n\omega)$ is the \emph{F.T.} evaluated at
the $n$th harmonic frequency $\Omega=n\omega$. Thus, in terms of
the $F.T.$ components, the HHG amplitude $A_{j}(\Omega)$ becomes,
\begin{eqnarray}
A_{j}(\Omega) & = & \sum_{n}-2\pi i\delta(\Omega-n\omega)e^{-i(\bm{K}_{\Omega}-n\bm{k}_{\omega})\cdot\bm{X}_{j}}\nonumber \\
 &  & \times N_{\Omega}\bm{\epsilon}_{\Omega}\cdot\tilde{\bm{D}}_{i,i}(n\omega)\label{HhgAmp-2}\end{eqnarray}
 where we have carried out the time integration over $dt$, in terms
of the Dirac delta-function.

\subsection{Coherent Sum of HHG Amplitudes: Many-molecule vs. One-molecule Signal}

It is interesting also to consider the {\textit{total}} amplitude
$A_{tot.}(\Omega)$ of HHG emission from all the molecules interacting
with the (probe laser) field. This is given by the \textit{coherent}
sum of the individual amplitudes emitted by the molecules at the positions
${\bm{X}_{j}}$ for all $j=1,2,3,\cdots{\cal {N}}$, where ${\cal {N}}$
is the number of molecules in the interaction volume, or \begin{eqnarray}
A_{tot.}(\Omega) & \equiv & \sum_{j=1}^{{\cal {N}}}A_{j}(\Omega)\nonumber \\
 & = & \sum_{n}\left\{ \sum_{j=1}^{{\cal {N}}}e^{-i(\bm{K}_{\Omega}-n\bm{k}_{\omega})\cdot\bm{X}_{j}}\right\} _{1}\nonumber \\
 &  & \times\left\{ -2\pi i\sum_{n}\delta(\Omega-n\omega)T_{i,i}(\Omega)\right\} _{2}\label{SumHhgAmp}\end{eqnarray}
 where we may identify the basic HHG transition matrix element for
the emission of the $n$th harmonic per molecule as, \begin{eqnarray}
T_{i,i}(\Omega) & = & N_{\Omega}\bm{\epsilon}_{\Omega}\cdot\tilde{\bm{D}}_{i,i}(\Omega)\label{TraMat}\end{eqnarray}
 It can be seen from Eq. (\ref{SumHhgAmp}) that the $n$th harmonic
emission amplitude in fact factorizes into two parts, the first factor
corresponds to the sum of the macroscopic space dependent phases associated
with the random positions $\bm{X}_{j}$ of the $C.M.$s of the molecules
in the interaction volume, and the second factor corresponds to the
fundamental {}``one molecule'' emission amplitude, independent of
the position of the $C.M.$s of the molecules. The macroscopic phase
factor is explicitly given by \begin{equation}
\left\{ ...\right\} _{1}\equiv\left\{ \sum_{j=1,{\cal {N}}}e^{-i(\bm{K}_{\Omega}-n\bm{k}_{\omega})\cdot\bm{X}_{j}}\right\} _{1}\label{PhaSum}\end{equation}
 For a large number of molecules in the interaction volume, ${\cal {N}}\gg1$,
the phase factor oscillates greatly and thus tends to average out
to zero, \textit{except} when the condition, \begin{equation}
(\bm{K}_{\Omega}-n\bm{k}_{\omega})=0\label{MomCon}\end{equation}
 is fulfilled; in that case it yields the phase sum $\left\{ ...\right\} _{1}={\cal {N}}$.
It is readily understood that the condition (\ref{MomCon}) corresponds
exactly to the momentum conservation between the final momentum of
the emitted harmonic photon, $\hbar\bm{K}_{\Omega}$, and the sum
of the momenta of $n$ laser photons, $n\hbar\bm{k}_{\omega}$. This
is the phase-matching condition in the forward direction \cite{foot-20}.

The probability of emission of the harmonics is given as usual by
the absolute square of the total amplitude Eq. (\ref{SumHhgAmp}).
Under the phase-matching condition the latter is therefore coherently
\textit{amplified} by a (generally large) factor of ${\cal {N}}^{2}$.
This is also the origin of the \textit{quadratic} pressure dependence
of the high harmonic signals, as well as their unusual strengths,
that had been found in the very first experimental observations (e.g.
\cite{fer-87,rho-87}).

The second factor $\{...\}_{2}$ gives the fundamental {}``one-molecule''
quantum emission amplitude. We also note that if the absolute probability
of the harmonic emission is needed then the proportionality factor
$N_{\Omega}=\sqrt{\frac{2\pi\hbar\omega}{\L^{3}}}$ becomes essential,
and that for a given polarization direction of the emitted photon,
$\bm{\epsilon}_{\Omega}$, the projection of the dipole expectation
value must be taken in that direction.

\subsection{Continuous Medium and the Phase-matching Function}

If one assumes that the gas molecules are distributed effectively
continuously with a distribution function ${\cal {N}}\rho(\bm{R})d^{3}R$,
where $\rho(\bm{R})$ is the so-called {}``density function per molecule'',
then one may replace the sum over $j$ in Eq. (\ref{PhaSum}) by the
integration over the interaction volume. Clearly, in this case the
square of the macroscopic phase factor, $\left|[...]_{1}\right|^{2}$,
takes the form \begin{equation}
\left|[...]_{1}\right|^{2}={\cal {N}}^{2}{\cal {F}}\left(\bm{K}_{\Omega}-n\bm{k}_{\omega}\right)\label{PhaMatFac}\end{equation}
 where, \begin{equation}
{\cal {F}}\left(\bm{K}_{\Omega}-n\bm{k}_{\omega}\right)=\left|\int d^{3}X\rho(\bm{\bm{X}})e^{i\left(\bm{K}_{\Omega}-n\bm{k}_{\omega}\right)\cdot\bm{X}}\right|^{2}\label{PhaMatFun}\end{equation}
 which is the so-called phase-matching function. It peaks for its
argument near zero (near the forward direction), but falls off rapidly
away from it.

\subsection{\label{elastic}Coherent Elastic vs. Incoherent Inelastic Transitions}

Eqs. (\ref{MomCon}) and (\ref{SumHhgAmp}) show, respectively, that
both the phase matching condition (momentum conservation) and the
frequency matching condition (energy conservation) in the process
ought to be fulfilled {\textit{simultaneously}} in order that the
macroscopic signal to be coherently amplified in space and time. As
already noted earlier, the former condition leads to the directional
coherence (forward propagation) of the HHG emission, while the latter
implies the \textit{elastic} nature of the accompanying molecular
transitions for which the final ({}``recombination'') state $f$
of the molecular system is the {\textit{same}} as the initial state
$i$, with $E_{i}=E_{f}$. In contrast, for an {\textit{inelastic}}
transition, $i\rightarrow f$, when $E_{i}\ne E_{f}$, there would
be in general only {}``hyper-Raman'' emissions, with frequencies
$\Omega_{if}=\left(n'\omega-\left|E_{i}-E_{f}\right|\right)$, that
are generally incommensurate with the incident laser frequency or
its multiple, or the HHG frequency, $\Omega=n\omega$. Thus the non-vanishing
relative phase difference $\Delta\phi\equiv\left(\left|E_{i}-E_{f}\right|-n\omega\right)t$
would fail to stimulate the hyper-Raman transitions by the incident
field, unlike the stimulated spontaneous nature of the associated
HHG. Also the non-vanishing momentum difference between hyper-Raman
radiation and the the multiple of the laser photons $\bm{K}_{\Omega_{if}}-n\bm{k}_{\omega}\ne0$
makes the former macroscopically and directionally incoherent.

Finally, we note that the probability of the electronically inelastic
processes associated with the transitions into the continuum (e.g.
ionization) or between continua (e.g. inverse Bremsstrahlung \cite{ehl-01})
that are commensurate with the emission of the $n$th harmonic at
the {}``one-molecule'' level, will be incoherent spatially, and
therefore would enhance only proportional to the total number of molecules,
${\cal {N}}$, in the interaction volume. This is in stark contrast
to the coherent amplification of the HHG emission at the $n$th harmonic,
that is proportional to ${\cal {N}}^{2}$.

\subsection{Differential Rate of Coherent High Harmonic Generation}

To derive the explicit expression for the probability of HHG per unit
time i.e. the \textit{rate} of generation of coherent high harmonics,
we take the absolute square of the total HHG amplitude Eq. (\ref{SumHhgAmp})
and divide by the long observation time $T$, use a useful representation
of the square of the delta-function (\cite[p. R12]{bec-05}),\begin{equation}
\delta^{2}(\Omega-n\omega)=\lim_{T\rightarrow\infty}\frac{T}{2\pi}\delta(\Omega-n\omega),\label{DelSqu}\end{equation}
 and sum over the emitted photon modes (with $\sum_{\bm{K}_{\Omega}}\equiv L^{3}\int d{\hat{\bm{K}}_{\Omega}}\int dK_{\Omega}K_{\Omega}^{2}$)
and get:\begin{eqnarray}
W(\Omega) & = & \lim_{T\rightarrow\infty}\sum_{\bm{K}}\frac{\left|A_{tot.}(\Omega)\right|^{2}}{T}\nonumber \\
 & = & {\cal {N}}^{2}\sum_{n}\int{d\hat{\bm{K}}_{\Omega}}{\cal {F}}\left({\bm{K}}_{\Omega}-n{\bm{k}}_{\omega}\right)\nonumber \\
 &  & \times\int dW(n\omega)\end{eqnarray}
 where,\begin{eqnarray}
dW(n\omega) & = & 2\pi\delta(\Omega-n\omega)L^{3}\nonumber \\
 &  & \times\left|N_{\Omega}\bm{\epsilon}_{\Omega}\cdot\tilde{\bm{D}}_{i,i}(n\omega)\right|^{2}{K}_{\Omega}^{2}d{K}_{\Omega}\end{eqnarray}
 is the differential rate of HHG per molecule. Noting that the main
contribution arises from the phase matching condition along the forward
direction, we may carry out the mode-integrations to get:\begin{equation}
W(n\omega)=2\pi\left|T_{i,i}^{(n)}\right|^{2}\frac{(n\omega)^{2}}{c^{3}}\label{OneMolHhgRat}\end{equation}
 where we have used, $K_{\Omega}\equiv\frac{\Omega}{c}$, $\bm{K}_{\Omega}\equiv K_{\Omega}\hat{\bm{K}}_{\Omega}$,
and the fundamental transition matrix element for the emission of
the $n$th harmonic, $T_{i,i}^{(n)}$, is given in terms of the $F.T.$
of the dipole expectation value $\tilde{\bm{D}}_{i,i}(n\omega)$ by:
\begin{equation}
T_{i,i}^{(n)}=\sqrt{2\pi(n\omega)}\bm{\epsilon}_{\Omega}\cdot\tilde{\bm{D}}_{i,i}(n\omega)\label{TraMatPri}\end{equation}
 for, $L^{3}\left|N_{\Omega}\right|^{2}={2\pi(n\omega)}$.

\section{Evaluation of {}``One Molecule'' HHG Amplitude}

Clearly the dynamical properties of the HHG signal are given by the
rate of HHG emission per molecule, Eq. (\ref{OneMolHhgRat}), while
the total signal is the same to within a proportionality constant
given by the square of the number of molecules in the interaction
volume, ${\cal {N}}^{2}$, and the phase-matching constant $\int{d\hat{\bm{K}}_{\Omega}}{\cal {F}}\left({\bm{K}}_{\Omega}-n{\bm{k}}_{\omega}\right)$
that peaks in the forward direction. We therefore proceed to evaluate
the dynamical signal per molecule (in a relative scale) as follows:
(i) solve the Schrödinger equation for the nuclear and the electronic
motions of the interacting laser-molecule system, (ii) construct a
complete set of orthonormal reference states, $\left|i\right\rangle \equiv\left|\chi_{i}(t)\right\rangle $,
of the molecule, created by the pump pulse, (iii) determine their
statistical weights according to the \textit{one-to-one} correspondence
with the thermally occupied rotational eigenstates of the ensemble,
(iv) calculate the {}``one molecule'' probability amplitude for
HHG for each member of the ensemble of linearly independent reference
states $\left|i\right\rangle $, using Eq. (\ref{OneMolHhgRat}),
and finally, (v) obtain the (scaled) signal {}``per molecule'' by
thermally averaging the \textit{probabilities} of HHG emission from
each member of the ensemble of the reference states, using the distribution
of their statistical weights.

In the Born-Oppenheimer approximation and non-overlapping pump and
probe pulse condition, we may consider the evolution of the wavefunctions
of the nuclear and the electronic parts separately and combine them
together to obtain the wavefunction of the interacting system to evaluate
the transition matrix elements of interest.

\subsection{Pump Pulse Interaction and Rotational Wavepackets as Reference States}

The nuclear rotational motion under the action of the pump pulse is
determined by the Schrödinger equation governed by the partial Hamiltonian
\begin{equation}
H_{N}(t)+V_{N-L1}(t),\label{HamNucMot}\end{equation}
 i.e. \begin{equation}
i\frac{\partial}{\partial t}\Phi_{JM}(t)=\left(H_{N}+V_{N-L_{1}}(t)\right)\Phi_{JM}(t).\label{NucSchEqu}\end{equation}
 We first construct the fundamental set of linearly independent solutions
of Eq. (\ref{NucSchEqu}), each evolving independently from each of
the occupied rotational eigenstates $\left\{ \left|J_{0}M_{0}\right\rangle \right\} $.
We expand it on the basis of the eigenstates $\left\{ \left|JM\right\rangle \right\} $,
as \begin{equation}
\Phi_{J_{0}M_{0}}(t)=\sum_{JM}C_{JM}^{(J_{0}M_{0})}(t)\left|JM\right\rangle e^{-iE_{JM}t}\label{RotWavPac}\end{equation}
 The coefficients $C_{JM}^{(J_{0}M_{0})}(t)$ satisfy the system of
coupled linear differential equations \begin{equation}
i\frac{\partial}{\partial t}C_{JM}^{(J_{0}M_{0})}(t)=\sum_{J'M'}\left\langle JM\left|V_{N-L_{1}}(t)\right|J'M'\right\rangle C_{J'M'}^{(J_{0}M_{0})}(t)\label{CoeEqu}\end{equation}
 This set of equations can be easily obtained (e.g. \cite{sei-01})
by projecting on a given eigenstate from the left. In practice we
obtain the set of the fundamental solutions $\left|\Phi_{J_{0}M_{0}}\right\rangle $
by numerical integration using the well-known Runge-Kutta method \cite{pre-nr},
starting with the following independent initial conditions: \begin{equation}
C_{JM}^{(J_{0}M_{0})}\left(t_{i}\right)=\delta_{J,J_{0}}\delta_{M,M_{0}}\end{equation}
 We may note explicitly here that (a) each independent wavepacket-state
$\left|\Phi_{J_{0}M_{0}}(t)\right\rangle $ evolves in \textit{one-to-one}
correspondence with the initially occupied rotational eigenstate $\left|J_{0}M_{0}\right\rangle $.
Taken together they form a complete set of orthonormal rotational
wavepacket-states (linear superposition of rotational eigenstates):\begin{equation}
\sum_{J_{0}M_{0}}\left|\Phi_{J_{0}M_{0}}(t)\right\rangle \left\langle \Phi_{J_{0}M_{0}}(t)\right|=\bm{1}\label{RotComSet}\end{equation}
 In general a gas jet of molecules in a pump-probe experiment at a
finite temperature $T$, is not in a pure quantum state but rather
is in a state of thermal mixture of of the rotational eigenstates,
$\left\{ \left|J_{0}M_{0}\right\rangle \right\} $. We therefore introduce
the quantum statistical mechanical device of a hypothetical ensemble
of mutually independent and identical reference molecules, each of
which occupies the electronic ground state and the rotational eigenstates
$\left\{ \left|J_{0}M_{0}\right\rangle \right\} $, the latter with
statistical weights $\rho\left(J_{0}M_{0}\right)$, given by the Boltzmann
distribution: \begin{equation}
\rho\left(e,J_{0}M_{0}\right)=(1)_{e}\times Z_{P}e^{-E_{J_{0}M_{0}}/kT},\end{equation}
 where \begin{equation}
Z_{P}=\sum_{J_{0}}\left(2J_{0}+1\right)e^{-E_{J_{0}}/kT}\end{equation}
 is the rotational partition function; $E_{J_{0}.M_{0}}=J_{0}\left(J_{0}+1\right)hBc$,
for all $M_{0}$; $B$ stands for the rotational constant. We shall
assume for the present purpose that the pump pulse is not too strong
so that the change in the occupation probability of the ground electronic
state after the pump pulse interaction is negligible and hence the
ground electronic state at a time $t$ before the interaction with
the probe pulse evolves simply to $\left|\phi_{e}(t)\right\rangle =e^{-iE_{e}t}\left|\phi_{e}(0)\right\rangle $,
where $E_{e}$ is the ground state energy. (We may assume that electronically
only the ground electronic state $\left|\phi_{e}(0)\right\rangle $
is occupied initially). Thus, the linearly independent reference states
of the molecule, after the interaction of the pump pulse and immediately
before the interaction with the probe pulse, can be written as the
direct product of the nuclear rotational wavepacket states and the
electronic ground state: \begin{equation}
\left|\chi_{i}(t)\right\rangle \equiv\left|\Phi_{J_{0}M_{0}}(t)\right\rangle \left|\phi_{e}(t)\right\rangle \,\,\,\,\, i\equiv\left(e,J_{0}M_{0}\right).\label{RefSta}\end{equation}
 The reference density matrix describing the molecular ensemble prepared
by the pump pulse takes the form: \begin{eqnarray}
{\rho}_{mol}(e,J_{0}M_{0}) & = & \sum_{i}\left|\chi_{i}(t)\right\rangle \rho\left(e,J_{0}M_{0}\right)\left\langle \chi_{i}(t)\right|\nonumber \\
 & = & \left|\phi_{e}(t)\right\rangle \left|\Phi_{J_{0}M_{0}}(t)\right\rangle \rho\left(j_{0}M_{0}\right)\nonumber \\
 &  & \times\left\langle \Phi_{J_{0}M_{0}}(t)\right|\left\langle \phi_{e}(t)\right|\label{MolDenMat}\end{eqnarray}
 where, $i\equiv\left(e,J_{0}M_{0}\right)$. The above ensemble of
molecular states describes the effective {}``initial'' condition
of the system after the pump pulse, when the probe pulse arrives at
the molecular at $\bm{X}_{j}$. To avoid any possible confusion regarding
the presence of the {}``mixed-state'' of the ensemble, and the {}``rotational
coherence'', we may already point out explicitly that while the ensemble
is characterized by the statistical occurrence of the orthonormal
reference states $\left\{ \left|\chi_{i=e,J_{0}M_{0}}(t)\right\rangle \right\} $,
each one of these states carries the information of the rotational
coherence induced by the pump pulse, as coded in the individual rotational
wavepackets $\left\{ \left|\Phi_{J_{0}M_{0}}(t)\right\rangle \right\} $.
Thus, when the thermal average of the HHG emission signal must be
taken with respect to the \textit{probability} of emission from each
member of the ensemble (as required by quantum statistical mechanics),
it can not, and will not, wash out the rotational coherence that is
present within each of them individually.

\subsection{Interaction with Probe-Pulse and Evolution of the Electronic State}

To proceed further, we next consider the evolution of the electronic
state, governed by the partial Hamiltonian \begin{equation}
H_{e}+V_{e-L_{2}}\left(t-t_{d}\right).\end{equation}
 It is obtained conveniently from the knowledge of the electronic
Green's function $G_{e}(t,t')$ \cite{fai-08} associated with the
above Hamiltonian and defined by the inhomogeneous equation: \begin{equation}
\left\{ i\frac{\partial}{\partial t}-\left(H_{e}+V_{e-L_{2}}\left(t-t_{d}\right)\right)\right\} G_{e}(t,t')=\delta(t-t')\bm{1}.\label{EleGreEqu}\end{equation}
 A solution of the above equation can be written as (in the strong-field
KFR-approximation): \begin{eqnarray}
G_{e}\left(t,t'\right) & = & -i\theta\left(t-t'\right)\sum_{j,\bm{p}}\left|\phi_{j}^{(+)}\right\rangle e^{-iE_{j}^{+}t}\left|\bm{p}\left(t-t_{d}\right)\vphantom{\phi_{j}^{(+)}}\right\rangle \nonumber \\
 &  & \times e^{-\frac{i}{2}\int_{t'-t_{d}}^{t-t_{d}}p^{2}(u)du}\nonumber \\
 &  & \times\left\langle \bm{p}\left(t'-t_{d}\right)\vphantom{\phi_{j}^{(+)}}\right|e^{iE_{j}^{+}t'}\left\langle \phi_{j}^{(+)}\right|\label{EleGreFun}\end{eqnarray}
 where $j$ runs over all the ionic electronic states $\left|\phi_{j}^{(+)}\right\rangle $,
with eigenvalues $E_{j}^{+}$, of the molecular ion and $\boldsymbol{p}$
is the free momentum of the electron; $\bm{p}(t)$ stands for the
instantaneous momentum in the presence of the field, defined as $\bm{p}(t)\equiv\left(\bm{p}+\frac{\bm{A}(t)}{c}\right)$.
The validity of Eq. (\ref{EleGreFun}) (within the Born-Oppenheimer
and KFR approximation) can be verified by substituting it in Eq. (\ref{EleGreEqu})
and using the completeness relation \begin{equation}
\sum_{\bm{p}}\left\langle \bm{r}\left|\phi_{\bm{p}}(t)\right\rangle \left\langle \phi_{p}(t)\right|\bm{r}\right\rangle =\bm{1}\label{VolComSet}\end{equation}
 of the of Volkov wavefunctions defined by: \begin{equation}
\left\langle \bm{r}\left|\phi_{\bm{p}}(t)\right.\right\rangle =e^{i\bm{p}(t)\cdot\bm{r}}e^{-\frac{i}{2}(\int^{t}(p^{2}(u))du)}\label{VolWavFun}\end{equation}
 as well as the completeness relation of the ionic states \begin{equation}
\sum_{j}\left|\phi_{j}^{(+)}\right\rangle \left\langle \phi_{j}^{(+)}\right|=1\label{IonComSet}\end{equation}
 We should note that the ionic states are generally much more tightly
bound than the active electron in the highest occupied molecular orbital
(HOMO). Thus in deriving $G_{e}$ above, we have further neglected
the change in the ionic states due the interaction with the probe
pulse, which we may refer to as {}``bare-ion'' approximation.

Finally, using Eqs. (\ref{RotComSet}) and (\ref{EleGreFun}) we obtain
the total Green's function $G_{0}(t,t')$ of the interacting system:
\begin{eqnarray}
G_{0}(t,t') & = & -i\theta(t-t')\sum_{j\bm{p}JM}\left|\phi_{j}^{(+)}\right\rangle \left|\phi_{\bm{p}}(t-t_{d})\right\rangle \nonumber \\
 &  & \times\left|\Phi_{JM}(t)\vphantom{\phi_{j}^{(+)}}\right\rangle e^{-iE_{j}^{+}(t-t')}\left\langle \Phi_{JM}(t')\vphantom{\phi_{j}^{(+)}}\right|\nonumber \\
 &  & \times\left\langle \phi_{\bm{p}}(t'-t_{d})\right|\left\langle \phi_{j}^{(+)}\right|\label{TotGreFun}\end{eqnarray}
The above Green's function of the system (\ref{TotGreFun}) therefore
holds 
under (a) the adiabatic Born-Oppenheimer, (b) the strong-field KFR
and (c) the {}``bare-ion'' approximations.

\subsection{The Total Wavefunction in Strong-Field Molecular KFR-approximation}

Combining the Eqs. (\ref{KFRWavFun1-a}) and (\ref{RefSta}), we obtain
the total intense-field molecular wavefunction at the lowest order
strong field KFR-approximation: \begin{eqnarray}
\left|\Psi_{i}(t)\right\rangle  & = & \left|\chi_{i}(t)\right\rangle +\int_{-\infty}^{\infty}dt'\nonumber \\
 &  & \times G_{0}(t,t')V_{e-L_{2}}(t'-t_{d})\left|\chi_{i}(t')\right\rangle \label{KFRWavFun1-b}\end{eqnarray}
 where, \begin{equation}
\left|\chi_{i}(t)\right\rangle =\left|\phi_{e}(t)\right\rangle \left|\Phi_{J_{0}M_{0}}(t)\right\rangle \label{IniWavFun}\end{equation}
 is a member of the ensemble of reference states of interest.

\subsection{Evaluation of the Dipole Expectation Value}

In the above we have obtained the necessary ingredients for evaluating
the expectation of the dipole operator Eq. (\ref{DipExpVal}) explicitly.
Substituting Eqs. (\ref{TotGreFun}) and (\ref{KFRWavFun1-b}) in
Eq. (\ref{DipExpVal}), we get:\begin{eqnarray}
\bm{D}_{i,i}(t) & = & \left\langle \chi_{i}(t)\left|\vphantom{\Psi_{i}^{(1)}}\hat{\bm{d}}_{e}\right|\Psi_{i}^{(1)}(t)\right\rangle +c.c.\nonumber \\
 & = & \left\{ -i\int_{-\infty}^{t}dt'\left\langle \phi_{e}(t)\right|\left\langle \Phi_{J_{0}M_{0}}(t)\right|\hat{\bm{d}}_{e}\right.\nonumber \\
 &  & \times\sum_{j\bm{p}JM}\left|\phi_{j}^{(+)}\right\rangle \left|\phi_{\bm{p}}(t-t_{d})\right\rangle \nonumber \\
 &  & \times\left|\Phi_{JM}(t)\vphantom{\phi_{j}^{(+)}}\right\rangle e^{-iE_{j}^{+}(t-t')}\left\langle \Phi_{JM}(t')\vphantom{\phi_{j}^{(+)}}\right|\nonumber \\
 &  & \times\left\langle \phi_{\bm{p}}(t'-t_{d})\right|\left\langle \phi_{j}^{(+)}\right|\nonumber \\
 &  & \times\left.V_{e-L_{2}}(t'-t_{d})\left|\Phi_{J_{0}M_{0}}(t')\right\rangle \left|\phi_{e}(t')\right\rangle \vphantom{\int_{-\infty}^{t}}\right\} \nonumber \\
 &  & +c.c.\end{eqnarray}
 To simplify the expression further we (i) change the variable $t'\rightarrow t'+t_{d}$,
and similarly, $t\rightarrow t+t_{d}$; (ii) note that the free evolution
of the rotational wavepacket after the interaction with the pump pulse
at $\left(t+t_{d}\right)$ is $\Phi_{J_{0}M_{0}}\left(t+t_{d}\right)=e^{-iH_{N}t}\Phi_{J_{0}M_{0}}\left(t_{d}\right)$,
and similarly, at $\left(t'+t_{d}\right)$, $\Phi_{J_{0}M_{0}}\left(t'+t_{d}\right)=e^{-iH_{N}t'}\Phi_{J_{0}M_{0}}\left(t_{d}\right)$,
(iii) the time dependence of the unperturbed initial electronic state
at $\phi_{e}\left(t+t_{d}\right)=\left|\phi_{e}\right\rangle e^{-iE_{i}(t_{+}t_{d}}$,
and similarly for the ionic states, $\phi_{j}^{(+)}\left(t+t_{d}\right)=\left|\phi_{j}^{(+)}\right\rangle e^{-iE_{j}^{+}(t-t_{d})}$,
(v) introduce the overlaps ({}``Dyson-orbitals''), \begin{equation}
\left|\phi_{e}^{(j)}\right\rangle =\left\langle \left.\phi_{j}^{(+)}\left(1,2,...N_{e}-1\right)\right|\phi_{e}(1,2,...N-1,N)\right\rangle \label{DysOrb}\end{equation}
 and retain only the (dominant) contribution from the lowest lying
ionic state ($j=0$), to obtain:

\begin{eqnarray}
\bm{D}_{i,i}(t) & = & -i\sum_{jJM,\bm{p}}\left\langle \Phi_{J_{0}M_{0}}(t)\vphantom{\phi_{e}^{(0)}}\right|\left\langle \phi_{e}^{(0)}\left|\hat{\bm{d}}_{e}\right|\right.\nonumber \\
 &  & \times\left|\bm{p}(t-t_{d})\vphantom{\phi_{e}^{(0)}}\right\rangle \left|\phi_{j}^{+}\vphantom{\phi_{e}^{(0)}}\right\rangle \left|\Phi_{JM}(t)\vphantom{\phi_{e}^{(0)}}\right\rangle \nonumber \\
 &  & \times\int_{-\infty}^{t}dt'e^{-i(E_{j}^{+}-E_{0})(t-t')}\nonumber \\
 &  & \times e^{-i\int_{t'-t_{d}}^{t-t_{d}}(\bm{p}(u)^{2}/2)du}\nonumber \\
 &  & \times\left\langle \Phi_{JM}(t')\vphantom{\phi_{e}^{(0)}}\right|\left\langle \phi_{j}^{+}\vphantom{\phi_{e}^{(0)}}\right|\left\langle \bm{p}(t'-t_{d})\vphantom{\phi_{e}^{(0)}}\right|\nonumber \\
 &  & \times\left|V_{e-L_{2}}\left(t'-t_{d}\right)\vphantom{\phi_{e}^{(0)}}\right.\left|\phi_{e}^{(0)}\right\rangle \left|\Phi_{J_{0}M_{0}}(t')\vphantom{\phi_{e}^{(0)}}\right\rangle \nonumber \\
 &  & +c.c.\label{DipExpVal-1}\end{eqnarray}
 Or, \begin{eqnarray}
\bm{D}_{i,i}(t) & = & \left\langle \Phi_{J_{0}M_{0}}(t)\right|\bm{D}_{e}(t)\left|\Phi_{J_{0}M_{0}}(t)\right\rangle ,\end{eqnarray}
 where clearly the electronic part of the expectation value is given
by the integral \begin{eqnarray}
\bm{D}_{e}(t) & = & \left\{ -i\sum_{\bm{p}}\left\langle \phi_{e}^{(0)}\left|\hat{\bm{d}}_{e}\vphantom{\phi_{e}^{(0)}}\right|\bm{p}(t)\right\rangle \right.\nonumber \\
 &  & \times\int_{-\infty}^{t_{d}+t}dt'e^{-i\int_{t'}^{t}(\bm{p}(u)^{2}/2+E_{B})du}\nonumber \\
 &  & \times\left.\left\langle \bm{p}(t')\left|-\bm{F}(t')\cdot\hat{\bm{d}}_{e}\vphantom{\phi_{e}^{(0)}}\right|\phi_{e}^{(0)}\right\rangle \right\} +c.c.\label{EleDipExpVal}\end{eqnarray}
 where, $\bm{F}(t)$ is the probe field. Finally, by using the rate
of emission of the $n$th harmonic 
as given by Eq. (\ref{OneMolHhgRat}), we obtain the dynamic HHG signal,
for a pump-probe delay time $t_{d}$, \begin{eqnarray}
S^{(n)}\left(t_{d},\alpha\right) & =2\pi & \sum_{J_{0}M_{0}}\rho(J_{0})\left|\left\langle \Phi_{J_{0}M_{0}}\left(t_{d}\right)\left|T_{e}^{(n)}(\theta,\phi;\alpha)\right|\right.\right.\nonumber \\
 &  & \times\left.\left|\vphantom{T_{e}^{(n)}}\Phi_{J_{0}M_{0}}\left(t_{d}\right)\right\rangle \right|^{2}\frac{(n\omega)^{2}}{c^{3}}\label{HhgSig1}\end{eqnarray}
 where $(\theta,\phi)\equiv\hat{\bm{R}}_{N}$ is the orientation of
the molecular axis in space and $T_{e}^{(n)}(\theta,\phi;\alpha)$
is a \emph{HHG transition operator}.

\subsection{Derivation of the HHG operator $T^{(n)}(\theta,\phi;\alpha)$}

We shall now proceed to derive an explicit expression of the HHG transition
operator $T_{e}^{(n)}(\theta,\phi;\alpha)$ for an arbitrary relative
angle $\alpha$ between the linear polarization directions of the
pump and the probe pulse. To this end, we first consider the most
common experimental geometry in which the pump and probe polarizations
are chosen to be parallel.

\subsubsection{HHG operator: Parallel polarization $\alpha=0$}

We recall that for a linearly polarized probe pulse $\boldsymbol{F}_{2}(t)\,=\mathbf{\hat{\boldsymbol{\epsilon_{\Omega}}}}\, F(t)\cos{\omega t}$,
the corresponding vector potential is \begin{equation}
\boldsymbol{A}(t)\,=-\hat{\mathbf{\boldsymbol{\epsilon_{\Omega}}}}\,\left(\frac{cF(t)}{\omega}\right)\sin{\omega t}\label{eq:33}\end{equation}
 It is convenient in this case to take the space fixed polar axis
($z$-axis) along the common direction of the polarizations $\bm{\epsilon}_{1}\parallel\bm{\epsilon}_{2}\parallel\hat{\bm{z}}$.
To evaluate the tripe-integral over the intermediate momenta $\boldsymbol{p}$
in Eq. (\ref{EleDipExpVal}) we employ the stationary phase method
\cite{lew-94}, with the stationary values\begin{equation}
\boldsymbol{p}_{st}(t,t')=\frac{1}{t-t'}\int_{t'}^{t}\boldsymbol{A}(t'')dt'',\label{StaValPee}\end{equation}
 for which the derivative of the action $S(t,t')$ with respect to
$t'=t-\tau$ is equal to zero. The corresponding stationary value
of the action is, \begin{equation}
S_{st}(t,t')=\int_{t'}^{t}\left\{ \frac{1}{2}\left(\boldsymbol{p}_{st}(t,t'')-\frac{1}{c}\boldsymbol{A}(t'')\right)^{2}+E_{B}\right\} dt'',\label{StaAct}\end{equation}
 where, $\boldsymbol{p}(t)=\boldsymbol{p}_{st}(t,t')-\frac{1}{c}\boldsymbol{A}(t)$
and $\boldsymbol{p}(t')=\boldsymbol{p}_{st}(t,t')-\frac{1}{c}\boldsymbol{A}(t')$.
Thus, projecting the resulting value of $D_{e}(t)$ on to the polarization
direction $\epsilon_{\Omega}$ of the emitted harmonic we get, \begin{eqnarray}
D_{e}\left(t\right) & = & \left\{ i\int_{0}^{t}dt'\left(\frac{\pi}{\epsilon+i(t-t')/2}\right)^{3/2}\right.\nonumber \\
 &  & \times\left\langle \phi_{e}^{(0)}\left|\varepsilon_{\Omega}.\boldsymbol{r}\right|\boldsymbol{p}\left(t\right)\right\rangle e^{-iS_{st}(t,t')}\nonumber \\
 &  & \times\left.\left\langle \boldsymbol{p}\left(t'\right)\left|\boldsymbol{F}(t').\boldsymbol{r}\right|\phi_{e}^{(0)}\right\rangle \right\} +c.c.\label{EleDipSta}\end{eqnarray}
 We may note that the first matrix element in this expression (reading
from the right to the left) corresponds to the {}``ionization''
transition at time $t'$, $d_{ion}(t')\equiv\left\langle \boldsymbol{p}\left(t'\right)\left|\boldsymbol{F}(t').\boldsymbol{r}\right|\phi_{e}^{(0)}\right\rangle $,
whereas the last matrix element corresponds to a {}``recombination''
transition of the electron back to the same initial state at a time
$t$, $d_{rec}(t)\equiv\left\langle \phi_{e}^{(0)}\left|\boldsymbol{\epsilon}_{\Omega}.\boldsymbol{r}\right|\boldsymbol{p}\left(t\right)\right\rangle $.
The interval $(t-t')$ corresponds to the intermediate time that the
electron spends in the continuum Volkov states, between the absorption
of $n$ photons in the first step and the emission of the harmonic
frequency $\Omega=n\omega$ in the last step. We have assumed that
there was no significant depletion of the ground state population
during the process. However, if needed, a weak depletion due to ionization
could be accounted for without difficulty by introducing in the above
expression the exponential decay factor: $e^{-(\gamma/2)(t+t')}$,
where $\gamma$ is the total ionization rate.

\subsubsection{The {}``ionization'' and {}``recombination'' matrix elements}

To evaluate the matrix elements of {}``ionization'' and {}``recombination''
in Eq. (\ref{EleDipSta}), we assume that the wavefunction of the
active electron may be given by the highest occupied molecular orbital
or HOMO. This can be written either in the multi-center LCAO-MO form,
or, by transforming it into an equivalent single-center- MO (by appropriate
transformations, e.g. \cite{har-65,fai-70}). It is useful also to
note that in the latter form, it often suffices for problems of interaction
of molecules with long-wavelength laser fields, and in conjunction
with the {}``length gauge'', to retain only the asymptotic limit
of the orbitals at distances away from the molecular center.

In general, let the unperturbed MO of the active electron of a linear
molecule, in the Born-Oppenheimer approximation, be given in the body
fixed frame by the single-center expansion:\begin{equation}
\phi_{e}^{(0)}(\boldsymbol{r})=\sum_{l}C_{l}^{(m)}R_{l}(r)Y_{lm}(\hat{\boldsymbol{r}})\label{SinCenMolOrb}\end{equation}
 where $C_{l}^{(m)}$ are the expansion coefficients (at a given internuclear
separation R), $R_{l}(r)$ are the radial waves of angular momentum
$l$, $Y_{lm}(r)$ are the spherical harmonics, and $m$ is the projection
of the angular momentum of the active electron along the molecular
axis, that is conserved and hence also characterizes the MO.

Next, we transform the molecular orbitals Eq. (\ref{SinCenMolOrb})
from the body fixed frame to the space fixed frame by using the Wigner
transformation $\bm{D}$,\begin{eqnarray}
\phi_{e}^{(0)}(\boldsymbol{r}) & = & \hat{\boldsymbol{D}}\phi_{e}^{(0)}(\boldsymbol{r})\nonumber \\
 & = & \sum_{l}C_{l}^{(m)}R_{l}(r)\nonumber \\
 &  & \times\sum_{\mu}D_{\mu m}^{l}(\phi,\theta,\chi)Y_{l\mu}(\widehat{\boldsymbol{r}})\label{MolOrbSpaFix}\end{eqnarray}
 Above $D_{\mu m}^{l}(\phi,\theta,\chi)=e^{-i\mu\phi}d_{\mu m}^{l}(\theta)e^{-im\chi}$
is the Wigner rotation matrix where $(\phi,\theta,\chi)$ are the
Euler's angles which define the orientation of the molecular axis
to the space fixed coordinate frame \cite{var-88}. The middle term
of the Wigner matrix, $d_{\mu m}^{l}(\theta)$, have been tabulated
e.g. in reference \cite{var-88,zar-88}. The matrix element of the
dipole along the direction of the probe pulse, appearing in Eq. (\ref{EleDipSta}),
then reads:\begin{eqnarray}
d_{ion}(t') & = & F(t')\sum_{l}C_{l}^{(m)}\sum_{\mu}D_{\mu m}^{l}(\phi,\theta,\chi)\nonumber \\
 &  & \times\left\langle e^{i\boldsymbol{p}_{t}.\boldsymbol{r}}\left|\mathbf{\epsilon}_{\omega}.\boldsymbol{r}\right|R_{l}(r)Y_{l\mu}(\widehat{\boldsymbol{r}})\right\rangle \label{eq:36}\end{eqnarray}
 Further, we expand $e^{i\boldsymbol{p}_{t'}.\boldsymbol{r}}$ in
spherical harmonics,\begin{eqnarray}
e^{i\boldsymbol{p}_{t'}.\boldsymbol{r}} & = & \frac{2\pi}{\sqrt{p_{t'}r}}\sum_{l'm'}(i)^{l'}\, J_{l'+1/2}\left(p_{t'}r\right)\nonumber \\
 &  & \times Y_{l'm'}\left(\hat{\boldsymbol{p}}_{t'}\right)\, Y_{l'm'}\left(\widehat{\boldsymbol{r}}\right)\label{eq:37}\end{eqnarray}
 and note that in this system of axes we have,\begin{equation}
\mathbf{\epsilon}_{\omega}.\boldsymbol{r}=r\sqrt{\frac{4\pi}{3}}\, Y_{10}(\widehat{\boldsymbol{r}})\label{eq:38}\end{equation}
 Note also that the instantaneous momentum $\boldsymbol{p}_{t'}$
can be either parallel or anti-parallel with respect to the direction
of the field so that $\theta_{p_{t}}=0,\pi$ and $\phi_{p_{t}}=0$.
Therefore, the spherical harmonics with the argument $\hat{\boldsymbol{p}}_{t'}$
can be simplified to $Y_{l'm'}(\hat{\boldsymbol{p}}_{t'})=(\sigma)^{l'}\sqrt{\frac{2l'+1}{4\pi}}\delta_{m',0}$,
with $\sigma=1$ for $\theta_{p_{t}}=0$ and $\sigma=-1$ for $\theta_{p_{t}}=\pi$.
Substituting Eqs. (\ref{eq:37}) and (\ref{eq:38}) in Eq. (\ref{eq:36})
we obtain (with $l'=l_{i}$),\begin{eqnarray}
d_{ion}(t') & = & F(t')\sum_{l_{i}}C_{l_{i}}^{(m)}D_{m0}^{l_{i}}(\phi,\theta,\chi)\beta_{ion}(l_{i},m;t')\hphantom{--}\label{IonDip}\end{eqnarray}
 \begin{eqnarray}
\beta_{ion}(l_{i},m;t') & = & \frac{(2\pi)^{3/2}}{\sqrt{3p_{t'}}}\,\sum_{l}(i\sigma)^{l}\,\sqrt{(2l_{i}+1)}\nonumber \\
 &  & \times\left\langle l_{i}0\left|10\right|l\mu\right\rangle I_{l_{i},l}(t')\label{eq:39}\end{eqnarray}
 where, we have defined the radial integrals ($m$ fixed) by\begin{equation}
I_{l_{i},l}(t')=\int_{0}^{\infty}J_{l+\frac{1}{2}}(p_{t'}r)\, R_{l_{i}}^{m}(r)r^{-1/2}\, r\, r^{2}dr\label{eq:40}\end{equation}
 Using the Slater orbitals representation of the single center radial
functions, $R_{l}(r)=r^{\eta-1}e^{-\alpha r}$, the radial integrals
($I$'s) appearing in $d_{ion}$ (Eq. (\ref{IonDip})) can be evaluated
explicitly by using the formula \cite{gra-65}\begin{eqnarray}
\int_{0}^{\infty}e^{-\alpha x}J_{\nu}(\beta x)\, x^{\mu-1}dx & = & \frac{\left(\frac{\beta}{2}\right)^{\nu}\Gamma(\nu+\mu)}{\sqrt{(\alpha^{2}+\beta^{2})^{\nu+\mu}}\Gamma(\nu+1)}\nonumber \\
 &  & \times F\left(\frac{\nu+\mu}{2},\frac{1-\mu+\nu}{2},\vphantom{\frac{\beta^{2}}{\alpha^{2}+\beta^{2}}}\right.\nonumber \\
 &  & \left.\nu+1;\frac{\beta^{2}}{\alpha^{2}+\beta^{2}}\right)\label{ForRadInt}\end{eqnarray}
 where $F(a,b,c;x)$ is a hypergeometric function. Note that since
the argument $x\equiv\frac{\beta^{2}}{\alpha^{2}+\beta^{2}}<1$ the
hypergeometric function is guaranteed to converge for all values of
$a$, $b$, and $c$. For ionization step, the radial integration
reads\begin{eqnarray}
I_{l_{i},l_{i}+1}(t') & = & \frac{\left(\frac{p_{t'}}{2}\right)^{l+3/2}\Gamma(l_{i}+Z_{c}/p_{B}+4)}{\sqrt{(P_{B}^{2}+p_{t'}^{2})^{l_{i}+Z_{c}/p_{B}+4}}\Gamma(l_{i}+\frac{5}{2})}\nonumber \\
 &  & \times F\left(\frac{l_{i}+Z_{c}/p_{B}+4}{2},\frac{l_{i}-Z_{c}/p_{B}}{2},\right.\nonumber \\
 &  & \left.l_{i}+\frac{5}{2};\frac{p_{t'}^{2}}{p_{B}^{2}+p_{t'}^{2}}\right)\nonumber \\
I_{l_{i},l_{i}-1}(t') & = & \frac{\left(\frac{p_{t'}}{2}\right)^{l_{i}-1/2}\Gamma(l_{i}+Z_{c}/p_{B}+2)}{\sqrt{(p_{B}^{2}+p_{t'}^{2})^{l_{i}+Z_{c}/p_{B}+2}}\Gamma(l_{i}+\frac{1}{2})}\nonumber \\
 &  & \times F\left(\frac{l_{i}+Z_{c}/p_{B}+2}{2},\frac{l_{i}-Z_{c}/p_{B}-2}{2},\right.\nonumber \\
 &  & \left.l_{i}+\frac{1}{2};\frac{p_{t'}^{2}}{p_{B}^{2}+p_{t'}^{2}}\right)\label{AsyRadInt}\end{eqnarray}
 Exactly the same expressions hold for the radial integrals appearing
in the recombination dipole $d_{rec}(l_{r},m;t)$ (except that $l_{i}$
is changed to $l_{r}$, and $t'$ to $t$, in Eq. (\ref{AsyRadInt})),
throughout.

The Clebsch-Gordan coefficient in Eq. (\ref{eq:39}) implies that
only the $\mu=0$ term and $l=l_{i}\pm1$ terms survive in the sums
and we get for the ionization dipole,\begin{equation}
d_{ion}(t')=F(t')\sum_{l_{i}}C_{l_{i}}^{(m)}D_{m0}^{(l_{i})}(\phi,\theta,\chi)\beta_{ion}\left(l_{i},m;t'\right)\label{IonMat}\end{equation}
 with\begin{eqnarray}
\beta_{ion}(l_{i},m;t') & = & \frac{2\pi}{\sqrt{2p_{t'}}}\,{\sqrt{(2l_{i}+1)}}\nonumber \\
 &  & \times\left((i\sigma)^{l_{i}+1}\,(l_{i}+1)\, I_{l_{i},l_{i}+1}(t')\right.\nonumber \\
 &  & +\left.(i\sigma)^{l_{i}-1}\, l_{i}\, I_{l_{i},l_{i}-1}(t')\vphantom{(i\sigma)^{l_{i}+1}}\right)\label{IonDip-1}\end{eqnarray}
 In the above we have introduced the angular momentum notation $l_{i}$
for the initial bound state in the {}``ionization'' matrix element
(and $l_{r}$, in the {}``recombination'' matrix element).

We may assume that the emitted harmonic is observed with its polarization
along the same direction as the probe pulse polarization. (There is
no difficulty, except lengthier algebra, to obtain the expression
for the polarization direction orthogonal to it, but the former would
give the dominant contribution under phase matching condition. Following
an analogous calculation as above we get the {}``recombination''
matrix element as:\begin{equation}
d_{rec}(t)=\sum_{l_{r}}{C_{l_{r}}^{(m)*}\, D_{0m}^{l_{r}*}(\phi,\theta,\chi)}\beta_{rec}\left(l_{r},m;t'\right)\label{RecMat}\end{equation}
 with \begin{eqnarray}
\beta_{rec}(l_{r},m;t) & = & \frac{2\pi}{\sqrt{2p_{t}}}\,\sqrt{(2l_{r}+1)}\nonumber \\
 &  & \times\left((i\sigma)^{l_{r}+1}\,(l_{r}+1)\, I_{l_{r},l_{r}+1}(t)\right.\nonumber \\
 &  & \left.+(i\sigma)^{l_{r}-1}\, l_{r}\, I_{l_{r},l_{r}-1}(t)\vphantom{(i\sigma)^{l_{r}+1}}\right)\label{RecDip-1}\end{eqnarray}
 Substituting Eqs. ( \ref{IonMat}-\ref{IonDip-1}) and (\ref{RecMat}-\ref{RecDip-1})
in Eq. (\ref{EleDipSta}), we obtain \begin{eqnarray}
D_{e}\left(t\right) & = & i\sum_{l_{i},l_{r}}d_{0m}^{l_{r}}(\theta){d_{0m}^{l_{i}}(\theta)}\nonumber \\
 &  & \times\left[C_{l_{r}}^{(m)*}C_{l_{i}}^{(m)}M_{e}(t)+c.c.\right]\label{FinEleDip}\end{eqnarray}
 where we have used the relation \begin{equation}
D_{m0}^{l_{r}*}(\phi,\theta,\chi)D_{m0}^{l_{i}}(\phi,\theta,\chi)=d_{0m}^{l_{r}}(\theta)d_{0m}^{l_{i}}(\theta)\label{SimWigPro}\end{equation}
 and defined the radial integral, \begin{eqnarray}
M_{e}(t) & = & i\int_{-\infty}^{t_{d}+t}dt\left(\frac{\pi}{(\epsilon+i(t-t')/2)}\right)^{3/2}\nonumber \\
 &  & \times{\beta_{rec}(l_{r},m;t)}e^{-iS_{st}(t,t')}\nonumber \\
 &  & \times F(t'){\beta_{ion}(l_{i},m;t')}\label{elePar}\end{eqnarray}
 Next, by integrating over $t'$, taking the Fourier transform with
respect to $t$, we obtain (cf. Eq.(\ref{TraMat})), the HHG operator
$T_{e}^{(n)}(\theta,\phi;0)$ for the $n$th harmonic generation:

\noindent \begin{eqnarray}
T_{e}^{(n)}(\theta,\phi;0) & = & \sqrt{2\pi(n\omega)}\tilde{D}_{e}(n\omega)\nonumber \\
 & = & \sqrt{2\pi(n\omega)}\sum_{l_{i},l_{r}}d_{0,m}^{l_{r}}(\theta)d_{0,m}^{l_{i}}(\theta)\nonumber \\
 &  & \times\tilde{\alpha}_{zz}^{(n)}\left(l_{r},l_{i};m\right)\label{FinHhgOpe}\end{eqnarray}
 where, $\tilde{\alpha}_{zz}^{(n)}\left(l_{r},l_{i};m\right)$ is
given by the $n$th Fourier coefficient of $D_{e}(t)$: \begin{eqnarray}
\tilde{\alpha}_{zz}^{(n)}\left(l_{r},l_{i};m\right)\equiv\hphantom{\left[C_{l_{r}}^{(m)*}C_{l_{i}}^{(m)}M_{e}(t)+c.c.\right]}\nonumber \\
{\textit{ F.T.}}\left[C_{l_{r}}^{(m)*}C_{l_{i}}^{(m)}M_{e}(t)+c.c.\right] & (n\omega)\label{EleFouTra}\end{eqnarray}

\noindent Next, by substituting Eq. (\ref{FinHhgOpe}) in Eq. (\ref{HhgSig1}),
we obtain the rotational matrix elements, \begin{equation}
\left\langle \Phi_{J_{0}M_{0}}(t_{d})\left|d_{0m}^{l_{r}}(\theta)d_{0m}^{l_{i}}(\theta)\right|\Phi_{J_{0}M_{0}}(t_{d})\right\rangle ,\end{equation}
 which can be evaluated directly by using the tabulated values of
the $d_{0m}^{l}(\theta)$ given by elementary trigonometric functions
(see, Tab. \ref{tab:01}). Alternatively, we may first combine the
product \begin{eqnarray}
d_{0m}^{l_{r}}(\theta)d_{0m}^{l_{i}}(\theta) & = & \sqrt{\frac{4\pi}{2l_{r}+1}}(-1)^{m}Y_{l_{r},-m}(\theta,\phi)\nonumber \\
 &  & \sqrt{\frac{4\pi}{2l_{i}+1}}Y_{l_{i},m}(\theta,\phi)\nonumber \\
 & = & \sum_{L=|l_{r}-l_{i}}^{l_{r}+l_{i}}(-1)^{m}\left\langle l_{r},l_{i},-m,m|L,0\right\rangle \nonumber \\
 &  & \times\left\langle l_{r},l_{i},0,0|L,0\right\rangle P_{L}(\cos{\theta})\end{eqnarray}
 and we obtain \begin{equation}
T_{e}^{(n)}(\theta,\phi;0)=\sum_{l_{i},l_{r},L}{\tilde{a'}}_{zz}^{(n)}\left(l_{r},l_{i},L;m\right)P_{L}(\cos{\theta})\label{SpeTraMatTwo}\end{equation}
 where,\begin{eqnarray}
{\tilde{a'}}_{zz}^{(n)}\left(l_{r},l_{i},L;m\right) & = & \sqrt{2\pi(n\omega)}\tilde{\alpha}_{zz}^{(n)}\left(l_{r},l_{i};m\right)\nonumber \\
 &  & \times(-1)^{m}\left\langle l_{r},l_{i},-m,m|L,0\right\rangle \nonumber \\
 &  & \times\left\langle l_{r},l_{i},0,0|L,0\right\rangle .\label{GenDynCoe}\end{eqnarray}
 Thus, the expectation value of the transition operator with respect
to the rotational wavepacket can be obtained more elegantly in terms
of the Legendre polynomials moments: \begin{equation}
\left\langle P_{L}\right\rangle _{J_{0}M_{0}}\left(t_{d}\right)\equiv\left\langle \Phi_{J_{0}M_{0}}\left(t_{d}\right)\right|P_{L}(\theta)\left|\Phi_{J_{0}M_{0}}\left(t_{d}\right)\right\rangle \label{LegMom}\end{equation}
 Finally, by substituting the above relations (Eq. (\ref{SpeTraMatTwo}))
in Eq. (\ref{HhgSig1}), and taking the statistical average over the
ensemble of the emission probabilities from the ensemble of rotational
wavepackets, we obtain the HHG signal (i.e. the rate per unit time
of generation of the $n$th harmonic per molecule) in the special
case of parallel polarizations (cf. \cite{fai-07}):

\noindent \begin{eqnarray}
S^{(n)}\left(t_{d};0\right) & = & 2\pi\sum_{J_{0}M_{0}}\rho(J_{0})\left|\sum_{L,l_{r},l_{i}}\tilde{a'}_{zz}^{(n)}\left(l_{r},l_{i},L;m\right)\right.\nonumber \\
 &  & \times\left.\left\langle P_{L}\right\rangle _{J_{0}M_{0}}(t_{d})\vphantom{\sum_{L,l_{r},l_{i}}\tilde{a'}_{zz}^{(n)}}\right|^{2}\frac{(n\omega)^{2}}{c^{3}}\label{SpeHhgSig}\end{eqnarray}
\begin{figure}[H]
\begin{centering}
\includegraphics[scale=0.45]{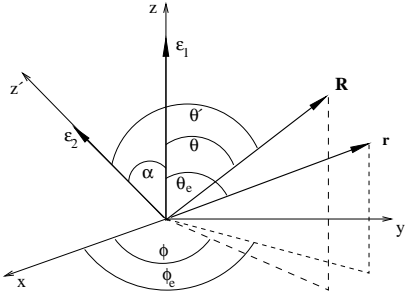} 
\par\end{centering}

\caption{\label{fig:SchDia} A schematic diagram defining: molecular axis
$\boldsymbol{R}$, electron position $\boldsymbol{r}$, pump polarization
$\boldsymbol{\epsilon}_{1}$, and probe polarization $\boldsymbol{\epsilon}_{2}$;
$\alpha$ is the operational laboratory angle.}
\end{figure}

\subsection{General Polarization Geometry: Arbitrary $\alpha$}

So far we have assumed that the pump and the probe polarizations are
parallel and that they point along the space fixed polar axis $\hat{\bm{z}}$.
In the general case, we may define, without loss of generality, the
relative angle between the polarizations, $\alpha$, to lie in the
($z-z'-x$)-plane (cf. Fig. \ref{fig:SchDia}). From the figure, it
can be seen that we simply need to re express the direction of the
molecular axis, $(\theta,\phi)$, given with respect to the pump polarization
$\bm{\epsilon}_{1}\parallel\hat{\bm{z}}$), in terms of the direction
$(\theta',\phi')$ with respect to the probe polarization direction
$\bm{\epsilon}_{2}\parallel\hat{\bm{z}}$. This is readily achieved
by simply replacing $\cos{\theta}\rightarrow\cos{\theta'}$, and using
the well known relation \begin{equation}
\cos{\theta'}=\cos{\alpha}\cos{\theta}+\sin{\theta}\sin{\alpha}\cos{\phi}\label{Cos3D}\end{equation}
 or the vector addition coefficients and the addition theorem \begin{equation}
P_{L}(\cos{\theta'})=\frac{4\pi}{2L+1}Y_{L,M}(\theta,\phi)Y_{L,M}^{*}(\alpha,0).\label{AddThe}\end{equation}
 Thus, we obtain the general expression of the HHG operator for any
$\alpha$ \cite{fai-08}: \begin{eqnarray}
T_{e}^{(n)}(\theta',\phi';\alpha) & = & \sqrt{2\pi(n\omega)}\sum_{l_{i},l_{r}}d_{0,m}^{l_{r}}(\theta')d_{0,m}^{l_{i}}(\theta')\nonumber \\
 &  & \times\tilde{a}_{zz}^{(n)}\left(l_{r},l_{i};m\right).\label{EleTraMat}\end{eqnarray}
 Or,

\begin{eqnarray}
T_{e}^{(n)}(\theta',\phi';\alpha) & = & \sum_{l_{r},l_{i},L,M}\tilde{a'}_{zz}^{(n)}\left(l_{r},l_{i},L;m\right)P_{L}(\cos{\theta'})\nonumber \\
 & = & \sum_{LM}\sum_{l_{r},l_{i}}\tilde{a'}_{zz}^{(n)}\left(l_{r},l_{i},L;m\right)\nonumber \\
 &  & \times\frac{4\pi}{2L+1}Y_{L,M}(\theta,\phi)Y_{L,M}^{*}(\alpha,0)\end{eqnarray}
 where $\tilde{a'}_{zz}^{(n)}\left(l_{r},l_{i},L;m\right)$ is given
by Eq. (\ref{GenDynCoe}).

It is should be noted that, in general, if the molecular orbital coefficients
were assumed to be complex, \begin{equation}
C_{l}^{(m)}\equiv\left|C_{l}^{(m)}\right|e^{i\phi_{l}},\label{OrbCoe}\end{equation}
 then we should rewrite the dynamic parameters $\tilde{\alpha}_{zz}^{(n)}\left(l_{r},l_{i};m\right)$

(Eq.(\ref{EleFouTra})) as:\begin{eqnarray}
\tilde{\alpha}_{zz}^{(n)}\left(l_{r},l_{i};m\right) & = & {\textit{F.T.}}\left[C_{l_{r}}^{(m)*}C_{l_{i}}^{(m)}M_{e}(t)+c.c.\right]\nonumber \\
 & = & 2\left|C_{l_{r}}^{(m)}\right|\left|C_{l_{i}}^{(m)}\right|\left[\cos{\phi_{l_{i}l_{r}}}\tilde{u}_{l_{r},l_{i},m}^{(n)}\right.\nonumber \\
 &  & \left.-\sin{\phi_{l_{i}l_{r}}}{\tilde{v}}_{l_{r},l_{i},m}^{(n)}\right]\end{eqnarray}
 where, \begin{equation}
\phi_{l_{i}l_{r}}\equiv\left(\phi_{l_{i}}-\phi_{l_{r}}\right).\label{CoePha}\end{equation}
 $M_{e}(t)$ is given by Eq. (\ref{elePar}), and we have defined,
\begin{eqnarray}
\tilde{u}_{l_{r},l_{i},m}^{(n)} & = & {\textit{F.T.}}\left[Re\left\{ M_{e}(t)\right\} \right](n\omega)\label{ReaCoe}\\
\tilde{v}_{l_{r},l_{i},m}^{(n)} & = & {\textit{F.T.}}\left[Im\left\{ M_{e}(t)\right\} \right](n\omega)\label{InaCoe}\end{eqnarray}
 Thus, finally, we can express the general transition matrix element
for the $n$th order harmonic as an expansion in Legendre polynomials
in $\cos{\alpha}$, and the corresponding Legendre moments of the
time-dependent axis distribution of the molecule:

\begin{eqnarray}
T^{(n)}\left(t_{d},\alpha\right) & = & \left\langle \Phi_{J_{0}M_{0}}\left(t_{d}\right)\right|T^{(n)}(\theta,\phi;\alpha)\left|\Phi_{J_{0}M_{0}}\left(t_{d}\right)\right\rangle \nonumber \\
 & = & \sum_{L,l_{r},l_{i}}\sqrt{2\pi(n\omega)}\left[\cos\left(\phi_{l_{i},l_{r}}\right){\tilde{u}}_{l_{i},l_{r},m}^{(n)}\right.\nonumber \\
 &  & \left.-\sin\left(\phi_{l_{i}l_{r}}\right){\tilde{v}}_{l_{i},l_{r},m}^{(n)}\right]2|C_{l_{r}}^{(m)}||C_{l_{i}}^{(m)}|\nonumber \\
 &  & \times(-1)^{m}\left\langle l_{r},l_{i},-m,m;L,0\right\rangle \nonumber \\
 &  & \times\left\langle l_{r},l_{i},0,0;L,0\right\rangle \nonumber \\
 &  & \times\left\langle P_{L}\right\rangle _{J_{0}M_{0}}\left(t_{d}\right)P_{L}(\cos{\alpha})\label{GenTraMat}\end{eqnarray}
 where, we have taken the expectation value of the HHG operator with
respect to the rotational wavepacket $\left|\Phi\left(t_{d}\right)\right\rangle $,
to obtain \begin{eqnarray}
\frac{4\pi}{2L+1}\left\langle \Phi_{J_{0}M_{0}}\left(t_{d}\right)\right|Y_{LM}(\theta,\phi)\left|\Phi_{J_{0}M_{0}}\left(t_{d}\right)\right\rangle \nonumber \\
=\left\langle P_{L}\right\rangle _{J_{0}M_{0}}\left(t_{d}\right)\, P_{L}(\cos{\alpha})\delta_{M,0}\end{eqnarray}
 This follows from the observation that the magnetic quantum numbers
of all the rotational eigenstates in the individual wavepackets have
the same value $M_{0}$. It holds when the interaction operator $V_{e-L_{2}}(t)$
does not depend on the azimuth angle of the molecular axis in the
body fixed frame. In the above expression, we have also used the relation
$\sqrt{\frac{4\pi}{2L+1}}Y_{L0}(\theta,\phi)=P_{L}(\cos{\theta})$,
and an analogous relation with respect to the angle $\alpha$, to
simplify.

It is useful to note also that, if the orbital expansion coefficients
are real, as is often the case, then $\phi_{l_{i},l_{r}}=(0,\pi)$
and therefore the quantity in the square brackets in Eq. (\ref{GenTraMat})
simplifies to $[...]=\left[{\cos{(\phi_{l_{i},l_{r}})}\tilde{u}}_{l_{r},l_{i},m}^{(n)}\right]$
only.

\subsection{A General Formula for the HHG Signal}

Thus, finally, we substitute Eq. (\ref{GenTraMat}) in Eq. (\ref{SpeHhgSig})
and obtain the desired general expression (cf. \cite{fai-08}) for
the $n$th harmonic signal from a linear molecule, for any value of
$t_{d}$ and $\alpha$:

\begin{eqnarray}
S^{(n)}(t_{d},\alpha) & = & \sum_{J_{0}M_{0}}\rho_{J_{0}M_{0}}2\pi\left|\sum_{L,l_{r},l_{i}}\left[\cos{\phi_{l_{i},l_{r}}}{\tilde{u}}_{l_{r},l_{i},m}^{(n)}\right.\right.\nonumber \\
 &  & -\left.\sin{\phi_{l_{i}l_{r}}}{\tilde{v}}_{l_{r},l_{i},m}^{(n)}\right]2|C_{l_{r}}^{(m)}||C_{l_{i}}^{(m)}|\nonumber \\
 &  & \times(-1)^{m}\left\langle l_{r},l_{i},-m,m;L,0\right\rangle \nonumber \\
 &  & \times\left\langle l_{r},l_{i},0,0;L,0\right\rangle \nonumber \\
 &  & \times\left.\left\langle P_{L}\right\rangle _{J_{0}M_{0}}\left(t_{d}\right)P_{L}(\cos{\alpha})\vphantom{{\tilde{u}}_{l_{i},l_{r},m}^{(n)}}\right|^{2}\nonumber \\
 &  & \times\frac{(n\omega)^{3}}{c^{3}}\label{GenHhgSig}\end{eqnarray}

We may conclude the section by noting that for the special case of
parallel polarizations, $\alpha=0$, $P_{L}(\cos{0})=P_{L}(1)=1$,
Eq. (\ref{GenHhgSig}) correctly goes over to the signal obtained
for that special case, Eq. (\ref{SpeHhgSig}) (cf. \cite{fai-07}).
\begin{table}

\caption{\label{tab:01}Explicit form of $d_{0m}^{l}(\theta)$ required for
evaluating Eq. (\ref{FinHhgOpe}) \cite{var-88,zar-88}}

\begin{centering}
\begin{tabular}{ccc}
\hline 
$l$&
$\mathrm{N_{2}}\,\,\,\,\,\,(m=0)$&
$\mathrm{O_{2}}\,\,\,\,\,\,(m=1)$\tabularnewline
\hline 
0&
1&
-\tabularnewline
2&
$\frac{1}{2}\left(3\cos^{2}\theta-1\right)$&
$\sqrt{\frac{3}{2}}\,\sin\theta\cos\theta$ \tabularnewline
4&
$\frac{1}{8}(3-30\cos^{2}\theta+35\cos^{4}s)$~&
~$-\frac{\sqrt{5}}{4}\,\sin\theta\cos\theta\left(3-\cos^{2}\theta\right)$\tabularnewline
\hline
\end{tabular}
\par\end{centering}
\end{table}

\section{Applications to Diatomic Molecules $\mathrm{N_{2}}$ and $\mathrm{O_{2}}$}

\subsection{Parallel Geometry, $\alpha=0$: Elementary Expression of $T_{e}^{(n)}(\theta,\alpha=0)$
for $\mathrm{N_{2}}$}

$\mathrm{N_{2}}$ has $\sigma_{g}$ symmetry, and we approximate its
MO by the asymptotic approximation from single center molecule (Eq.
(\ref{SinCenMolOrb})) with $m=0$ and $l=0,2,4$ \cite{ton-02,kje-05}
whose angular coefficient are given in table \ref{tab:02}. The radial
part of electronic wave function is given by\begin{equation}
R_{l}(r)=r^{\eta-1}e^{-p_{B}r}\label{RadWavEle}\end{equation}
 with $\eta\equiv Z_{c}/p_{B}$; $Z_{c}$ is the core charge and $p_{B}=\sqrt{2\left|E_{B}\right|}$
with $E_{B}$ is binding energy.

Evaluating Eq. (\ref{FinHhgOpe}) for $m=0$ and $l_{i},l_{r}=0,2,4$
give us the HHG operator for $\mathrm{N_{2}}$\begin{equation}
T_{e}^{(n)}(\theta)=\sqrt{2\pi(n\omega)}\sum_{l_{i},l_{r}=0,2,4}d_{00}^{l_{r}}(\theta)\tilde{a}_{zz}^{n}(l_{r},l_{i};0)d_{00}^{l_{i}}(\theta)\label{eq:47}\end{equation}
 Using the expressions for the reduced rotation matrices from Tab.
\ref{tab:01} and simplifying, we may rewrite the operator as a sum
of powers of $\cos^{2}\theta$ only,\begin{eqnarray}
T_{e}^{(n)}(\theta) & = & \sqrt{2\pi(n\omega)}\left[b_{0}^{(n)}+b_{1}^{(n)}\cos^{2}\theta+b_{2}^{(n)}\cos^{4}\theta\right.\nonumber \\
 &  & \left.\mathrm{+b_{3}^{(n)}\cos^{6}\theta+b_{4}^{(n)}\cos^{8}\theta}\right]\label{FinNitHhgOpe}\end{eqnarray}
 where the coefficients $b_{j}^{(n)}$s reads,\begin{eqnarray}
b_{0}^{(n)} & = & \tilde{a}_{zz}^{(n)}(0,0;0)-\frac{1}{2}\tilde{a}_{zz}^{(n)}(2,2;0)+\frac{3}{8}\tilde{a}_{zz}^{(n)}(4,4;0)\nonumber \\
 &  & -\frac{1}{2}\left(\tilde{a}_{zz}^{(n)}(0,2;0)+\tilde{a}_{zz}^{(n)}(2,0;0)\right)\nonumber \\
 &  & +\frac{3}{8}\left(\tilde{a}_{zz}^{(n)}(0,4;0)+\tilde{a}_{zz}^{(n)}(4,0;0)\right)\nonumber \\
 &  & -\frac{3}{16}\left(\tilde{a}_{zz}^{(n)}(2,4;0)+\tilde{a}_{zz}^{(n)}(4,2;0)\right)\nonumber \\
b_{1}^{(n)} & = & -\frac{3}{2}\tilde{a}_{zz}^{(n)}(2,2;0)\nonumber \\
 &  & +\frac{3}{2}\left(\tilde{a}_{zz}^{(n)}(0,2;0)+\tilde{a}_{zz}^{(n)}(2,0;0)\right)\nonumber \\
 &  & -\frac{15}{4}\left(\tilde{a}_{zz}^{(n)}(0,4;0)+\tilde{a}_{zz}^{(n)}(4,0;0)\right)\nonumber \\
 &  & -\frac{21}{16}\left(\tilde{a}_{zz}^{(n)}(2,4;0)+\tilde{a}_{zz}^{(n)}(4,2;0)\right)\nonumber \\
b_{2}^{(n)} & = & \frac{35}{8}\left(\tilde{a}_{zz}^{(n)}(0,4;0)+\tilde{a}_{zz}^{(n)}(4,0;0)\right)\nonumber \\
 &  & -\frac{125}{16}\left(\tilde{a}_{zz}^{(n)}(2,4;0)+\tilde{a}_{zz}^{(n)}(4,2;0)\right)\nonumber \\
b_{3}^{(n)} & = & \frac{105}{16}\left(\tilde{a}_{zz}^{(n)}(2,4;0)+\tilde{a}_{zz}^{(n)}(4,2;0)\right)\nonumber \\
b_{4}^{(n)} & = & \frac{1225}{16}\tilde{a}_{zz}^{(n)}(4,4;0)\label{eq:49}\end{eqnarray}
 Thus, by using Eq. (\ref{HhgSig1}), the $n$th harmonic signal for
$\mathrm{N_{2}}$ becomes,\begin{eqnarray}
S^{(n)}(t_{d}) & = & {\cal {C}}\sum_{j=0}^{3}\sum_{j'\geq j}^{3}c_{jj'}^{(n)}\left\langle \left\langle \cos^{2j}\theta\right\rangle \left\langle \cos^{2j'}\theta\right\rangle \right\rangle \nonumber \\
 & = & {\cal {C}}\left\{ c_{00}^{(n)}+c_{01}^{(n)}\left\langle \left\langle \cos^{2}\theta\right\rangle \right\rangle \left(t_{d}\right)\right.\nonumber \\
 &  & +c_{11}^{(n)}\left\langle \left\langle \cos^{2}\theta\right\rangle ^{2}\right\rangle \left(t_{d}\right)+c_{02}^{(n)}\left\langle \left\langle \cos^{4}\theta\right\rangle \right\rangle \left(t_{d}\right)\nonumber \\
 &  & \left.+\cdots+c_{44}^{(n)}\left\langle \left\langle \cos^{8}\theta\right\rangle ^{2}\right\rangle \left(t_{d}\right)\right\} \label{NitHhgSig}\end{eqnarray}
 where ${\cal {C}}=\left(\sqrt{2\pi n\omega}\right)^{2}2\pi\frac{\left(n\omega\right)^{2}}{c^{3}}=(2\pi)^{2}\left(\frac{n\omega}{c}\right)^{3}$.
The coefficients $c_{jj'}^{(n)}$ are related to $b_{j}^{(n)}$ as
follows \begin{equation}
c_{j,j'}^{(n)}=\left\{ \begin{array}{cc}
\left|b_{j}^{(n)}\right|^{2} & \mathrm{for}\, j=j'\\
2Re\left(b_{j}^{(n)}b_{j'}^{(n)*}\right) & \mathrm{for}\, j\neq j'\end{array}\right.\label{eq:51}\end{equation}

The leading two terms of the signal for $\mathrm{N_{2}}$, Eq. (\ref{NitHhgSig}),
consist of a constant term proportional to $c_{00}^{(n)}$, that arises
from the leading angular momentum term $l=0$ of the active molecular
orbital of $\mathrm{N_{2}}$, and a term proportional to the second
moment $\left\langle \left\langle \cos^{2}\theta\right\rangle \right\rangle \left(t_{d}\right)$
that corresponds to the usual {}``degree of alignment'' $A\left(t_{d}\right)$.
We may note in passing that the above result does not support a recent
model calculation \cite{ram-07,ram-08} that emphasizes that the leading
contribution for HHG signal from $\mathrm{N_{2}}$ arises from the
fourth moment $\left\langle \cos^{4}{\theta}\right\rangle $; that
would require, for example, dropping the basic contribution of the
$l=0$ term i.e. $b_{0}^{(n)}$ in Eq. (\ref{FinNitHhgOpe}) -- for
the HHG operator for $\mathrm{N_{2}}$ -- that of course would not
be justifiable due to the $\sigma-symmetry$ of its active orbital.

\begin{table}

\caption{\label{tab:02}The molecular properties used in this work. $I_{p}$
is adiabatic ionization potential, $B$ is rotational constant of
molecule, $\alpha_{\parallel}$ and $\alpha_{\perp}$ are parallel
and perpendicular polarizability, and $C_{l}^{(m)}$'s are angular
coefficient of the electronic wave function. }

\begin{centering}
\begin{tabular}{cccc}
\hline 
&
$\mathrm{N_{2}}$&
$\mathrm{O_{2}}$&
Ref.\tabularnewline
\hline 
HOMO~~&
~~$\sigma_{g},\, m=0$~~&
~~$\pi_{g},\, m=1$&
~~\cite{jor-73,her-50}\tabularnewline
$I_{p}\,\mathrm{(\mathrm{eV})}$ &
15.58&
12.03&
\cite{ton-02}\tabularnewline
$B$ ($\mathrm{cm^{-1}})$ &
2.0&
1.4377&
\cite{jam-92}\tabularnewline
$\alpha_{\parallel}\left(\textrm{\AA}^{3}\right)$ &
2.38&
2.35&
\cite{hir-54}\tabularnewline
$\alpha_{\perp}\left(\textrm{\AA}^{3}\right)$ &
1.45&
1.21&
\cite{hir-54}\tabularnewline
$C_{0}^{(m)}$ &
2.02&
-&
\cite{ton-02}\tabularnewline
$C_{2}^{(m)}$ &
0.78&
0.62&
\cite{ton-02}\tabularnewline
$C_{4}^{(m)}$ &
0.04 &
0.03&
\cite{ton-02}\tabularnewline
\hline
\end{tabular}
\par\end{centering}
\end{table}

\subsection{Parallel Geometry, $\alpha=0$: Elementary Expression of $T_{e}^{(n)}(\theta;\alpha=0)$
for $\mathrm{O_{2}}$}

$\mathrm{O_{2}}$ has $\pi_{g}$ symmetry, and thus we approximate
its MO by the asymptotic approximation with $m=1$ and $l=2,4$ \cite{ton-02,kje-05}
whose angular coefficient are given in table \ref{tab:02}. The HHG
operator (Eq. (\ref{FinHhgOpe})) for $\mathrm{O_{2}}$ reads\begin{eqnarray}
T_{e}^{(n)}(\theta) & = & \sqrt{2\pi(n\omega)}\nonumber \\
 &  & \times\sum_{l_{i},l_{r}=2,4}d_{01}^{l_{r}}(\theta)\tilde{a}_{zz}^{(n)}(l_{r},l_{i};1)d_{01}^{l_{i}}(\theta)\hphantom{---}\label{OxyHhgOpe}\end{eqnarray}

\noindent By using the expressions for the reduced rotation matrices
from Tab. \ref{tab:01} and simplifying, we may rewrite the operator
as a sum of powers of $\sin^{2}\theta\cos^{2n}\theta$ only,\begin{eqnarray}
T_{e}^{(n)}(\theta) & = & \sqrt{2\pi(n\omega)}\left[b_{1}^{(n)}\,\sin^{2}\theta\cos^{2}\theta+b_{2}^{(n)}\sin^{2}\theta\cos^{4}\theta\right.\nonumber \\
 &  & \left.+b_{3}^{(n)}\sin^{2}\theta\cos^{6}\theta\right]\label{FinOxyHhgOpe}\end{eqnarray}
 where $b_{j}^{(n)}$ -coefficients are given by\begin{eqnarray}
b_{1}^{(n)} & = & \frac{3}{2}\tilde{a}_{zz}^{(n)}(2,2;1)+\frac{45}{16}\tilde{a}_{zz}^{(n)}(4,4;1)\nonumber \\
 &  & -\frac{3}{4}\sqrt{\frac{15}{2}}\left(\tilde{a}_{zz}^{(n)}(2,4;1)+\tilde{a}^{(n)}(4,2;1)\right)\nonumber \\
b_{2}^{(n)} & = & -\frac{105}{8}\tilde{a}_{zz}^{(n)}(4,4;1)\nonumber \\
 &  & +\frac{7}{4}\sqrt{\frac{15}{2}}\left(\tilde{a}_{zz}^{(n)}(2,4;1)+\tilde{a}_{zz}^{(n)}(4,2;1)\right)\nonumber \\
b_{3}^{(n)} & = & \frac{245}{16}\tilde{a}_{zz}^{(n)}(4,4;1)\label{eq:54}\end{eqnarray}
 Finally, substituting operator expression (Eq. (\ref{FinOxyHhgOpe}))
in Eq. (\ref{HhgSig1}) we obtain the $n$th signal of $\mathrm{O_{2}}$
reads\begin{eqnarray}
S^{(n)}(t_{d}) & = & {\cal {C}}\sum_{j=1}^{3}\sum_{j'\geq j}^{3}c_{jj'}^{(n)}\left\langle \left\langle \sin^{2}\theta\cos^{2j}\theta\right\rangle \left\langle \sin^{2}\theta\cos^{2j'}\theta\right\rangle \right\rangle \nonumber \\
 & = & {\cal {C}}\left\{ c_{11}^{(n)}\left\langle \left\langle \sin^{2}\theta\cos^{2}\theta\right\rangle ^{2}\right\rangle \left(t_{d}\right)\right.\nonumber \\
 &  & +c_{12}^{(n)}\left\langle \left\langle \sin^{2}\theta\cos^{2}\theta\right\rangle \left\langle \sin^{2}\theta\cos^{4}\theta\right\rangle \right\rangle \left(t_{d}\right)\nonumber \\
 &  & \left.+\cdots+c_{33}^{(n)}\left\langle \left\langle \sin^{2}\theta\cos^{6}\theta\right\rangle ^{2}\right\rangle \left(t_{d}\right)\right\} \label{OxyHhgSig}\end{eqnarray}
 Above, coefficients $c_{jj'}^{(n)}$ are related to $b_{j}^{(n)}$
coefficients of Eq. (\ref{eq:54}) through Eq. (\ref{eq:51}).

We note that, unlike in the case of $\mathrm{N_{2}}$ considered above,
now there is no constant leading term in the signal for $\mathrm{O_{2}}$,
Eq. (\ref{OxyHhgSig}). This is a consequence of the $\pi-symmetry$
of the active orbital for $\mathrm{O_{2}}$, which does not permit
the lowest $l=0$ angular momentum component for its active orbital.

\subsection{Arbitrary Relative Polarization Angle $\alpha$: HHG Signal}

We now consider the signals for $\mathrm{N_{2}}$ and $\mathrm{O_{2}}$
in the general case in which the probe and the pump polarizations
make an arbitrary angle $\alpha$ between them, as shown in Fig. \ref{fig:SchDia}.
Unlike alignment angle $\theta$, the pump-probe angle $\alpha$ can
be controlled in the laboratory and may provide a possible control
of HHG of molecule. To obtain the signals in terms of elementary trigonometric
functions in this more general case, we refer to Fig. \ref{fig:SchDia}.
The direction of the molecular axis is now denoted by ($\theta',\phi'$).
The same expression for the signal as in the parallel case now holds
in terms of the primed angles. The HHG signal \textit{\emph{(Eq. (\ref{NitHhgSig})
for $\mathrm{N_{2}}$ and Eq. (\ref{OxyHhgSig}) for $\mathrm{O_{2}}$)
for arbitrary angle $\alpha$ now can be written as}}:\begin{eqnarray}
S^{(n)}(t_{d};\alpha) & = & {\cal {C}}\left\{ c_{00}^{(n)}+c_{01}^{(n)}\left\langle \left\langle \cos^{2}\theta'\right\rangle \right\rangle \left(t_{d}\right)\right.\nonumber \\
 &  & +c_{11}^{(n)}\left\langle \left\langle \cos^{2}\theta'\right\rangle ^{2}\right\rangle \left(t_{d}\right)\nonumber \\
 &  & \left.+\cdots+c_{44}^{(n)}\left\langle \left\langle \cos^{8}\theta'\right\rangle ^{2}\right\rangle \left(t_{d}\right)\right\} \label{NitHhgSigAlf}\end{eqnarray}
 for $\mathrm{N_{2}}$ and\begin{eqnarray}
S^{(n)}(t_{d};\alpha) & = & {\cal {C}}\left\{ c_{11}^{(n)}\left\langle \left\langle \sin^{2}\theta'\cos^{2}\theta'\right\rangle ^{2}\right\rangle \left(t_{d}\right)\right.\nonumber \\
 &  & +c_{12}^{(n)}\left\langle \left\langle \sin^{2}\theta'\cos^{2}\theta'\right\rangle \left\langle \sin^{2}\theta'\cos^{4}\theta'\right\rangle \right\rangle \left(t_{d}\right)\nonumber \\
 &  & \left.+\cdots+c_{33}^{(n)}\left\langle \left\langle \sin^{2}\theta'\cos^{6}\theta'\right\rangle ^{2}\right\rangle \left(t_{d}\right)\right\} \label{OxyHhgSigAlf}\end{eqnarray}
 for $\mathrm{O_{2}}$. Above, $\left\langle \left\langle f(\theta')\right\rangle \right\rangle \left(t_{d}\right)=\sum_{J_{0}M_{0}}\rho(J_{0})\left\langle \Phi_{J_{0}M_{0}}\left(t_{d},\theta\right)\left|f(\theta')\right|\Phi_{J_{0}M_{0}}\left(t_{d},\theta\right)\right\rangle $
is an expectation value of function $f(\theta')$ given in probe frame
but evaluated with respect to the rotational wave packet obtained
in the pump frame. Before evaluating the above integral, it is convenient,
therefore, to transform the HHG operators in the variables $(\theta',\phi')$
in the angles $(\theta,\phi)$ of the pump-frame (i.e. with the $z$
along the pump polarization). This can be done by the simple transformations,
given by Eq. (\ref{Cos3D}), where $\phi$ is the angle between plane
of molecular axis - pump pulse and plane of pump - probe pulses. The
elementary expression for the expectation value of the alignment operator
$A\left(t_{d};\alpha\right)=\left\langle \cos^{2}\theta'\right\rangle $
in the case of non-zero angle $\alpha$ reads:\begin{eqnarray}
A(t_{d};\alpha) & = & \left\langle \cos^{2}\theta'\right\rangle \nonumber \\
 & = & \left(\cos^{2}\alpha-\frac{1}{2}\sin^{2}\theta\right)\left\langle \cos^{2}\theta\right\rangle +\frac{1}{2}\sin^{2}\alpha\nonumber \\
 &  & +\frac{1}{4}\sin^{2}\alpha\left(\left\langle \sin^{2}\theta e^{2i\phi}\right\rangle +c.c.\right)\nonumber \\
 &  & +\frac{1}{2}\sin2\alpha\left(\left\langle \sin\theta\cos\theta e^{i\phi}\right\rangle +c.c.\right)\label{eq:57}\end{eqnarray}
 where $\left\langle \sin\theta\cos\theta e^{\pm i\phi}\right\rangle $
couples the $J'$ states with $\Delta J=0,\pm2$ and $\Delta M=\pm1$
whereas $\left\langle \sin^{2}\theta e^{\pm2i\phi}\right\rangle $
couples the $J'$ states with $\Delta J=0,\pm2$ and $\Delta M=\pm2$.
We note that for the linearly polarized pump pulse of the present
interest, the interaction Hamiltonian is proportional to $cos^{2}\theta$,
which is independent of $M$ in the space fixed pump-frame. Thus the
$M$-quantum number of the rotational wave-packet remains constant,
or $M=M_{0}$, throughout the evolution. Hence, the expectation values
of $\left\langle \sin\theta\cos\theta e^{\pm i\phi}\right\rangle $
and $\left\langle \sin^{2}\theta e^{\pm2i\phi}\right\rangle $ vanish
and we get, \begin{equation}
\left\langle \cos^{2}\theta'\right\rangle =\frac{1}{2}\left(3\cos^{2}\alpha-1\right)\left\langle \cos^{2}\theta\right\rangle +\frac{1}{2}\sin^{2}\alpha\label{Cos2Alf}\end{equation}
 In a similar way, we obtain the expectation value for higher order\begin{eqnarray}
\left\langle \cos^{4}\theta'\right\rangle  & = & \frac{1}{8}\left(35\cos^{4}\alpha-30\cos^{2}\alpha+3\right)\left\langle \cos^{4}\theta\right\rangle \nonumber \\
 &  & +\frac{3}{8}\left(-10\cos^{4}\alpha+12\cos^{2}\alpha-2\right)\left\langle \cos^{2}\theta\right\rangle \nonumber \\
 &  & +\frac{3}{8}\sin^{4}\alpha\label{Cos4Alf}\end{eqnarray}
 We note in passing that for $\alpha=0$, $\left\langle \cos^{2}\theta'\right\rangle $
in Eq. (\ref{Cos2Alf}) and $\left\langle \cos^{4}\theta'\right\rangle $
in Eq. (\ref{Cos4Alf}) reduce to $\left\langle \cos^{2}\theta\right\rangle $
and $\left\langle \cos^{4}\theta\right\rangle $, respectively.

Thermal averaging Eq. (\ref{Cos2Alf}) gives us the {}``degree of
alignment'' or the alignment-moment: \begin{equation}
A(t_{d},\alpha)=\frac{1}{2}\left(3\cos^{2}\alpha-1\right)\left\langle \left\langle \cos^{2}\theta\right\rangle \right\rangle (t_{d})+\frac{1}{2}\sin^{2}\alpha\label{DegAli}\end{equation}
 which also appears in the second leading term of the signal for N$_{2}$,
for arbitrary angle $\alpha$, (see, Eq. (\ref{NitHhgSigAlf})). Squaring
and taking the thermal average of Eq. (\ref{Cos2Alf}) gives us $\left\langle \left\langle \cos^{2}\theta'\right\rangle ^{2}\right\rangle $,
which is the third term of HHG signal of $\mathrm{N_{2}}$. The thermal
average of Eq. (\ref{Cos4Alf}) gives us $\left\langle \left\langle \cos^{4}\theta'\right\rangle \right\rangle $
which appears in the fourth term of HHG signal of $\mathrm{N_{2}}$.
The difference of Eq. (\ref{Cos2Alf}) and Eq. (\ref{Cos4Alf}) gives
us \begin{eqnarray}
\left\langle \sin^{2}\theta'\cos^{2}\theta'\right\rangle  & = & \frac{1}{8}\left(-35\cos^{4}\alpha+30\cos^{2}\alpha-3\right)\left\langle \cos^{4}\theta\right\rangle \nonumber \\
 &  & +\frac{1}{8}\left(30\cos^{4}\alpha-24\cos^{2}\alpha+2\right)\left\langle \cos^{2}\theta\right\rangle \nonumber \\
 &  & +\frac{1}{8}\left(-3\sin^{4}\alpha+4\sin^{2}\alpha\right)\label{Sin2Cos2Alf}\end{eqnarray}
 Squaring and then thermally averaging Eq. (\ref{Sin2Cos2Alf}) yields
the leading term of HHG signal of $\mathrm{O_{2}}$, given by Eq.
(\ref{OxyHhgSigAlf}). In a similar way, we can explicitly exhibit
the $\alpha$-dependence of the higher order terms in the signal for
$\mathrm{O_{2}}$, Eq. (\ref{OxyHhgSigAlf}), as well.

\section{Results and Discussions }

\subsection{Signals in the Time Domain}

\begin{figure}

\noindent \begin{centering}
\includegraphics[scale=0.3]{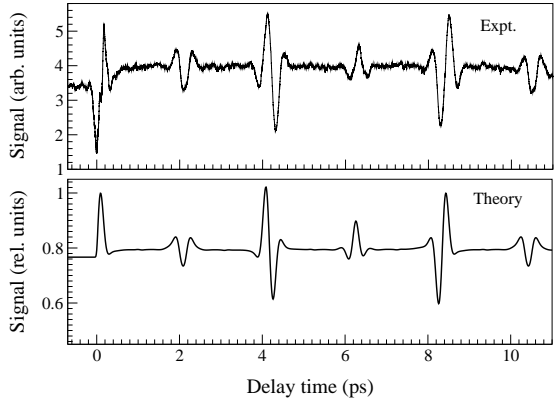} 
\par\end{centering}

\caption{\label{fig:DynN2}Comparison of the experimental \cite{miy-05,miy-06}
and the theoretical dynamic $19$th HHG signal for $\mathrm{N_{2}}$;
pump intensity $I=0.8\times10^{14}$ W/cm$^{2}$, probe intensity
$I=1.7\times10^{14}$ W/cm$^{2}$; duration $40\,\mathrm{fs}$, wavelength
$800\,\mathrm{nm}$, temperature $200\,\mathrm{K}$. }
\end{figure}
We now apply the theory to analyze the observed HHG signals from the
diatomic molecules, $\mathrm{N_{2}}$ and $\mathrm{\mathrm{O_{2}}}$.
In typical recent experiments (e.g. \cite{ita-05,miy-05,kan-05,miy-06})
an ensemble of $\mathrm{N_{2}}$ or $\mathrm{O_{2}}$ molecules is
first set into free rotation by a femtosecond pump pulse. The HHG
signals were detected by monitoring the emission due to a second more
intense femtosecond probe pulse, that was delayed with respect to
the first by successively increasing the time intervals, $t_{d}$,
in the picosecond domain, between them.

\begin{figure}

\noindent \begin{centering}
\includegraphics[scale=0.3]{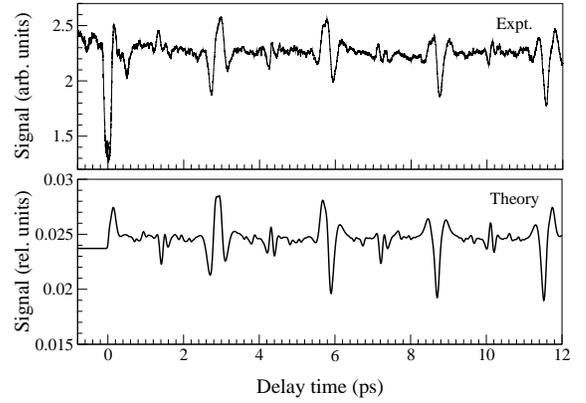} 
\par\end{centering}

\caption{\label{fig:DynO2}Comparison of the experimental \cite{miy-05,miy-06}
and the theoretical $19$th harmonic dynamic signal for $\mathrm{O_{2}}$;
pump intensity $I=0.5\times10^{14}$ W/cm$^{2}$, probe intensity
$I=1.2\times10^{14}$ W/cm$^{2}$; durations $40\,\mathrm{fs}$, wavelengths
$800\,\mathrm{nm}$, temperature $200\,\mathrm{K}$. }
\end{figure}

In the experiments for $\mathrm{N_{2}}$, for example by Miyazaki
et al. \cite{miy-05,miy-06}, a peak pump-intensity $I_{1}=0.8\times10^{14}\,\mathrm{W/cm^{2}}$,
a peak probe-intensity $I_{2}=1.7\times10^{14}\,\mathrm{W/cm^{2}}$
were used; the central wavelength $\lambda=800\,\mathrm{nm}$ and
the pulse duration $\tau=$40 fs were kept the same for both the pulses.
For the experiment with $\mathrm{O_{2}}$, the harmonic signal was
measured in a similar fashion for $I_{1}=0.5\times10^{14}\,\mathrm{W/cm^{2}}$
and $I_{2}=1.2\times10^{14}\,\mathrm{W/cm^{2}}$; the other parameters
were kept the same as in the case of $\mathrm{N_{2}}$. For the purpose
of a direct comparison, our calculations were performed for the same
parameter values as in these experiments \cite{miy-05,miy-06}. In
Fig. \ref{fig:DynN2} and Fig. \ref{fig:DynO2}, we compare the calculated
HHG signals as a function of $t_{d}$ for $\mathrm{N_{2}}$ and $\mathrm{O_{2}}$,
with the experimental data obtained for the $19$th order harmonic.
The effective ensemble temperature was taken to be T= 200 K, that
was estimated from the matching of the peak position of the spectral
distribution with that of the Boltzmann distribution as suggested
first in \cite{fai-07}. It can be seen from Fig. \ref{fig:DynN2}
that the experimental data for $\mathrm{N_{2}}$ show the {}``revival''
phenomenon with a full revival period $T_{rev}=8.4\,\mathrm{ps}$
(which is consistent with the rotational constant of $\mathrm{N_{2}}$
(cf. Tab. \ref{tab:02})) as well as a $\frac{1}{2}$ and a $\frac{1}{4}$
fractional-revival. The observed signal for $\mathrm{O_{2}}$ shows,
in addition to the full revival (period for $\mathrm{O_{2}}$ is $T_{rev}=11.6\,\mathrm{ps}$)
and two fractional revivals similar to the two seen for $\mathrm{N_{2}}$,
an additional $\frac{1}{8}$ revival. The calculated signals can be
seen to follow the same sequence of the full and the three fractional
revivals as seen in the experimental signal. We note that these observations
for $\mathrm{N_{2}}$ and $\mathrm{O_{2}}$ are also consistent with
the data of Itatani et al. \cite{ita-05} and Kanai et al.\cite{kan-05}.

To understand the similarities and the differences between the signals
for $\mathrm{N_{2}}$ and $\mathrm{O_{2}}$, we use the analytical
results of the present theory below. The properties of the HHG signal
of $\mathrm{N_{2}}$ are governed by Eq. (\ref{NitHhgSig}). The first
term gives a constant background. The second term $\left\langle \left\langle \cos^{2}\theta\right\rangle \right\rangle \left(t_{d}\right)$
is the dominant dynamic term and makes the signal to mimic the {}``degree
of alignment'' $A(t_{d})\equiv\left\langle \left\langle \cos^{2}\theta\right\rangle \right\rangle \left(t_{d}\right)$.
The third term $\left\langle \left\langle \cos^{2}\theta\right\rangle ^{2}\right\rangle \left(t_{d}\right)$
give unequal maxima and minima i.e. the difference between the maximum
signal and the average signal is greater than the difference between
the average signal and the minimum signal. Furthermore we point out
that at a lower initial temperature, the valley of $\left\langle \left\langle \cos^{2}\theta\right\rangle ^{2}\right\rangle \left(t_{d}\right)$
that occurs, for higher temperatures, at the $\frac{1}{4}T_{rev}$
revival, can \textit{split} into two valleys, due to this term, and
thus the third term can strongly affect 
the HHG spectrum, as can be seen in the experiment by Itatani, \emph{et
al.} \cite{ita-04,ita-05}. Another earlier puzzle regarding its dynamic
signal observed was the failure of the alignment measure $A(t_{d})=\left\langle \left\langle \cos^{2}\theta\right\rangle \right\rangle \left(t_{d}\right)$
to account for the dynamic HHG signal for $\mathrm{O_{2}}$, observed
by Itatani \emph{et al.} \cite{ita-05}. In fact, Itatani \emph{et
al.} found that their data behaved more closely to the expectation
value $B(t_{d})\equiv\left\langle \left\langle \sin^{2}2\theta\right\rangle \right\rangle \left(t_{d}\right)$.
From Eq. (\ref{OxyHhgSig}) it can be seen that indeed the \textit{leading}
term of the signal for $\mathrm{O_{2}}$ is given by $\left\langle \left\langle \sin^{2}\theta\cos^{2}\theta\right\rangle ^{2}\right\rangle \left(t_{d}\right)=\frac{1}{16}\left\langle \left\langle \sin^{2}2\theta\right\rangle ^{2}\right\rangle \left(t_{d}\right)$,
which is directly proportional to the observed signal. Moreover, the
present theory also predicts that there ought to be modifications
to this result due to the higher order terms in Eq. (\ref{OxyHhgSig}).
In contrast to $\mathrm{N_{2}}$ there is no significant difference
between minima and maxima for $\mathrm{O_{2}}$, since all terms in
Eq. (\ref{eq:39}) have similar minima and maxima. The present theory
also predicts that there ought to be modifications to this result
due to the higher order terms in Eq. (\ref{OxyHhgSig}). In contrast
to $\mathrm{N_{2}}$, however, there should not be a significant asymmetry
between the size of the maxima and the minima for $\mathrm{O_{2}}$,
since the terms in Eq. (\ref{eq:39}) have similar maxima and minima.
The can be, however, quantitative contributions from the higher order
terms predicted by the theory. In fact, as mentioned earlier, Kanai
\emph{et al}. \cite{kan-05} found empirically that their experimental
HHG signals for $\mathrm{N_{2}}$ and $\mathrm{O_{2}}$ demanded heuristic
introduction of operators involving \textit{higher} orders of $\cos^{2}\theta$
functions, or Legendre polynomials, as the dynamic signal could not
be well expressed in term of $\left\langle \left\langle \cos^{2}\theta\right\rangle \right\rangle \left(t_{d}\right)$
only for $\mathrm{N_{2}}$, or $\left\langle \left\langle \sin^{2}2\theta\right\rangle \right\rangle \left(t_{d}\right)$
only for $\mathrm{O_{2}}$. In fact, the present theory provides an
\textit{ab initio} derivation of the desired general expansion of
the HHG signal in terms of the moments of the Legendre polynomials
Eq. (\ref{GenHhgSig}) and/or of the powers of $\cos^{2}\theta$,
e.g. Eqs. (\ref{OxyHhgSig}) and (\ref{NitHhgSig}).

A related characteristic of interest first observed by Miyazaki \emph{et
al} (e.g. \cite{miy-06}) is the appearance of extra series and lines
in the Fourier spectrum of the dynamic HHG signal for \textit{both}
$\mathrm{N_{2}}$ and $\mathrm{O_{2}}$, that are Raman forbidden.
These extra lines can not be attributed to $A\left(t_{d}\right)=\left\langle \left\langle \cos^{2}\theta\right\rangle \right\rangle \left(t_{d}\right)$,
for $\mathrm{N_{2}}$, or to $B\left(t_{d}\right)=\left\langle \left\langle \sin^{2}2\theta\right\rangle \right\rangle \left(t_{d}\right)$
for $\mathrm{O_{2}}$. It will be seen below that the ${F.T.}$ of
the higher orders terms of Eq. (\ref{NitHhgSig}) for $\mathrm{N_{2}}$
and of Eq. (\ref{OxyHhgSig}) for $\mathrm{O_{2}}$, given by the
present theory can consistently account for their appearance. %
\begin{figure}
\begin{centering}
\includegraphics[scale=0.35]{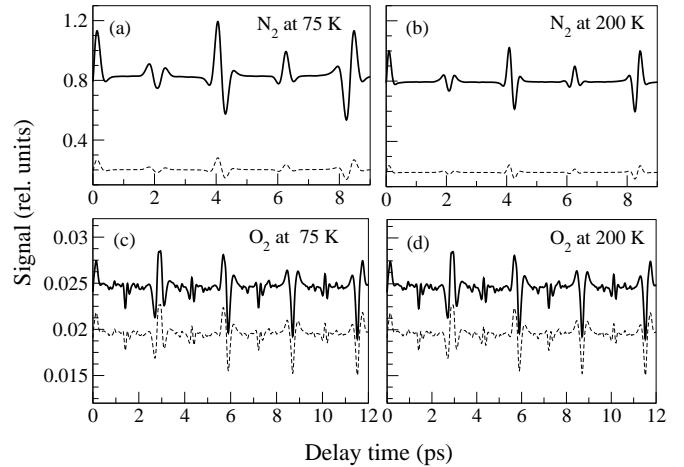} 
\par\end{centering}

\caption{\label{fig:TemOrdDep}The dependency of HHG signal on the harmonic
order and on the initial temperature of the gas. Solid and dashed
lines are for the $19$th and the $21$th order, respectively. }
\end{figure}

Comparing the expressions for the signals for $\mathrm{N_{2}}$ and
$\mathrm{O_{2}}$ and directly calculating the $c_{jj'}^{(n)}$ coefficients
in the respective signals it is found that the signal for $\mathrm{N_{2}}$
is much stronger than that for $\mathrm{O_{2}}$, as also observed
experimentally \cite{miy-06}.

We may briefly discuss here the dependency of the HHG signals on the
initial temperature, an example of which is shown in Fig. \ref{fig:TemOrdDep}.
It can be seen from the figure that the lower initial temperature
gives a greater amplitude of revival. This may be understood as follows.
A lower initial temperature gives a lower value of the maximum of
the statistically occupied $J_{0}$ levels and hence also a lower
value of the maximum initial value of $M_{0}$, than at a higher temperature.
As a result of interaction with the linearly polarized pump pulse
(quantization axis along the polarization axis) at a given intensity,
each wavepacket that evolves from a given initial $\left|J_{0}M_{0}\right\rangle $
state, can couple to the higher levels $J'>max.(J_{0})$ but can not
raise the initial maximum value of $M_{0}$. Therefore, for a given
intensity, the ratio of $J'$ to $M_{0}$ is higher for a lower temperature,
and as a consequence the degree of alignment $A\left(t_{d}\right)=\left\langle \left\langle cos^{2}\theta\right\rangle \right\rangle $
tends to be also higher, implying that the molecule becomes more strongly
aligned during a revival.

\subsection{Rotational Revivals: Periods and Phases}

If a linear molecule has a permanent dipole moment (e.g. hetero-nuclear
diatomics), then the interaction Hamiltonian of the (pump) laser with
the molecular frame depends on the first power of $\cos{\theta}$,
where $\theta$ is the angle of rotation of the molecular axis with
respect to the laser polarization axis. In contrast, the interaction
with the polarizability of the molecule (e.g. for homo- or hetero-nuclear
diatomics) depends on $\cos^{2}{\theta}$. Thus in general the interaction
may contain the operators $\cos^{n}{\theta}$ with $n=1$ and/or $2$.
Then in either case, the rotational wavepackets created by the later
can be written in the form:

\begin{eqnarray}
\Phi_{J_{0}M_{0}}(t) & = & \sum_{j=0,1,2,3\cdots}C_{J_{0}+nj,M_{0}}(t)\nonumber \\
 & \times & e^{-\frac{i}{\hbar}E_{J_{0}+nj}t}\left|J_{0}+nj,M_{0}\right\rangle \hphantom{---}\end{eqnarray}

This can be obtained, for example, from a consideration of the perturbative
solution of Eq. (\ref{NucSchEqu}) in successive power of the interaction
Hamiltonian, and noting that the rotational eigenstates couple either
by $P_{2}(\cos{\theta}$) (in the absence of a permanent dipole moment)
with a minimum (non-zero) $n=2$ or by $P_{1}(\cos{\theta})$ and
$P_{2}(\cos{\theta})$ (in the presence of a permanent dipole moment)
with a minimum $n=1$. It can be readily understood from the well
known properties of the vector addition coefficients that appear in
the integration over the product of three spherical harmonics (cf.
para below) that the expectation value of the $N$th cosine-moment
with respect to a rotational wavepacket at a time $t=t_{d}$, takes
the form:

\begin{eqnarray}
\left\langle \cos^{N}\theta\right\rangle _{J_{0}M_{0}}\left(t_{d}\right) & = & \left\langle \Phi_{J_{0}M_{0}}\left(t_{d}\right)\left|\cos^{N}\theta\right|\Phi_{J_{0}M_{0}}\left(t_{d}\right)\right\rangle \nonumber \\
 & = & \sum_{s}^{N}\sum_{p=-s}^{s}\sum_{j=0,1,2,3\cdots}\nonumber \\
 & \times & C_{J_{0}+nj+p,M_{0}}^{J_{0}M_{0}*}\left(t_{d}\right)C_{J_{0}+nj,M_{0}}^{J_{0}M_{0}}\left(t_{d}\right)\nonumber \\
 & \times & a_{s}\left\langle Y_{J_{0}+nj+p,M_{0}}\right|Y_{s,0}\left|Y_{J_{0}+nj,M_{0}}\right\rangle \nonumber \\
 & \times & \exp\left(-\frac{i}{\hbar}\left(E_{J_{0}+nj+p}-E_{J_{0}+nj}\right)t_{d}\right)\nonumber \\
 &  & \hphantom{--}\label{ali:27}\end{eqnarray}
 where, the integers $s$ and $p$ have the same parity (even or odd)
as the parity of $N$. This follows from the fact that $\cos^{N}{\theta}$
can be expressed as a linear combination: $\cos^{N}{\theta}=\sum_{s}a_{s}P_{s}(\cos{\theta})$,
for all $s$ up to $N$, and since the matrix elements $\left\langle Y_{J_{0}+nj',M_{0}}\right|Y_{s,0}\left|Y_{J_{0}+nj,M_{0}}\right\rangle =0$,
unless, $J_{0}+nj'=J_{0}+nj+(p-s)\ge0$, and $J_{0}+nj+(p-s)+J_{0}+nj+s=\mathrm{even}$.
Thus, the phase of each individual term of Eq. (\ref{ali:27}), for
any given value of the integers $J_{0},n,N,j$, is given by\begin{eqnarray}
\Delta\phi_{n,N}^{J_{0}}(t_{d}) & = & \frac{1}{\hbar}\left(E_{J_{0}+nj+p}-E_{J_{0}+nj}\right)t_{d}\nonumber \\
 &  & \times2\pi hBct\left(\pm2pJ_{0}\pm2npj+p^{2}\pm p\right)\nonumber \\
 & = & 2\pi\frac{t_{d}(hBc)2}{\hbar}\left(J_{0}+nj+\frac{p^{2}+p}{2}\right)\label{PhaDif}\end{eqnarray}
\begin{figure}
\noindent \begin{centering}
\includegraphics[scale=0.5]{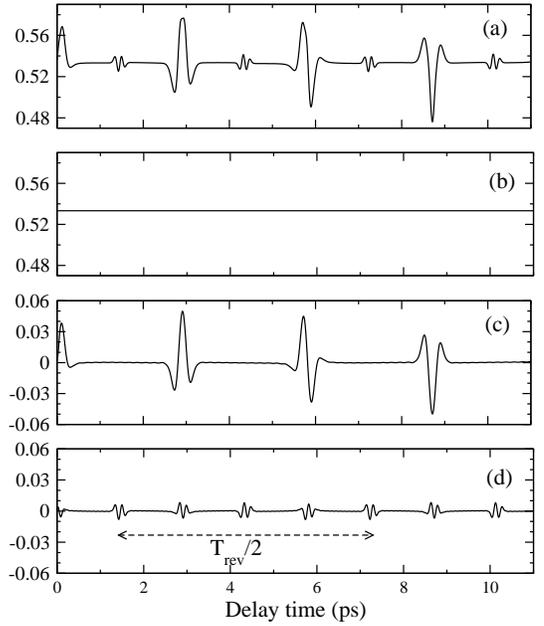} 
\par\end{centering}

\caption{\label{fig:FraRevFre}Revival structure of the moment $\left\langle \left\langle \sin^{2}2\theta\right\rangle \right\rangle (t_{d})$,
for the case of O$_{2}$, on the beat frequency: for all $\Delta J$
retained (panel a), with $\Delta J=0$ only (panel b), with $\Delta J=\pm2$
only (panel c), and with $\Delta J=\pm4$ only (panel d). It is clear
that the transition with $\Delta J=\pm4$ has the lowest fractional
revival at $\frac{1}{8}T_{rev}$, and has the shortest period.}
\end{figure}

where, we have used $E_{J,M}\equiv J(J+1)hBc$; $B$ is the rotational
constant, $T_{rev}\equiv\frac{1}{2Bc}$ is the rotational period,
and $h=2\pi\hbar$. We note first that the quantity in the last parentheses
above is an integer, independent of the value of $j$ and $J_{0}$.
We note that the maximum value of $s$ or $p$ above is $N$. The
phase difference (Eq. (\ref{PhaDif})) therefore equals to an even
or odd multiple of $\pi$, or odd multiple of $\frac{\pi}{2}$, depending
on the parity of the groups of rotational states. Therefore, the shortest
time period for which the phases of {\textit{all}} terms or all
terms within a parity group 
become equal in Eq. (\ref{PhaDif}), and hence coherently enhance
the signal, is clearly \begin{equation}
T_{min}=\frac{1}{nN}T_{r}.\label{ali:30}\end{equation}
 For times between the successive coherent enhancements or {}``revivals'',
the individual phases in Eq. (\ref{PhaDif}) disperse away from one
another and the revival peaks tend to be washed out by destructive
interference, and the HHG signal reduces to the average or the back-ground
level. %
\begin{figure}
\begin{centering}
\includegraphics[scale=0.35]{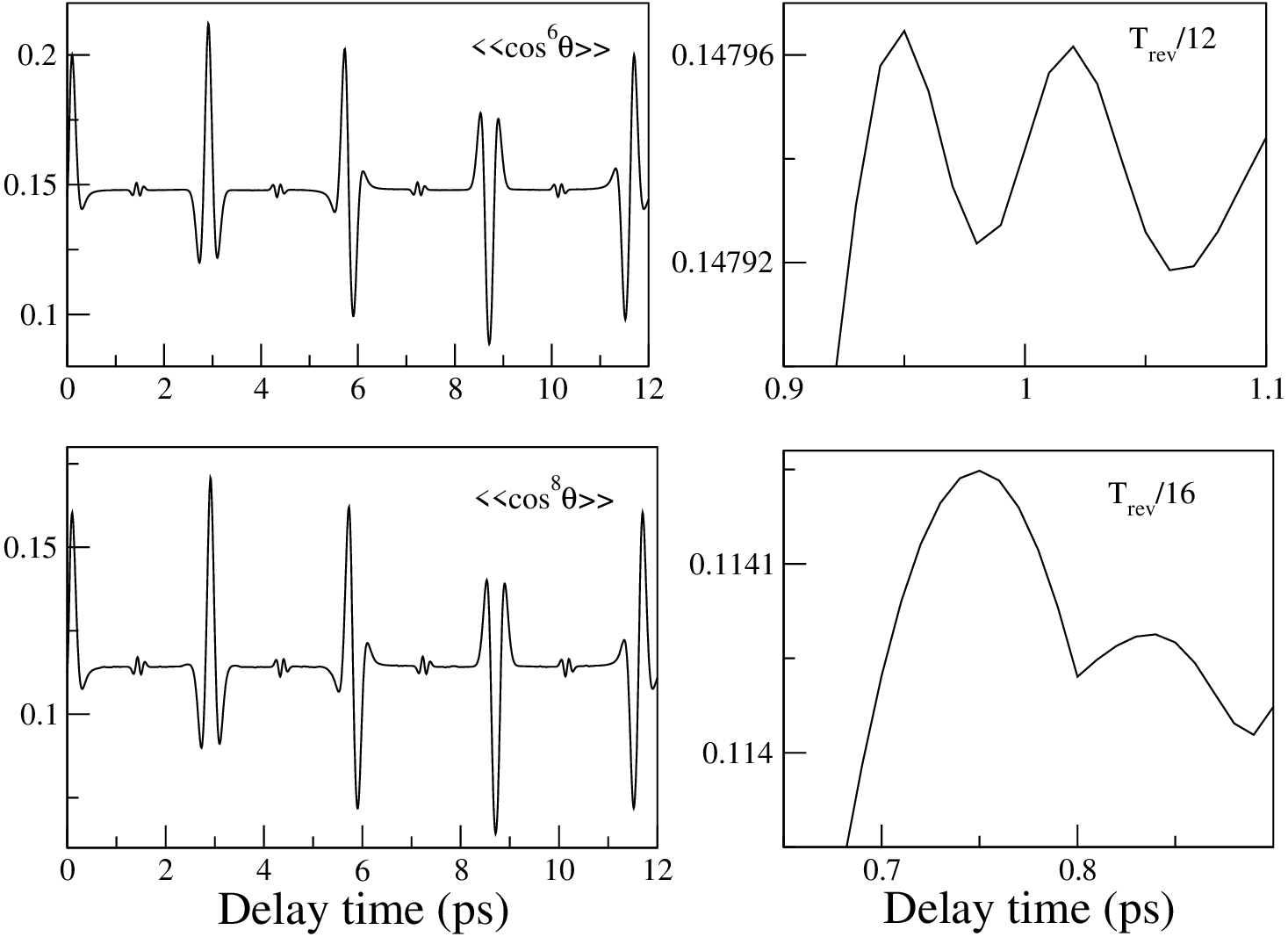} 
\par\end{centering}

\caption{\label{fig:FraRev} Weak components of the fractional revivals in
the dynamical alignment signal for the case of $\mathrm{O_{2}}$.
Contribution from $\left\langle \left\langle \cos^{6}\theta\right\rangle \right\rangle (t_{d})$,
near $T_{rev}/12$ (upper panel), and from $\left\langle \left\langle \cos^{8}\theta\right\rangle \right\rangle (t_{d})$,
near $T_{rev}/16$ (lower panel). They can hardy be detected in the
full delay-time signal that is dominated the leading lower order moments.
$I=0.5\times10^{14}\,\mathrm{W/cm^{2}}$, FWHM 40 fs, and initial
temperature 300 K.}
\end{figure}

We may summarize the above result as a {}``revival theorem'': If
the laser-molecule interaction Hamiltonian is characterized by the
lowest power $n$ of $\cos^{n}{\theta},\,\mathrm{with}\, n=1\,\mathrm{or}\,2$,
$\theta$ is the rotation angle, and if the highest discernible (numerically
significant) moment in the expression of the signal is $\left\langle \Phi_{J_{0}M_{0}}(t)\right|\cos^{N}{\theta}\left|\Phi_{J_{0}M_{0}}(t)\right\rangle $,
$N\ge1$, then the experimental signal would exhibit as many as $n\times N$
revivals within a full period $T_{r}=\frac{1}{2Bc}$, $B$ is the
rotational constant. Inversely, by counting the number of fractional
revivals in the observed HHG signal, one may determine the highest
order, $N$, and hence also the significant {}``cosine moments''
(up to the order $N$) that would be necessary to fit the observed
signal. We may note that the above theorem covers the well-known cases
of fractional revivals discussed earlier \cite{vra-02,ave-89,vra-96,blu-96}
as special cases.

For homonuclear diatomic molecules with no permanent dipole moment,
the lowest order pump pulse interaction is due to the polarizability
tensor with $n=2$. Thus for the standard alignment moment, $A\left(t_{d}\right)\equiv\left\langle \left\langle \cos^{2}{\theta}\right\rangle \right\rangle \left(t_{d}\right)$
with $N=2$, we get the lowest fractional period $T_{\frac{1}{4}}=\frac{1}{4}T_{r}$,
and the subsequent two fractional revivals $T_{\frac{1}{2}}$, $T_{\frac{3}{4}}$
(defined analogously) and the full revival at $T_{r}$, with in a
period. Thus the presence of the highest significant fourth cosine-moment
with $N=4$ would show the lowest $\frac{1}{nN}=\frac{1}{8}$ revival,
plus the subsequent six fractional revivals at ($\frac{1}{4}$, $\frac{3}{8}$,
$\frac{1}{2}$, $\frac{5}{8}$, $\frac{3}{4}$, $\frac{7}{8}$) $T_{r}$,
within a full period $T_{r}$. An example containing the effect of
the fourth cosine-moment is $B\left(t_{d}\right)=\left\langle \left\langle \sin^{2}2\theta\right\rangle \right\rangle \left(t_{d}\right)$,
which is illustrated in Fig. \ref{fig:FraRevFre}. For a heteromolecular
diatomic molecule with a permanent dipole moment, the lowest order
interaction Hamiltonian is characterized by the first power of $\cos{\theta}$
i.e. $n=1$. Thus the alignment measure, a cosine moment with $N=2$,
will show $n\times N=2$ revivals within in the full period. Higher
order revivals may occur since Eq. (\ref{ali:30}) in principle holds
for any combination $(N,n)$. We may recall, however, that for large
$N$, the expectation value might be too weak for the lowest fractional
revivals to be measured with sufficient resolution in practice. This
circumstance is illustrated in Fig. \ref{fig:FraRev} which shows
the high order fractional revivals for $N=6$ and $N=8$ cosine-moments,
$\left\langle \left\langle \cos^{6}\theta\right\rangle \right\rangle (t_{d})$
and $\left\langle \left\langle \cos^{8}\theta\right\rangle \right\rangle (t_{d})$,
along with their magnifications.

\subsection{Phase Relations of Fractional Revivals}

Can one predict the relative phases of the fractional revivals? We
may answer this question positively. From the phase difference (Eq.(\ref{PhaDif})),
one finds:\begin{equation}
\begin{array}{c}
\Delta\phi_{2,2}^{J_{0}+1}(T_{rev})\,-\,\Delta\phi_{2,2}^{J_{0}}(T_{rev})\,=\,4\pi\\
\Delta\phi_{2,2}^{J_{0}+1}(T_{rev}/2)\,-\,\Delta\phi_{2,2}^{J_{0}}(T_{rev}/2)\,=\,2\pi\\
\Delta\phi_{2,2}^{J_{0}+1}(T_{rev}/4)\,-\,\Delta\phi_{2,2}^{J_{0}}(T_{rev}/4)\,=\,\pi\end{array}\label{ali:32}\end{equation}

\begin{figure}

\begin{centering}
\includegraphics[scale=0.45]{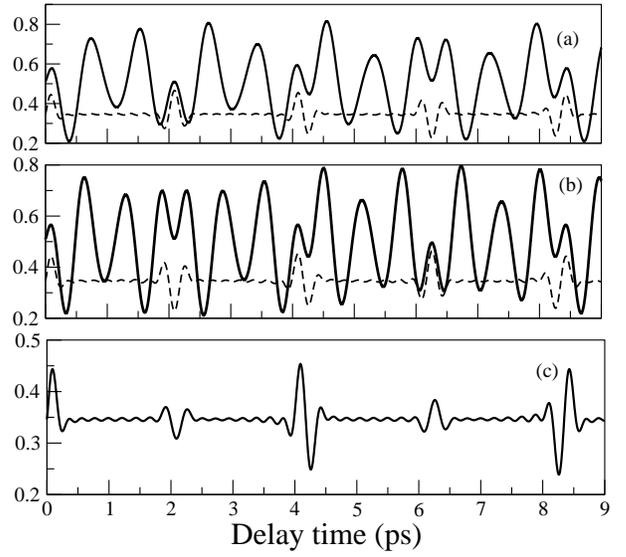} 
\par\end{centering}

\caption{\label{fig:OddEveCon}Dependence of alignment moment $\left\langle \left\langle \cos^{2}\theta\right\rangle \right\rangle (t_{d})$
(for the case of $\mathrm{N_{2}}$) on $\left|J_{0},M_{0}\right\rangle $.
Panel (a): calculated with a single state $\left|5,0\right\rangle $
only (solid line), and retaining all odd $J_{0}$ with $J_{max}=19$
(dashed line). Panel (b): with $\left|6,0\right\rangle $ only (solid
line), and all even $J_{0}$ with $J_{max}=20$ (dashed line). Panel
(c): calculated with all $J$ upto $J_{max}=20$. Results are for
pulse intensity $0.8\,\times10^{14}\,\mathrm{W/cm^{2}}$ and with
FWHM 40 fs, for initial temperature 300 K. }
\end{figure}

\noindent Eq. (\ref{ali:32}) predicts that at $\frac{1}{4}T_{r}$
the phase for $J_{even}$ is an exact mirror image of the phase for
$J_{odd}$, as in fact is the case in Fig. \ref{fig:OddEveCon}, calculated
for $\mathrm{N_{2}}$. From the above, we may further predict that:
\\
 (a) for $\mathrm{O_{2}}$, which posses $J_{odd}$ levels only, will
show a {}``peak'' at $T_{r}/4$,\\
 (b) $\mathrm{CO_{2}}$, which possesses $J_{even}$ levels only,
will show a {}``valley'' at $T_{r}/4$, and\\
 (c) $\mathrm{N_{2}}$, which possesses both the majority $J_{even}$
levels and the and the minority $J_{odd}$ levels in the ratio $2:1$
(due to the nuclear statistics of the molecule \cite{her-50}), will
show the the revival at $T_{r}/4$ that would be a {}``valley''
like the one for the $J_{even}$ levels only, but with only half its
normal {}``depth'', due to the counter contribution from the minority
$J_{odd}$ levels.

We note that one may also predict the nuclear statistics of such molecules
by comparing the revival shape at $T_{rev}/2$ and $T_{rev}/4$. Let
us first define a modulation amplitude at half-revival to be equal
to the difference between peak and the base (or average) signal: ($A_{1/2}=S_{1/2}^{top}-S_{1/2}^{av.}$).
Similarly, a modulation amplitude at quarter revival is equal to the
difference between the top and the base (average) signal: ($A_{1/4}=S_{1/4}^{top}-S_{1/4}^{av.}$).
The amplitude at half-revival is a sum of even and odd $J$ contributions,
and therefore $A_{1/2}$ is always positive. In contrast, the amplitude
at the quarter-revival arises from their difference, and therefore
$A_{1/4}$ can be positive (if it makes a {}``top'' alignment) or
negative (if it makes an {}``anti-top'' alignment). Therefore, the
existence of a {}``top'' signal at the quarter-revival is a sign
that even $J$ levels are dominant. Similarly the presence of an {}``anti-top''
signal at the quarter-revival signal is a sign of dominant odd $J$
levels. From this observation, one can deduce the nuclear statistics
from the ratio between the effective (finite) number of even and the
odd $J$ levels ($J_{even}$ and $J_{odd}$, respectively) excited:
\begin{equation}
\frac{J_{even}}{J_{odd}}=\frac{A_{1/2}-A_{1/4}}{A_{1/2}+A_{1/4}}\label{ali:NucRat}\end{equation}

Thus, for example, the dynamic signal of $\mathrm{O_{2}}$ shows $A_{1/2}=A_{1/4}$
indicating the absence of the even $J$ levels. In contrast, $A_{1/2}=-A_{1/4}$
for $\mathrm{CO_{2}}$, indicating the absence odd $J$ levels. For
$\mathrm{N_{2}}$, we have $A_{1/4}=-\frac{1}{3}A_{1/2}$, and hence
we have $J_{even}:J_{odd}=2:1$. This property might be used for detecting
the existence of isotopes of a molecular sample, as has been suggested
recently \cite{flei-06}.

\subsection{Beat Frequencies}

\begin{figure}
\begin{centering}
\includegraphics[scale=0.42]{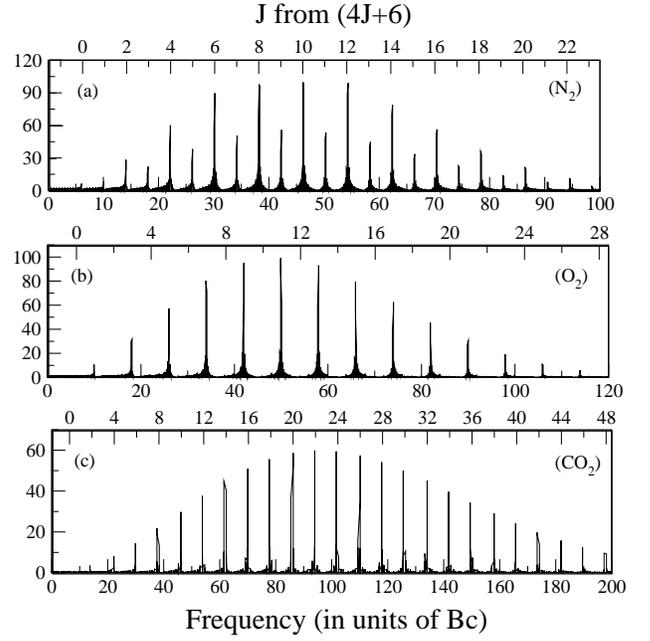} 
\par\end{centering}

\caption{\label{fig:FouSpeCos}Fourier transform of the alignment moment $A(t_{d})\equiv\left\langle \left\langle \cos^{2}\theta\right\rangle \right\rangle (t_{d})$,
plotted using $Bc$ as the basis frequency (lower scale). Following
the $(4J+6)$-rule for the line positions from $\left\langle \left\langle \cos^{2}\theta\right\rangle \right\rangle $,
the peak frequencies are seen to occur for odd $J$ only, for the
case of $\mathrm{O_{2}}$, even $J$ only for the case of $\mathrm{CO_{2}}$,
and for both even and odd $J$, for the case of $\mathrm{N_{2}}$.
The corresponding $J$ values are shown in upper scale. }
\end{figure}

From Eq. (\ref{ali:27}), it is seen that the phase difference associated
with $\left\langle \cos^{2}\theta\right\rangle $ is $(B/\hbar)(4J+6)$.
For $B$ in $\mathrm{cm^{-1}}$, the phase difference reads\begin{equation}
\Delta\phi\,(J\rightarrow J\pm2)=\,2\pi Bc(4J+6)\label{ali:f20}\end{equation}

\noindent with $c$ in $\mathrm{cm/second}$. According to Eq. (\ref{ali:f20}),
one can make a Fourier transform of $\left\langle \cos^{2}\theta\right\rangle $
using $Bc$ as basis frequency and find a series of peaks at \textbf{$(4J+6)$}.
Fig. \ref{fig:FouSpeCos} shows the Fourier transform of $\left\langle \left\langle \cos^{2}\theta\right\rangle \right\rangle $
of $\mathrm{N_{2}}$, $\mathrm{O_{2}}$, and $\mathrm{CO_{2}}$. The
spectrum of $\mathrm{O_{2}}$ has peak series at $(10,18,26,..)Bc=(4J_{odd}+6)Bc$,
showing that $\mathrm{O_{2}}$ has odd $J$ levels only. In contrast,
the peak series of $\mathrm{CO_{2}}$ are located at $(6,14,22,30,...)Bc=(4J_{even}+6)$,
showing that $\mathrm{CO_{2}}$ has even $J$ levels only. For $\mathrm{N_{2}}$,
we obtain a series $(6,14,22,30,...)Bc=(4J_{even}+6)$ that is twice
as strong as the series $(10,18,26,..)Bc=(4J_{odd}+6)Bc$. It implies
that both even and odd $J$ levels are present in $\mathrm{N_{2}}$,
in the ratio $J_{even}:J_{odd}=2:1$. These conclusion are consistent
with the analysis based on the dynamic signals. %

\begin{figure}
\begin{centering}
\includegraphics[scale=0.45]{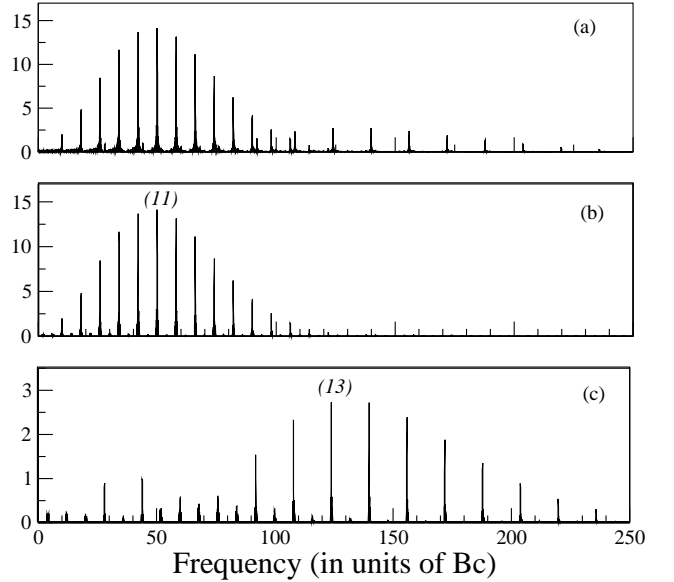} 
\par\end{centering}

\caption{\label{fig:FouSpeSinO2}Fourier transform of $\left\langle \left\langle \sin^{2}2\theta\right\rangle \right\rangle $
of $\mathrm{O_{2}}$ : all lines (panel a), line from transitions
with $\Delta J=\pm2$ only (panel b), and lines from transition with
$\Delta J=\pm4$ only (panel c); pulse of intensity $0.5\times10^{14}\,\mathrm{W/cm^{2}}$,
FWHM = 40 fs, initial temperature 300 K.}
\end{figure}

For $\left\langle \left\langle \sin^{2}2\theta\right\rangle \right\rangle $,
there are two kinds of difference or beat frequency. The first one
is related to the transitions with $\Delta J=\pm2$ and is expressed
by Eq. (\ref{ali:f20}). The second one is related to the transitions
with $\Delta J=\pm4$ and can be expressed as\begin{equation}
\Delta\phi(J\rightarrow J\pm4)=2\pi Bc(8J+20)\label{ali:f21}\end{equation}

\noindent As a results, in addition to the series of lines \textbf{$(4J+6)$},
the Fourier transform of $\left\langle \left\langle \sin^{2}2\theta\right\rangle \right\rangle $
also has another series of lines at \textbf{$(8J+20)$}, with $\Delta J=4$.
Fig. \ref{fig:FouSpeSinO2} shows the calculated Fourier transform
of $\left\langle \left\langle \sin^{2}2\theta\right\rangle \right\rangle $
of $\mathrm{O_{2}}$. It is seen from Fig. \ref{fig:FouSpeSinO2},
that the first series ($\Delta J=\pm2$) reaches its maximum at $J_{max}=11$,
while the second one ($\Delta J=\pm4$) at $J_{max}=13$. This difference
comes from the fact that the $\Delta J=4$ transition requires $\Delta J=2$
as an intermediate transition. As a result, a $\Delta J=4$ transition
can occur one step after the $\Delta J=2$ transition; for $\mathrm{O_{2}}$
with only $J_{odd}$ levels present, this implies a a shift in $J$
by 2, from $J_{max}=11$ to $J_{max}=13$, as seen above. From Fig.
\ref{fig:FouSpeSinO2} one also finds that the intensity of the second
transition is smaller than of the first one. This arises from the
circumstance that the allowed matrix element of the second transition
with the greater separation in $J$ is weaker than the one with the
lesser separation.

\subsection{Signals in the Frequency Domain}

\begin{figure}
\noindent \begin{centering}
\includegraphics[scale=0.5]{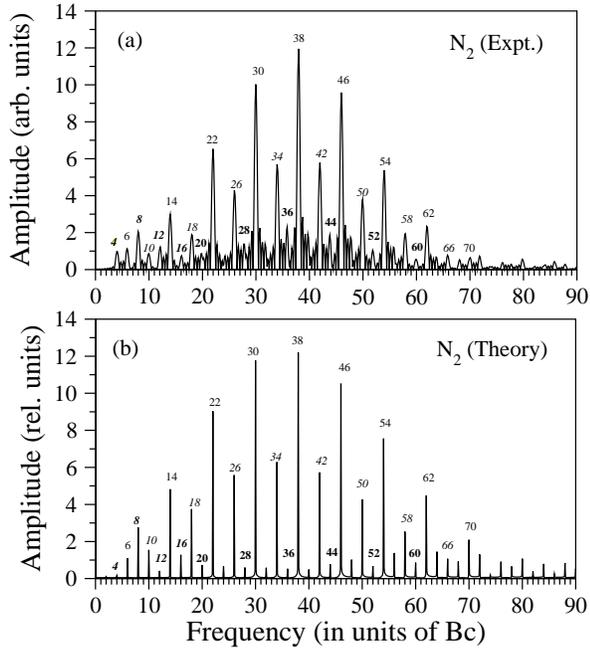} 
\par\end{centering}

\caption{\label{fig:FouSpeN2}Comparison of the experimental \cite{miy-05}
vs. theoretical Fourier spectrum for the dynamic $19$th harmonic
signal for $\mathrm{N_{2}}$ (Fig. \ref{fig:DynN2}). Both spectra
show: series I: $(6,14,22,30,38,..)\, Bc$, series II: $({\it 10,18,26,34,42,..})\, Bc$,
series III: $(\mathbf{20,28,36,44,52,60,}..)Bc$, and series IV: $(\boldsymbol{{\it 4,8,12,16,..}})Bc$.}
\end{figure}

To further compare with experimental data, we Fourier transform the
calculated dynamic signals to get their spectra in the frequency domain.
They may then be compared with the $F.T.$ of the experimental data.
The results for the $19$th harmonic signal for $\mathrm{N_{2}}$
is compard with the experimental data in Fig. \ref{fig:FouSpeN2}.
It can be seen that the experimental spectrum (panel a) exhibits two
prominent series I: $(6,14,22,30,..)\, Bc$ and II: $({\it 10,18,26,34,..})\, Bc$,
which are also present in the theoretical spectrum (panel b). They
can be easily understood to arise from the $F.T.$ of the $\left\langle \left\langle \cos^{2}\theta\right\rangle \right\rangle \left(t_{d}\right)$
term in Eq. (\ref{NitHhgSig}) which vanishes unless $\Delta J=0,\pm2$;
this produces a sequence of lines $(E_{J+2}-E_{J})/2\pi=(4J+6)\, Bc$,
and gives the series I and II, for the even and the odd $J$ levels,
respectively. The relative prominence of the series I over the series
II, from both experiment and theory, seen in the two panels in Fig.
\ref{fig:FouSpeN2}, could be understood as the $2:1$ ratio of the
$J$ even over $J$ odd levels, a well-known consequence of the nuclear
spin statistics of $\mathrm{N_{2}}$ (e.g. \cite{doo-03,her-50}).
The weakly resolved series III: $(\mathbf{20},\mathbf{28},\mathbf{36},\mathbf{44},..)Bc$
and series IV: $(\boldsymbol{{\it 4,8,12,16,..}})Bc$ in Fig. \ref{fig:FouSpeN2}(a)
are the unexpected series that could not be produced by the $F.T.$
of the leading term $\left\langle \left\langle \cos^{2}\theta\right\rangle \right\rangle \left(t_{d}\right)$.
We note that the series III and IV, although weak, are certainly also
present in the theoretical spectrum in Fig. \ref{fig:FouSpeN2}(b).
To interpret their origin, we consider the two higher order terms
involving $\left\langle \left\langle \cos^{2}\theta\right\rangle ^{2}\right\rangle \left(t_{d}\right)$
and $\left\langle \left\langle \cos^{4}\theta\right\rangle \right\rangle \left(t_{d}\right)$
in the signal for N$_{2}$, Eq. (\ref{NitHhgSig}). Because of the
presence of the square of the second moment, the expected beat frequency
from $\left\langle \left\langle \cos^{2}\theta\right\rangle ^{2}\right\rangle \left(t_{d}\right)$
not only includes the frequencies $(4J+6)Bc$ but also their sum and
difference frequencies, as indicated below: %
\begin{figure}
\noindent \begin{centering}
\includegraphics[scale=0.5]{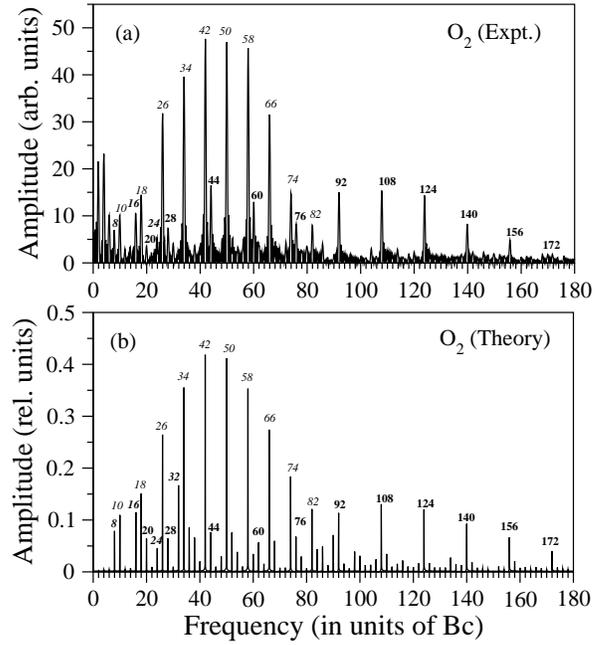} 
\par\end{centering}

\caption{\label{fig:FouSpeO2}Comparison of the experimental \cite{miy-05}
vs. theoretical Fourier spectrum of the dynamic $19$th harmonic signal
for $\mathrm{O_{2}}$ (Fig. \ref{fig:DynO2}). Both the spectra show:
series II: $({\it 10,18,26,34,42,..})\, Bc$, series III: $(\mathbf{20,28,36,44,52,60},..)\, Bc$,
and series V: $(\boldsymbol{{\it 8,16,24,..}})\, Bc$}
\end{figure}

\begin{table*}

\caption{\label{tab:03} List of all the predicted series arising from the
moments $\left\langle \left\langle \cos^{2}\theta\right\rangle \right\rangle $
, $\left\langle \left\langle \cos^{2}\theta\right\rangle ^{2}\right\rangle $,
and $\left\langle \left\langle \cos^{4}\theta\right\rangle \right\rangle $
that are present in the expression for the signal for $\mathrm{N_{2}}$
(for which both odd and even $J$ 's are allowed).}

\begin{centering}
\begin{tabular}{cccccc}
\hline 
No.&
Group freq.&
Weighting factor&
Formula&
Peak series (in $Bc$)&
Expt. series\tabularnewline
\hline 
\multicolumn{6}{l}{$\left\langle \left\langle \cos^{2}\theta\right\rangle \right\rangle $}\tabularnewline
\hline 
1&
-&
$a$&
-&
$0$&
-\tabularnewline
2&
$\omega_{1}$&
$b$&
$4J+6$&
$\begin{array}{c}
10,18,26,....\,\,\,\mathrm{for\, odd\, J}\\
6,14,22,....\,\,\,\,\mathrm{for\, even\, J}\end{array}$&
$\mathrm{\begin{array}{c}
\mathrm{II}\\
\mathrm{I}\end{array}}$\tabularnewline
\hline
\hline 
\multicolumn{6}{l}{$\left\langle \left\langle \cos^{2}\theta\right\rangle ^{2}\right\rangle $}\tabularnewline
\hline 
3&
-&
$aa'$&
-&
$0$&
-\tabularnewline
4&
$\omega_{1}$ and $\omega'_{1}$&
$a'b$ and $ab'$ &
$4J+6$&
$\begin{array}{c}
10,18,26,....\,\,\,\mathrm{for\, odd\, J}\\
6,14,22,.....\,\,\,\,\mathrm{for\, even}\,\mathrm{J}\end{array}$&
$\mathrm{\begin{array}{c}
\mathrm{II}\\
\mathrm{I}\end{array}}$\tabularnewline
5&
$\omega_{1}+\omega'_{1}$&
$\frac{bb'}{2}$&
$4(J+J')+12$&
$20,28,36,..$&
III\tabularnewline
6&
$\omega_{1}-\omega'_{1}$&
$\frac{bb'}{2}$&
$4(J-J')>0$&
$4,8,12,..$&
IV\tabularnewline
\hline
\hline 
\multicolumn{6}{l}{$\left\langle \left\langle \cos^{4}\theta\right\rangle \right\rangle $}\tabularnewline
\hline 
7&
-&
$a$&
-&
$0$&
-\tabularnewline
8&
$\omega_{1}$&
$b$&
$4J+6$&
$\begin{array}{c}
10,18,26,....\,\,\,\mathrm{for\, odd\, J}\\
6,14,22,.....\,\,\,\,\mathrm{for\, even\, J}\end{array}$&
$\mathrm{\begin{array}{c}
\mathrm{II}\\
\mathrm{I}\end{array}}$\tabularnewline
9&
$\omega_{2}$&
$c$&
$8J+20$&
$28,44,60..$&
III\tabularnewline
\hline
\end{tabular}
\par\end{centering}
\end{table*}

\begin{table*}

\caption{\label{tab:04}All possible frequency arising from $\left\langle \left\langle \sin^{2}\cos^{2}\theta\right\rangle ^{2}\right\rangle $
for $\mathrm{O_{2}}$ whose only odd $J$ 's are allowed. The weak
frequencies are noticed with ({*}). }

\begin{centering}
\begin{tabular}{cccccc}
\hline 
No.&
Group freq.&
Weighting factor&
Formula&
Peak series (in $Bc$)&
Expt. series\tabularnewline
\hline 
1&
-&
$aa'$&
-&
$0$&
-\tabularnewline
2&
$\omega_{1}$ and $\omega'_{1}$&
$a'b$ and $ab'$&
$4J+6$&
$10,18,26,...$&
II\tabularnewline
3&
$\omega_{2}$ and $\omega'_{2}$&
$a'c$ and $ac'$&
$8J+20$&
$28,44,60,...$&
III\tabularnewline
4&
$\omega_{1}+\omega'_{1}$&
$\frac{bb'}{2}$&
$4(J+J')+12$&
$20,28,36,..$&
III\tabularnewline
5&
$\omega_{1}-\omega'_{1}$&
$\frac{bb'}{2}$&
$4(J-J')>0$&
$8,16,24,..$&
V\tabularnewline
6&
$\omega_{2}+\omega'_{2}$&
$\frac{cc'}{2}$&
$8(J+J')+40$&
$56,72,88,..$&
I{*}\tabularnewline
7&
$\omega_{2}-\omega'_{2}$&
$\frac{cc'}{2}$&
$8(J-J')>0$&
$16,32,48,..$&
V{*}\tabularnewline
8&
$\omega_{1}+\omega'_{2}$ and $\omega_{2}+\omega'_{1}$ &
$\frac{bc'}{2}$ and $\frac{b'c}{2}$ &
$4(J+2J')+26$&
$38,46,54,..$&
VI{*}\tabularnewline
9&
$\omega_{1}-\omega'_{2}$&
$\frac{bc'}{2}$&
$4(J-2J')-14>0$&
$6,14,22,...$&
VI{*}\tabularnewline
10&
$\omega_{2}-\omega'_{1}$&
$\frac{b'c}{2}$&
$4(-J+2J')+14>0$&
$2,10,18,...$&
II{*}\tabularnewline
\hline
\end{tabular}
\par\end{centering}
\end{table*}

\begin{eqnarray}
 &  & \left(a+b\cos\omega_{1}t\right)\left(a'+b'\cos\omega'_{1}t\right)\nonumber \\
 &  & \,\,\,\,\,=aa'+a'b\cos\omega_{1}t+ab'\cos\omega'_{1}t+bb'\cos\omega_{1}t\cos\omega'_{1}t\nonumber \\
 &  & \,\,\,\,\,=aa'+a'b\cos\omega_{1}t+ab'\cos\omega'_{1}t\nonumber \\
 &  & \,\,\,\,\,\,\,\,\,+\frac{bb'}{2}\cos\left(\omega_{1}+\omega'_{1}\right)t+\frac{bb'}{2}\cos\left(\omega_{1}-\omega'_{1}\right)t\label{FreCosSqr}\end{eqnarray}

Above, the term $a$ arises from transition with $\Delta J=0$ with
$\omega_{0}=0$. The frequency $\omega_{1}$ arises from transition
with $\Delta J=\pm2$. The sum frequency $\left(\omega_{1}+\omega_{1}'\right)$
yields $\left(E_{J+2}-E_{J}\right)/2\pi\equiv\left(4(J+J')+12\right)Bc$
series whereas the difference $\left(\omega_{1}-\omega_{1}'\right)$
produces $\left(E_{J+2}-E_{J}\right)/2\pi\equiv\left(4(J-J')Bc\right)>0$.
For integer $J$ and $J'$ they yield the series IV: $(\boldsymbol{{\it 4,8,12,16,..}})Bc$.
The next term $\left\langle \left\langle \cos^{4}\theta\right\rangle \right\rangle \left(t_{d}\right)$
vanishes unless $\Delta J=0,\pm2$, and $\pm4$ produces not only
$(E_{J+2}-E_{J})/2\pi=(4J+6)\, Bc$ sequences lines but also $(E_{J+4}-E_{J})/2\pi=(8J+20)Bc$
gives series III $(\mathbf{20},\mathbf{28,36},\mathbf{44},..)Bc$.
All the possible series arising from these three leading terms and
their grouping according to those observed experimentally are shown
in table \ref{tab:03}. Note that series III is identical, and overlap,
with the series IV: $(\boldsymbol{{\it 4,8,12,16,..}})Bc$ and adds
to its strength. Moreover, the remaining lines at $(\boldsymbol{{\it 4,8,12,16}},\boldsymbol{{\it 24,32,...}})Bc$
found in the experimental spectrum in Fig. \ref{fig:FouSpeN2}(a)
as well as in the theoretical spectrum in Fig. \ref{fig:FouSpeN2}(b),
confirm the existence of the series IV which is distinct from the
series III. The existence of series III and IV is a prove of the fact
that the dynamic signal of $\mathrm{N_{2}}$ can not be described
in term of $\left\langle \left\langle \cos^{2}\theta\right\rangle \right\rangle \left(t_{d}\right)$
only.

In Fig. \ref{fig:FouSpeO2} we compare the experimental spectrum (panel
a) for $\mathrm{O_{2}}$ \cite{miy-05} with the theoretical spectrum
(panel b) calculated from Eq. (\ref{OxyHhgSig}). Both the experimental
and the theoretical spectra in Fig. \ref{fig:FouSpeO2} show the Raman-allowed
series II: $({\it 10,18,26,34,42,..})\, Bc$, but not the series I:
$(6,14,22,30,38,..)\, Bc$, seen for N$_{2}$. The anomalous series
III: $(\mathbf{20,28,36,44,}..)\, Bc$, discussed in the case of $\mathrm{N_{2}}$
above, however, appears for $\mathrm{O_{2}}$ as well. Finally, another
anomalous sequence V: $(\boldsymbol{{\it 8,16,24,..}})\, Bc$ can
be seen to be present in the data for $\mathrm{O_{2}}$ in Fig. \ref{fig:FouSpeO2}a,
that, we point out, can not be generated by $F.T.$ of $\left\langle \left\langle \sin^{2}2\theta\right\rangle \right\rangle $
term. To interpret the origin of the observed series in $\mathrm{O_{2}}$
we first consider the leading term given by Eq. (\ref{OxyHhgOpe}),
$\left\langle \left\langle \sin^{2}\theta\cos^{2}\theta\right\rangle ^{2}\right\rangle $.
The matrix element $\left\langle \sin^{2}\theta\cos^{2}\theta\right\rangle $
vanishes unless $\Delta J=0,\pm2,\pm4$ corresponds to frequency $\omega_{0}$,
$\omega_{1}$, and $\omega_{2}$. Thus, there will be the various
sum and difference frequencies that arise from the presence of the
squared moment, as follows:

\begin{eqnarray}
 &  & \left(a+b\cos\omega_{1}t+c\cos\omega_{2}t\right)\left(a'+b'\cos\omega'_{1}t+c'\cos\omega'_{2}t\right)\nonumber \\
 &  & \,\,\,\,=aa'+ab'\cos\omega'_{1}t+a'b\cos\omega_{1}t\nonumber \\
 &  & \,\,\,\,\,\,\,\,\,+ac'\cos\omega'_{2}t+a'c\cos\omega_{2}t\nonumber \\
 &  & \,\,\,\,\,\,\,\,\,+\frac{bb'}{2}\cos\left(\omega_{1}+\omega'_{1}\right)t+\frac{bb'}{2}\cos\left(\omega_{1}-\omega'_{1}\right)t\nonumber \\
 &  & \,\,\,\,\,\,\,\,\,+\frac{cc'}{2}\cos\left(\omega_{2}+\omega'_{2}\right)t+\frac{cc'}{2}\cos\left(\omega_{2}-\omega'_{2}\right)t\nonumber \\
 &  & \,\,\,\,\,\,\,\,\,+\frac{bc'}{2}\cos\left(\omega_{1}+\omega'_{2}\right)t+\frac{bc'}{2}\cos\left(\omega_{1}-\omega'_{2}\right)t\nonumber \\
 &  & \,\,\,\,\,\,\,\,\,+\frac{b'c}{2}\cos\left(\omega_{2}+\omega'_{1}\right)t+\frac{b'c}{2}\cos\left(\omega_{2}-\omega'_{1}\right)t\label{FreOxySqr}\end{eqnarray}
 with $a>b>c$. As discussed before, the frequency $\omega_{1}$ generates
the lines $(4J+6$) that for odd $J$ give the series II: $(10,18,26,..)Bc$.
The series I: $(6,14,22,..)\, Bc$ that would exist for even $J$
is absent from the O$_{2}$ spectrum. This is easily understood as
due to the nuclear spin of $\mathrm{O}$ atoms, which is $0$, that
strictly forbids any even $J$ rotational levels for $\mathrm{O_{2}}$
(as dictated by the overall symmetry of the total wavefunction for
$\mathrm{O_{2}}$). 
For odd $J$, the frequency $\omega_{2}$ produces the lines $(8J+20)Bc=(28,44,60,...)Bc$,
whereas the sum frequency $\left(\omega_{1}+\omega_{1}'\right)$ produces
the lines $\left(4(J+J')+12\right)Bc=(20,28,36,...)Bc$; taken together
they generate the series III: $(\boldsymbol{20,28,36,44},..)Bc$.
Similarly, the difference frequency $\omega_{1}-\omega_{2}$ gives
rise to the series V: $(\boldsymbol{{\it 8,16,24,..}})Bc$, as shown
in table \ref{tab:04}. All the above predicted series are observed
in the Fourier spectrum for O$_{2}$. It is also shown in table \ref{tab:04}
that frequency $\omega_{1}+\omega_{2}'$ and $\omega_{2}+\omega_{1}'$
produces the weak (strength of order order {}``bc'' ) series VI:
$(4(J+2J')+26)Bc=(38,46,54,...)Bc$. Despite its weakness the existence
of this series too is evidenced by the presence of the line at $38\, Bc$.
It is worth noting that the series V and VI can not be generated from
the moment $\left\langle \left\langle \sin^{2}2\theta\right\rangle \right\rangle $
alone. The remaining higher order terms in Eq. (\ref{OxyHhgSig})
contribute, generally very weakly, either to the lines in series above
or to some additional lines that can be seen in Fig. \ref{fig:FouSpeO2}(b),
but hardly resolved in Fig. \ref{fig:FouSpeO2}(a). Finally, we may
point out that the heights of the few lowest frequency lines in the
experimental data in Fig. \ref{fig:FouSpeO2}(a) for $\mathrm{O_{2}}$
are believed to be due to the fluctuation of the laser outputs in
the experiment (see, foot-note {[}20] of \cite{miy-05}). %
\begin{figure}[t]
\begin{centering}
\includegraphics[scale=0.5]{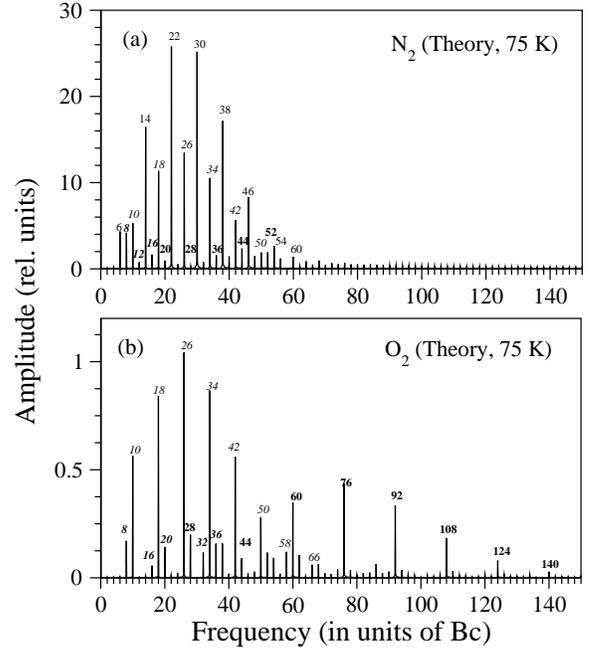} 
\par\end{centering}

\caption{\label{fig:FouSpeIniTem}Calculated spectra for $\mathrm{N_{2}}$
(panel a) and $\mathrm{O_{2}}$ (panel b) at a Boltzmann temperature
$75\,\mathrm{K}$; laser parameters are as in Fig. \ref{fig:DynN2}
and \ref{fig:FouSpeN2}, for $\mathrm{N_{2}}$, and, as in Fig. \ref{fig:DynO2}
and \ref{fig:FouSpeO2}, for $\mathrm{O_{2}}$.}
\end{figure}

We may point out that during the test calculations, the relative strengths
of the lines in a calculated spectrum were found to depend sensitively
(cf. Fig. \ref{fig:FouSpeIniTem} ) on the assumed molecular temperature,
which is rather difficult to determine experimentally. This sensitivity,
on the other hand, provides a way to estimate the temperature of the
molecular ensemble of interest in the experiment, by requiring that
the rotational line for the maximum height of the Fourier spectrum
of the dynamic signal to match with the peak of the Boltzmann distribution
of the initially occupied rotational levels, and adjusting the latter
to find the matching temperature.

Fig. \ref{fig:FouSpeIniTem} shows a calculated spectrum for initial
temperature 75 K whose peaks are shifted from one of 200 K (Figs.
\ref{fig:FouSpeN2} and \ref{fig:FouSpeO2}). We also point out that
our adiabatic theory produces series III in for $\mathrm{N_{2}}$
and series V and VI for $\mathrm{O_{2}}$. These terms arise from
the cross-term, and hence can not be produced from frozen nuclei approximation.
Thus, the spectrum in frequency domain gives more succinct and clearer
information of the HHG signal and therefore power full to test the
model \cite{fai-07}.

\subsection{Interplay of Polarization Geometry $\alpha$ and Delay Time $t_{d}$}

\begin{figure}
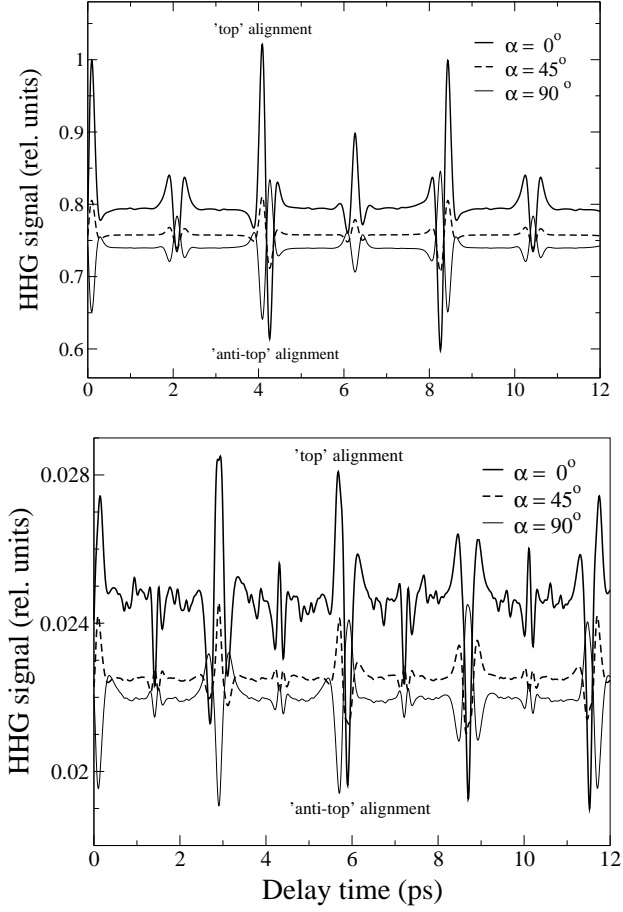

\begin{centering}
\includegraphics[scale=0.29]{fig-15a} 
\par\end{centering}

\begin{centering}
\includegraphics[scale=0.32]{fig-15b} 
\par\end{centering}

\caption{\label{fig:FouSpeN2O2}Calculated $19$th harmonic dynamic signal
for $\mathrm{N_{2}}$ (upper panel) and $\mathrm{O_{2}}$ (lower panel)
for various pump-probe polarization angles, i.e. $\alpha=0^{o}$,
$\alpha=45^{o}$, and $\alpha=90^{o}$. The laser parameters are similar
with one in Figs. \ref{fig:DynN2} and \ref{fig:DynO2} for $\mathrm{N_{2}}$
and $\mathrm{O_{2}}$, respectively. The initial temperature is 200
K.}
\end{figure}

So far we have limited our applications to the HHG signal for parallel
geometry of the pump and probe polarizations. We now consider the
more general case when probe polarization is rotated by a given angle
$\alpha$. Fig. \ref{fig:FouSpeN2O2} (upper panel) shows our computational
results of the HHG signals as a function of $t_{d}$, at three different
fixed $\alpha$, i.e. $\alpha=0^{o},\,45^{o}$, and $90^{o}$, for
$\mathrm{N_{2}}$. We note that the signal for $\alpha=90^{o}$ changes
its phase by $\pi$ with respect to the signal for $\alpha=0^{o}$,
a phenomenon that is also observed recently \cite{kan-05,miy-06,yos-07}.
In contrast, the signal for $\alpha=45^{o}$ is seen to remain rather
flat with change of $\alpha$. %

To see qualitatively the $\alpha$ dependence of HHG signal of $\mathrm{N_{2}}$,
we consider the leading term of Eq. (\ref{NitHhgSigAlf}) which is
given by \begin{eqnarray}
S^{(n)}(t_{d};\alpha) & = & c_{00}^{(n)}+c_{01}^{(n)}\left[\frac{1}{2}\sin^{2}\alpha+\frac{1}{2}\left(3\cos^{2}\alpha-1\right)\right.\nonumber \\
 &  & \left.\times\left\langle \left\langle \cos^{2}\theta\right\rangle \right\rangle \left(t_{d}\right)\vphantom{\frac{1}{2}}\right]+...\label{NitSigApp}\end{eqnarray}
 Thus, for the parallel polarizations we have, $S^{(n)}\left(t_{d};0^{o}\right)\approx c_{00}^{(n)}+c_{01}^{(n)}\left\langle \left\langle \cos^{2}\theta\right\rangle \right\rangle \left(t_{d}\right)$
and for the perpendicular polarizations, $S^{(n)}\left(t_{d};90^{o}\right)\approx c_{00}^{(n)}+\frac{c_{01}^{(n)}}{2}\left(1-\left\langle \left\langle \cos^{2}\theta\right\rangle \right\rangle \left(t_{d}\right)\right)$
which are clearly of opposite phase as a function of $t_{d}$. These
above expressions also show that the modulation depth for $\alpha=90^{o}$
is smaller than one for $\alpha=0^{o}$, that can not be obtained
by planar model \cite{kan-05}. Eq. (\ref{NitSigApp}) also implies
that the extrema of the signal would occur for $\sin\alpha\cos\alpha=0$,
or the maximum at $\alpha=0^{o}$ and the minimum for $\alpha=90^{o}$,
as seen in Fig. \ref{fig:FouSpeN2O2}(upper) and confirmed experimentally
\cite{kan-05,miy-06,kak-05}. Eq. (\ref{NitSigApp}) also implies
that at a critical angle $\alpha_{c}$ given by $\left(3\cos^{2}\alpha_{c}-1\right)=0$,
or $\alpha_{c}\approx55^{o}$, the signal essentially \textit{\emph{remains
constant and}} \textit{independent} of the delay $t_{d}$ between
the pulses. This geometry therefore can be used to generate a steady
state HHG signal from $\mathrm{N_{2}}$, with femtosecond pulses.%
\begin{figure}
\begin{centering}
\includegraphics[scale=0.3]{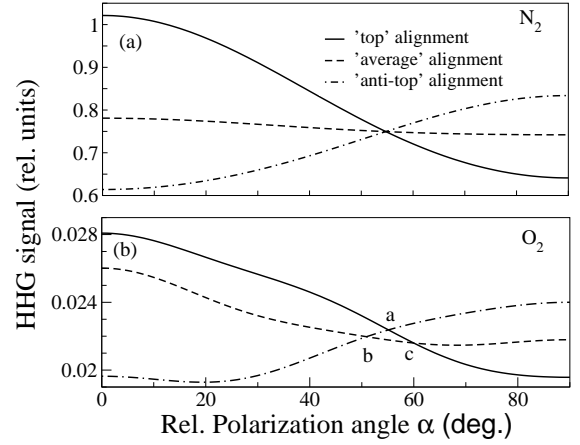} 
\par\end{centering}

\caption{\label{fig:AlfSig}Variation of the 19th HHG signal as a function
of pump-probe polarization angle $\alpha$, near the first half-revival,
for $\mathrm{N_{2}}$ (upper panel). The observation times are $t_{d}=4.090\,\mathrm{ps}$
for the {}``top'', $t_{d}=4.180\,\mathrm{ps}$, the {}``average'',
and $t_{d}=4.265\,\mathrm{ps}$ the {}``anti-top{}`` alignment times.
And similarly for $\mathrm{O_{2}}$ (lower panel). The pulse parameters
are the same as in Fig. \ref{fig:DynN2} for $\mathrm{N_{2}}$ and
Fig. \ref{fig:DynO2} for $\mathrm{O_{2}}$. The initial temperature
is 200 K. Note the existense of a {}``magic'' angle at $\alpha=\arctan{\sqrt{2}}\approx55^{o}$
for N$_{2}$, where the dynamical signals all coincide, and a {}``crossing
neighborhood'' near that angle for O$_{2}$. }
\end{figure}

The magic angle in fact is a generic signature for the $\sigma_{g}$
symmetry of the active molecular orbitals.

For $\mathrm{O_{2}}$, the leading term of HHG signal (Eq. (\ref{OxyHhgSigAlf}))
reads\begin{eqnarray}
S^{(n)}(t_{d};\alpha) & = & \frac{c_{11}^{(n)}}{64}\left\langle \left(\left(-35\cos^{4}\alpha+30\cos^{2}\alpha-3\right)\right.\right.\nonumber \\
 &  & \times\left\langle \cos^{4}\theta\right\rangle \left(t_{d}\right)\nonumber \\
 &  & +\left(30\cos^{4}\alpha-24\cos^{2}\alpha+2\right)\left\langle \cos^{2}\theta\right\rangle \nonumber \\
 &  & +\left.\left.\left(-3\sin^{4}\alpha+4\sin^{2}\alpha\right)\vphantom{\left(t_{d}\right)}\right)^{2}\right\rangle +...\label{OxySigApp}\end{eqnarray}
 Thus, for the parallel polarizations we have, $S^{(n)}\left(t_{d};0^{o}\right)\approx c_{11}^{(n)}\left\langle \left(-\left\langle \cos^{4}\theta\right\rangle +\left\langle \cos^{2}\theta\right\rangle \right)^{2}\right\rangle =c_{11}^{(n)}\left\langle \left\langle \sin^{2}\theta\cos^{2}\theta\right\rangle ^{2}\right\rangle $
and for the perpendicular polarizations, $S\left(t_{d};90^{o}\right)\approx\frac{c_{11}^{(n)}}{64}\left\langle \left(-3\left\langle \cos^{4}\theta\right\rangle +2\left\langle \cos^{2}\theta\right\rangle +1\right)^{2}\right\rangle $.
It was clear that the sign of $\left\langle \cos^{4}\theta\right\rangle $
does not change and hence the phase of eighth revival also remains
constant, as shown in Fig. \ref{fig:FouSpeN2O2} (lower panel) and
confirmed experimentally \cite{kan-05,miy-06,yos-07}. These above
expressions also can be expressed as $S\left(t_{d};90^{o}\right)\approx c_{11}^{(n)}\left\langle \left(\frac{3}{8}\left\langle \sin^{2}\theta\cos^{2}\theta\right\rangle -\frac{1}{8}\left\langle \cos^{2}\theta\right\rangle +\frac{1}{8}\right)^{2}\right\rangle $
shows that the modulation depth for $\alpha=90^{o}$ is smaller than
one for $\alpha=0^{o}$, that can not be obtained by planar model
\cite{kan-05}. %
\begin{figure}
\begin{centering}
\includegraphics[scale=0.4]{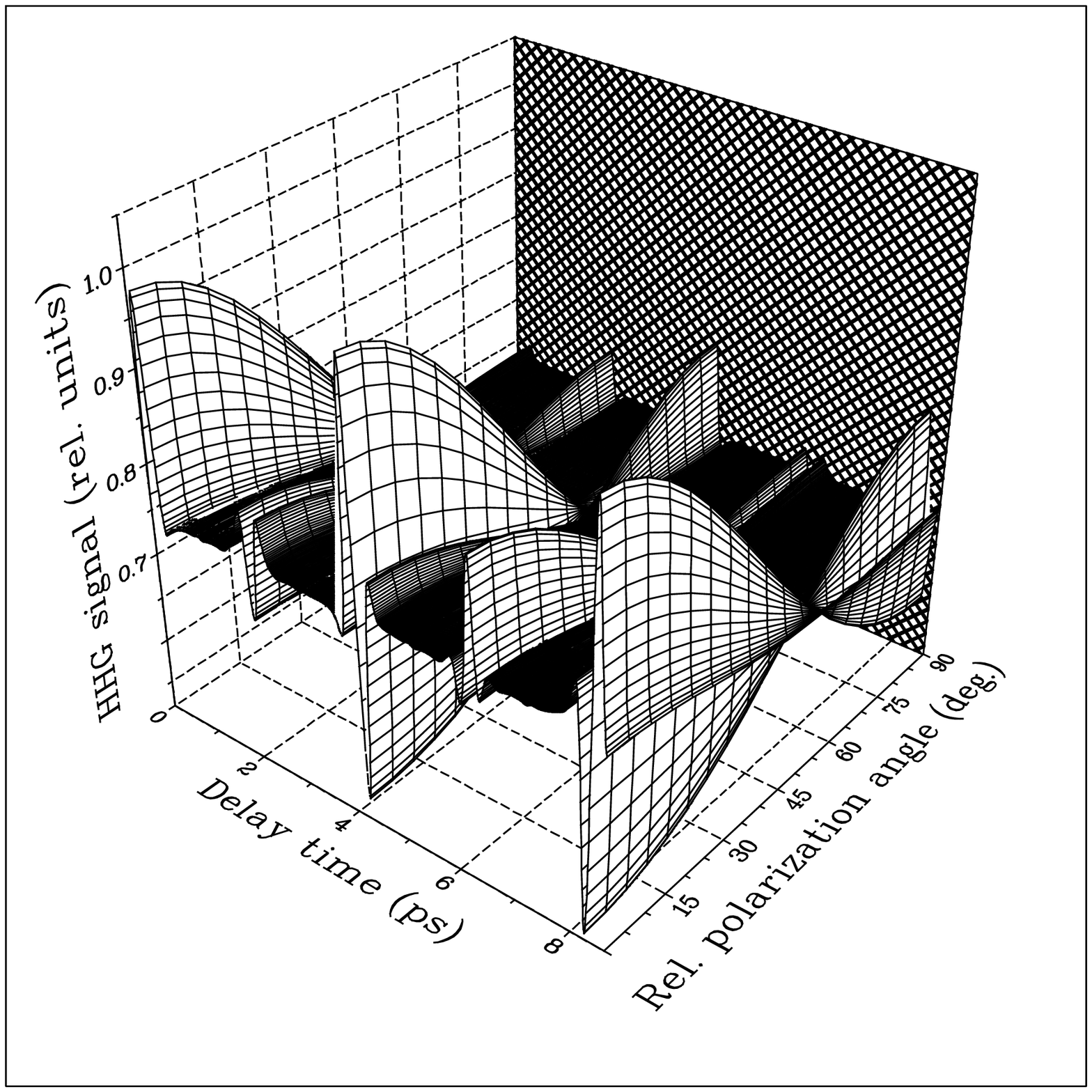} 
\par\end{centering}

\begin{centering}
\includegraphics[scale=0.4]{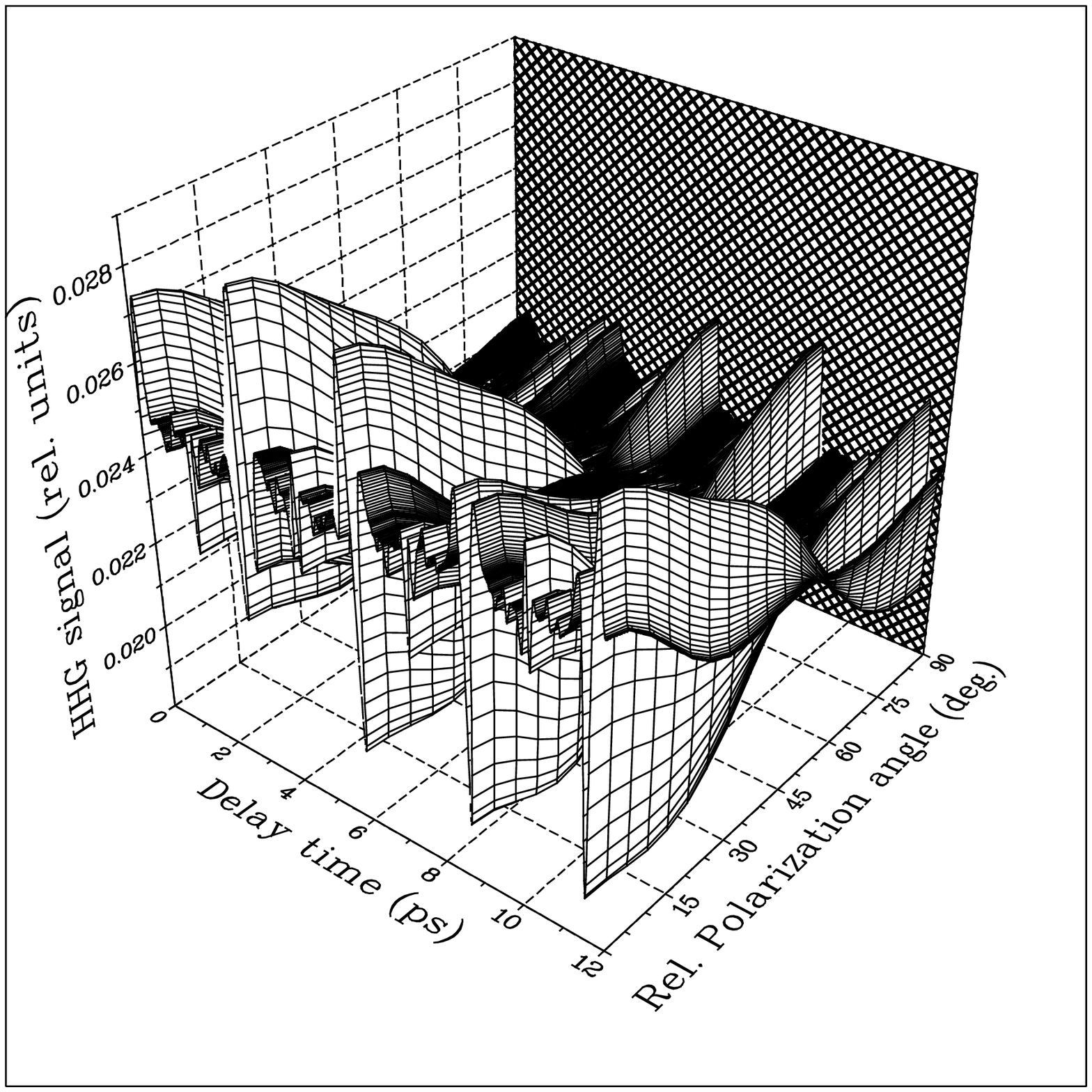} 
\par\end{centering}

\caption{\label{fig:SigN2O2-3D}HHG signals for $\mathrm{N_{2}}$ (upper panel)
and $\mathrm{O_{2}}$ (lower panel) as simultaneous function of delay
time $t_{d}$ and the relative angle $\alpha$ between the pump and
probe polarizations. The laser parameters are similar to that in \ref{fig:DynN2}
and \ref{fig:DynO2} for $\mathrm{N_{2}}$ and $\mathrm{O_{2}}$,
respectively. The initial temperature is 200 K.}
\end{figure}

We also calculated the signal at fixed time delay for various relative
polarization direction between pump and probe pulses. In Fig. \ref{fig:AlfSig}
we show the results for $\mathrm{N_{2}}$ (upper panel) and for $\mathrm{O_{2}}$
(lower panel), near the first revival $t_{d}\approx\frac{1}{2}T_{rev}$.
For the case of $\mathrm{N_{2}}$ a coincidence of the signals is
seen to occur at a critical angle $\alpha_{c}\approx55^{o}$. This
is fully consistent with the prediction of the near $t_{d}$-independence
of the signal for $\mathrm{N_{2}}$ at $\alpha_{c}$, made above.
The HHG signal (solid curve) at $t_{d}=4.090\,\mathrm{ps}$ (`top'-alignment)
lies above the signal at $t_{d}=4.265\,\mathrm{ps}$ (`anti-top' alignment),
for all angles $\alpha$ smaller than the critical $\alpha_{c}\approx55^{o}$;
the opposite relation holds above $\alpha_{c}$. For the case of $\mathrm{O_{2}}$
(lower panel), in contrast, there is no single critical value of $\alpha$
where the signals for all $t_{d}$'s could coincide. This is due mainly
to the different $\alpha$-dependence of $\left\langle \cos^{2}\theta\right\rangle $
and $\left\langle \cos^{4}\theta\right\rangle $ of Eq. (\ref{Sin2Cos2Alf}).
Nevertheless, it can be seen that the signal at the `top' alignment
becomes equal to the signal at the `anti-top' alignment not far from
$\alpha_{c}\approx55^{o}$ (Point a, Fig. \ref{fig:AlfSig}(b)), and
they reverse their relative strengths above it. During the calculation,
we get that the exact position of points $a$, $b$, and $c$ depend
on the initial temperature. These whole properties well agree with
the experimental data \cite{miy-06,kak-05}. We also note that the
calculated $\alpha$-dependent signal given by Zhou \emph{et al.}
gives the same properties for $\mathrm{N_{2}}$, but they predicted
the signal for $\mathrm{O_{2}}$ to be maximized at $\alpha\approx45^{o}$.

In Fig. \ref{fig:SigN2O2-3D}, we plot the calculated signals as function
of both the delay time $t_{d}$ and relative polarization angles $\alpha$
for both $\mathrm{N_{2}}$ (upper panel) and $\mathrm{N_{2}}$ (lower
panel). For both molecules, the modulation depths decrease by increasing
relative polarization angle, reach zero near critical angle $\alpha_{c}\approx55°$,
and increase again but in opposite phase above the critical angle.
The results for $\theta=90°-180°$ are exactly mirror image of the
results for $\theta=0°-90°$.

Before concluding this section it is also worthwhile to point out
that the $\alpha$-dependence of the HHG signals for the more complex
tri-atomic molecule CO$_{2}$ and the organic molecule acetylene,
$HC\equiv CH$, because of their active $\pi$ orbital symmetry, are
predicted from the general structure of the HHG signal given by Eq.
(\ref{GenHhgSig}) (even with out detailed calculations) to exhibit
a {}``cross-over'' neighborhood near $\alpha\approx55^{o}$; this
is indeed the case, as has been recently observed experimentally \cite{tor-07}.
Clearly, the presence of the {}``magic'' angle and the cross-over
neighborhood provide a signature of the symmetry of the active molecular
orbital, which can be useful in the context of the {}``inverse''
problem of molecular imaging \cite{lei-07} from the HHG data as suggested
first in \cite{fai-08}. Finally, the agreement between the present
results and experimental data provides a clear possibility to control
the HHG signals by varying both the time- delay \textit{and} the relative
pump-probe polarization angle, simultaneously.

\section{Some Problems of General Interest Related to Pump-Probe Signals for
HHG}

Before concluding this paper we report on the results of our investigations
of a number of pump-probe experiment related problems of interest
in the present context.

\subsection{Effect of Probe Pulse on the Alignment}

In pump-probe experiments it is generally assumed that the dynamical
alignment of the molecular axis is governed by the ultrashort pump
pulse, while the ultrashort probe pulse that leads to the HHG signal
does dot affect the alignment. To check the validity or otherwise
of this assumption, we directly compare here the dynamic alignment
moment, $A(t_{d};\alpha=0)$ calculated as usual assuming when only
the pump and when both pump and probe pulse couple to the molecular
polarizability, for $\mathrm{N_{2}}$. In the latter case, the total
field consists of the superposition of the two pulses with a displacement
$\Delta t$ in time between them:

\begin{eqnarray}
\bm{F}\left(t\right) & = & \bm{F}_{1}\cos\left(\omega_{1}t\right)\,+\,\bm{F}_{2}\cos\left(\omega_{2}\left(t-\Delta t\right)\right)\nonumber \\
 & = & \varepsilon_{10}\sqrt{g_{1}(t)}\cos\left(\omega_{1}t\right)\nonumber \\
 &  & +\,\varepsilon_{20}\sqrt{g_{2}\left(t-\Delta t\right)}\cos\left(\omega_{2}\left(t-\Delta t\right)\right)\label{ali:33}\end{eqnarray}

\noindent and\begin{eqnarray}
\left\langle \varepsilon^{2}\left(t\right)\right\rangle  & = & \frac{1}{2}\varepsilon_{10}^{2}g_{1}\left(t\right)\,+\,\frac{1}{2}\varepsilon_{10}^{2}g_{1}\left(t-\Delta t\right)\,\nonumber \\
 &  & +2\varepsilon_{10}\varepsilon_{20}g_{1}\left(t\right)g_{2}\left(t-\Delta t\right)\nonumber \\
 &  & \times\left\langle \left(\cos\left(\omega_{1}t\right)\right)\left(\cos\left(\omega_{2}\left(t-\Delta t\right)\right)\right)\right\rangle \hphantom{mm}\label{ali:34}\end{eqnarray}
 In the above the indices $1$ and $2$ stand for pump and probe pulse,
respectively. Suppose the data are recorded after the probe pulse
dies out, then the observing time is $t=\Delta t+\tau$, where $\tau$
is the duration of the probe pulse. Eq. (\ref{ali:34}) then reads

\noindent \begin{eqnarray}
\left\langle \varepsilon^{2}\left(\Delta t+\tau\right)\right\rangle  & = & \frac{1}{2}\varepsilon_{10}^{2}g_{1}\left(\Delta t+\tau\right)\,+\,\frac{1}{2}\varepsilon_{10}^{2}g_{1}\left(\tau\right)\,\nonumber \\
 &  & +2\varepsilon_{10}\varepsilon_{20}g_{1}\left(\Delta t+\tau\right)g_{2}\left(\tau\right)\nonumber \\
 &  & \times\left\langle \left(\cos\left(\omega_{1}\left(\Delta t+\tau\right)\right)\right)\left(\cos\left(\omega_{2}\tau\right)\right)\right\rangle \hphantom{mmm}\label{ali:34b}\end{eqnarray}
\begin{figure}
\begin{centering}
\includegraphics[scale=0.35]{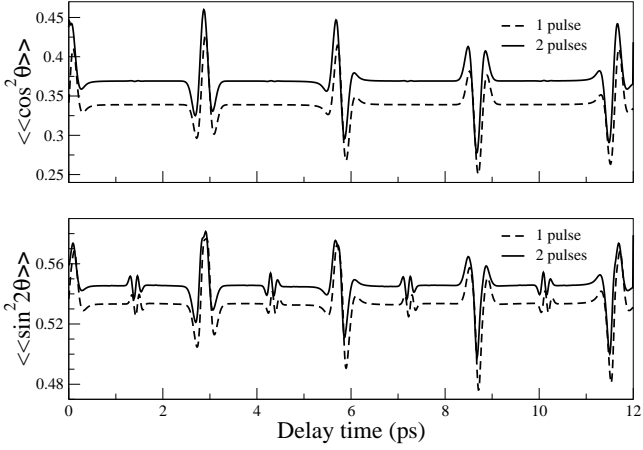} 
\par\end{centering}

\caption{\label{fig:EffPumPul} Shift of the alignment signal vs. delay time:
$\left\langle \left\langle \cos^{2}\theta\right\rangle \right\rangle $
for N$_{2}$ (upper panel), and $\left\langle \left\langle \sin^{2}2\theta\right\rangle \right\rangle $
for $\mathrm{O_{2}}$ (lower panel) at 300 K; $I_{pump}=0.8\times10^{14}\mathrm{W/cm^{2}}$
and $I_{probe}=1.7\times10^{14}\mathrm{W/cm^{2}}$ with FWHM $40\,\mathrm{fs}$.
See, text for further explanation.}
\end{figure}

\noindent showing its dependence on the delay between the two pulses
$\Delta t$ and the length of interaction of the probe pulse $\tau$.
In Fig. \ref{fig:EffPumPul}, we plot the alignment moment $\left\langle \left\langle \cos^{2}\theta\right\rangle \right\rangle \left(\Delta t+\tau\right)$
and $\left\langle \left\langle \sin^{2}2\theta\right\rangle \right\rangle \left(\Delta t+\tau\right)$
for $\mathrm{O_{2}}$, plotted as a function of delay between two
pulses $\Delta t$, for a fixed $\tau=40\,\mathrm{fs}$, as shown
by the solid curve . The results are compared with that obtained from
for the pump pulse alone (dashed curve), recorded at the same time.
The comparison clearly shows that the probe pulse changes the dynamic
alignment $\left\langle \left\langle \cos^{2}\theta\right\rangle \right\rangle (t_{d})$
in that the signal is shifted upward by the presence of the probe
pulse as may be expected from the enhanced intensity of the field
when both the pulses overlap significantly (before it dies out). Thus,
except perhaps when the two pulses overlap (or are separated only
negligibly) this do {\textit{not}} change the general characteristics
of the dynamical signals. Therefore, within the above mentioned exception,
one may neglect the effect of the probe pulse on the HHG signal.

\subsection{Effect of Initial Temperature}

We assume that the rotational eigenstates $\left|J_{0}M_{0}\right\rangle $
of the molecule are occupied thermally before the interaction with
the pump pulse. 
Unlike an upward transitions $(J_{0},M_{0})\rightarrow(J_{0}',M_{0})$
to the states with an arbitrarily high $J_{0}'$, the downward transition
toward $J_{0}'\geq M_{0}$ can be restricted. As a result, a wavepacket
state created by the pump pulse would consist of eigenstates with
higher occupation of $J_{0}'\ge M_{0}$, implying that the vector
of rotational angular momentum would tend to lie in a plane perpendicular
to the pump polarization direction. Since the rotational angular momentum
itself is perpendicular to the internuclear axis of a linear molecule,
the above condition, $J_{0}'\ge M_{0}$, means also that the molecular
axis would tend to align in the direction of the laser polarization.
This is the physical reason why the alignment angle of the molecular
axis with respect to the polarization direction after the laser interaction
is generally smaller after the interaction than before it, i.e. the
degree of alignment increases on interaction with the pump pulse.
Since at a lower initial temperature, the lower $M_{0}$ states are
relatively more occupied initially, the {}``degree of alignment''
$A\equiv\left\langle \left\langle \cos^{2}\theta\right\rangle \right\rangle $
would tend to be higher, allowing the molecules to be more readily
aligned at a lower initial temperature. %
\begin{figure}
\begin{centering}
\includegraphics[scale=0.35]{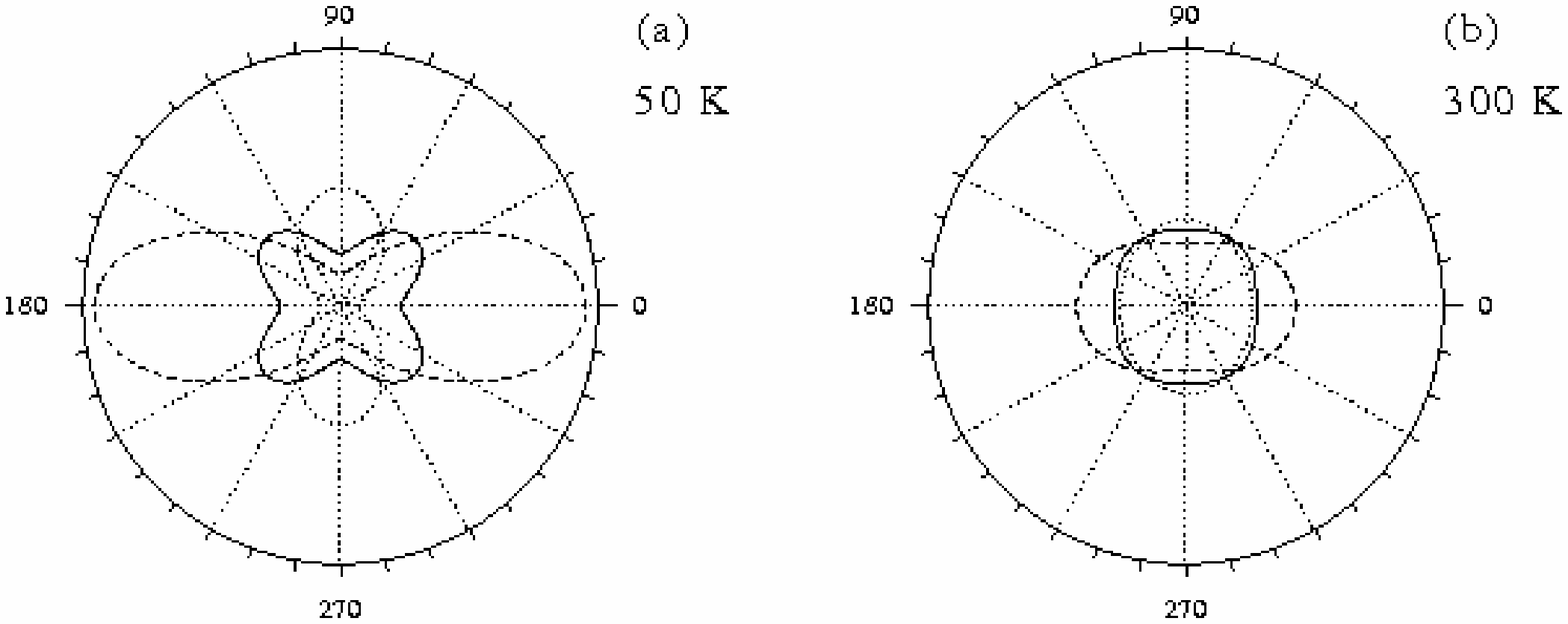} 
\par\end{centering}

\begin{centering}
$\frac{\frac{}{}}{\frac{}{}}$ 
\par\end{centering}

\begin{centering}
\includegraphics[scale=0.35]{fig-19b} 
\par\end{centering}

\caption{\label{fig:IniTem} Influence of the initial temperature, $50^{o}$
or $300^{o}$, on the molecular axis-distribution (panel a-b), on
the leading dynamic moment $\left\langle \left\langle \sin^{2}2\theta\right\rangle \right\rangle (t_{d})$
(panel c-d), and the Fourier spectrum (panel e-f) of the latter, for
the case of $\mathrm{O_{2}}$. The molecular axis-distribution (panel
a-b) are given at three values of the delay time $t_{d}$, i.e. at
the {}``top'' alignment $t_{d}=5.649\,\mathrm{ps}$ (dashed line),
at the {}``average'' alignment $t_{d}=5.812\,\mathrm{ps}$ (solid
line), and at the {}``anti-top'' alignment $t_{d}=5.975\,\mathrm{ps}$
(dotted line); the radii are in the same scale. I= $0.5\times10^{14}\,\mathrm{W/cm^{2}}$
and FWHM=$40\,\mathrm{fs}$. }
\end{figure}

\subsection{Mean Energy of the Molecule after the Pump Pulse}

\begin{figure}
\begin{centering}
\includegraphics[scale=0.35]{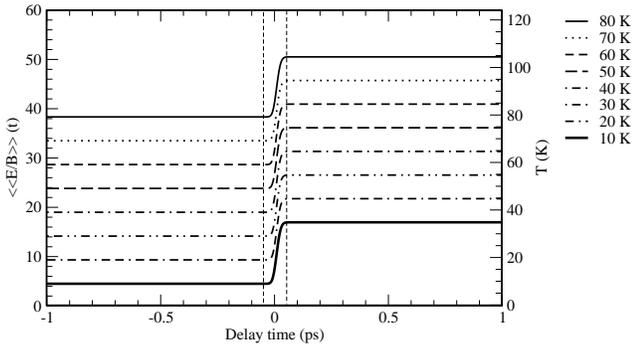} 
\par\end{centering}

\caption{\label{fig:MeaEne}Mean energy of $\mathrm{O_{2}}$ before and after
interaction with the laser pulse, for different initial temperatures.
The vertical dashed lines indicate the extent of the pulse duration;
I= $0.5\times10^{14}\,\mathrm{W/cm^{2}}$ FWHM=$40\,\mathrm{fs}$.
See, text for further explanation.}
\end{figure}

It is interesting also to examine the way the mean energy of the molecule
changes with increasing intensity of the pump pulse. Fig. \ref{fig:MeaEne}
shows the calculated mean energy $\left\langle E\right\rangle _{J_{0}M_{0}}(t)$
at a time $t$ , before and after the arrival of the peak of the pump
pulse (of length $t_{p}$). As expected, the figure shows that increasing
the peak pulse intensity, increases the mean energy of the molecule
or, the {}``effective temperature'' $T_{eff.}\equiv\left\langle E\right\rangle _{J_{0}M_{0}}\left(t>t_{p}\right)/k_{B}$,
where $k_{B}$ is the Boltzmann constant. However, it should be remembered
that after the pulse interaction, the molecular system is not a state
of thermal equilibrium, rather it is in a state of dynamical equilibrium
(or steady state) that can not be characterized thermodynamically.%

To estimate an effective {}``temperature'' of the rotational wavepacket
states $\left|\Phi_{J_{0}M_{0}}(t)\right\rangle $, in the steady
state regime, i.e. for $\left(t_{pump}\le t\le t_{probe}\right)$,
we note (a) that the rotational wavepacket states $\left|\Phi_{J_{0}M_{0}}(t)\right\rangle $
form a linearly independent set of states like the set of rotational
eigenstates $\left|J_{0}M_{0}\right\rangle $ from which they evolve,
(b) that the individual rotational wavepacket states evolve in {\textit{one-to-one}}
correspondence with the initially occupied rotational eigenstates
$\left|J_{0}M_{0}\right\rangle $, (c) that the mean energy of each
of the rotational wavepacket states reach a steady state, also one-to-one
of energy from $\left\langle E\right\rangle _{J_{0}M_{0}}\left(t>t_{pump}\right)=E_{J_{0}M_{0}}+\left\langle E\right\rangle _{J_{0}M_{0}}\left(t\ge t_{pump}\right)$.
If further the above change in the mean energy $\left\langle E\right\rangle _{J_{0}M_{0}}\left(t>t_{pump}\right)-\left\langle E\right\rangle _{J_{0}M_{0}}\left(t\ge t_{pump}\right)$
is {\textit{i}\emph{ndependent}} of the the individual states chosen
(indices $\left\{ J_{0}M_{0}\right\} $) then one might use it to
define an effective {}``temperature'' change, $\Delta T_{eff}$,
given by \begin{equation}
\Delta T_{eff}=\frac{\left\langle \Delta E\right\rangle \left(t\ge t_{pump}\right)}{k_{B}}\label{MeaCgaTem}\end{equation}
 We may note in Fig. \ref{fig:MeaEne}, that the change in the mean
energy in the steady state regime is indeed essentially independent
of the states of the system chosen. Thus, the effective {}``temperature''
of the system, at the end of the interaction with the pump pulse,
becomes \begin{equation}
T_{eff}=T_{0}+\Delta T_{eff}\label{EffTem}\end{equation}
 Note that $T_{eff.}$ is in general greater than the initial gas
(jet) temperature, $T_{0}$, and it tends to increase with the increase
of the pump intensity. Later on we shall describe a method of determining
this {}``effective temperature'' of the system from a theoretical
analysis of the experimental HHG data.

\subsection{Some Non-equivalent Definitions of the HHG Signal}

\begin{figure*}
\begin{centering}
\includegraphics[clip,scale=0.28]{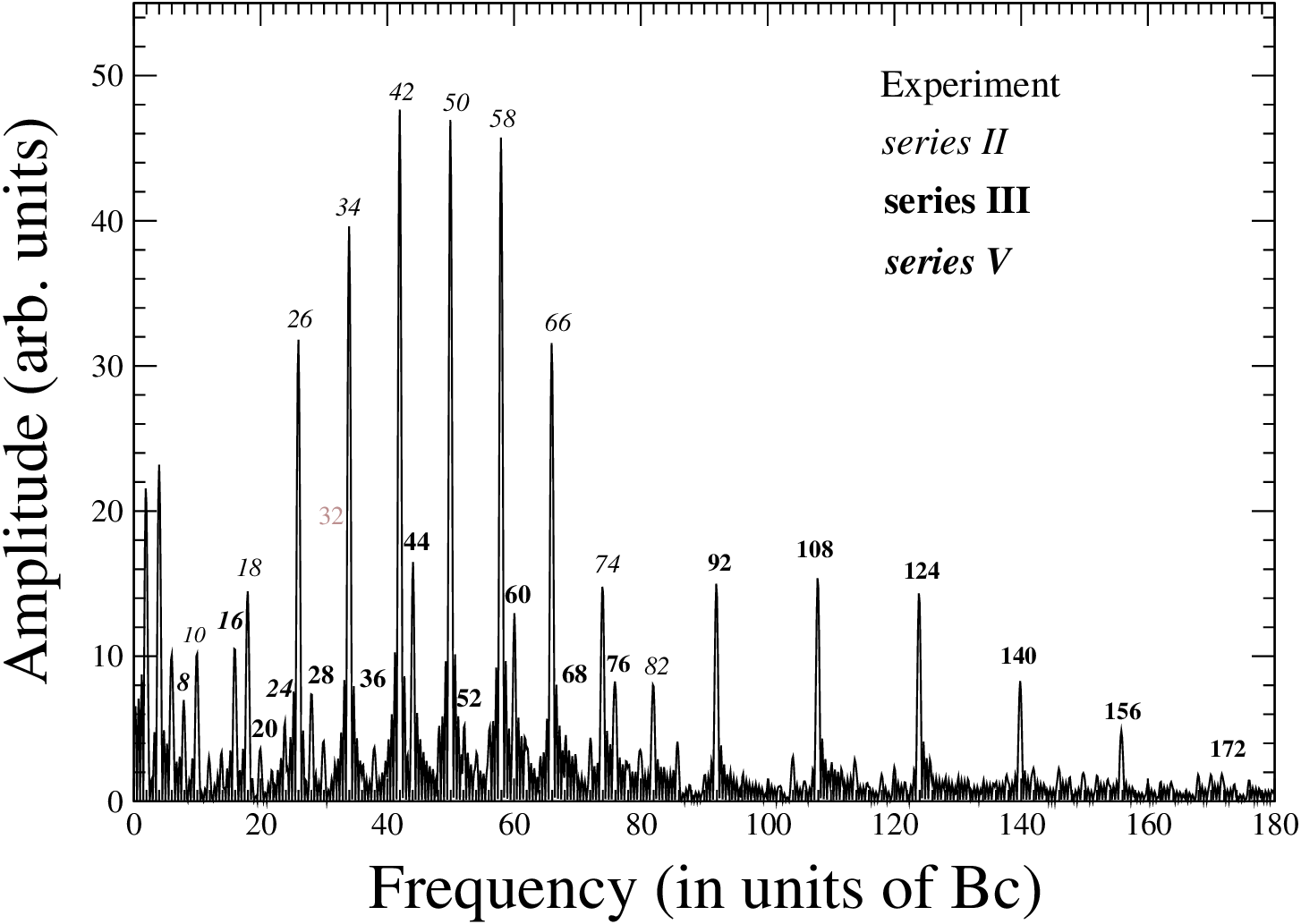}~~~~~\includegraphics[clip,scale=0.28]{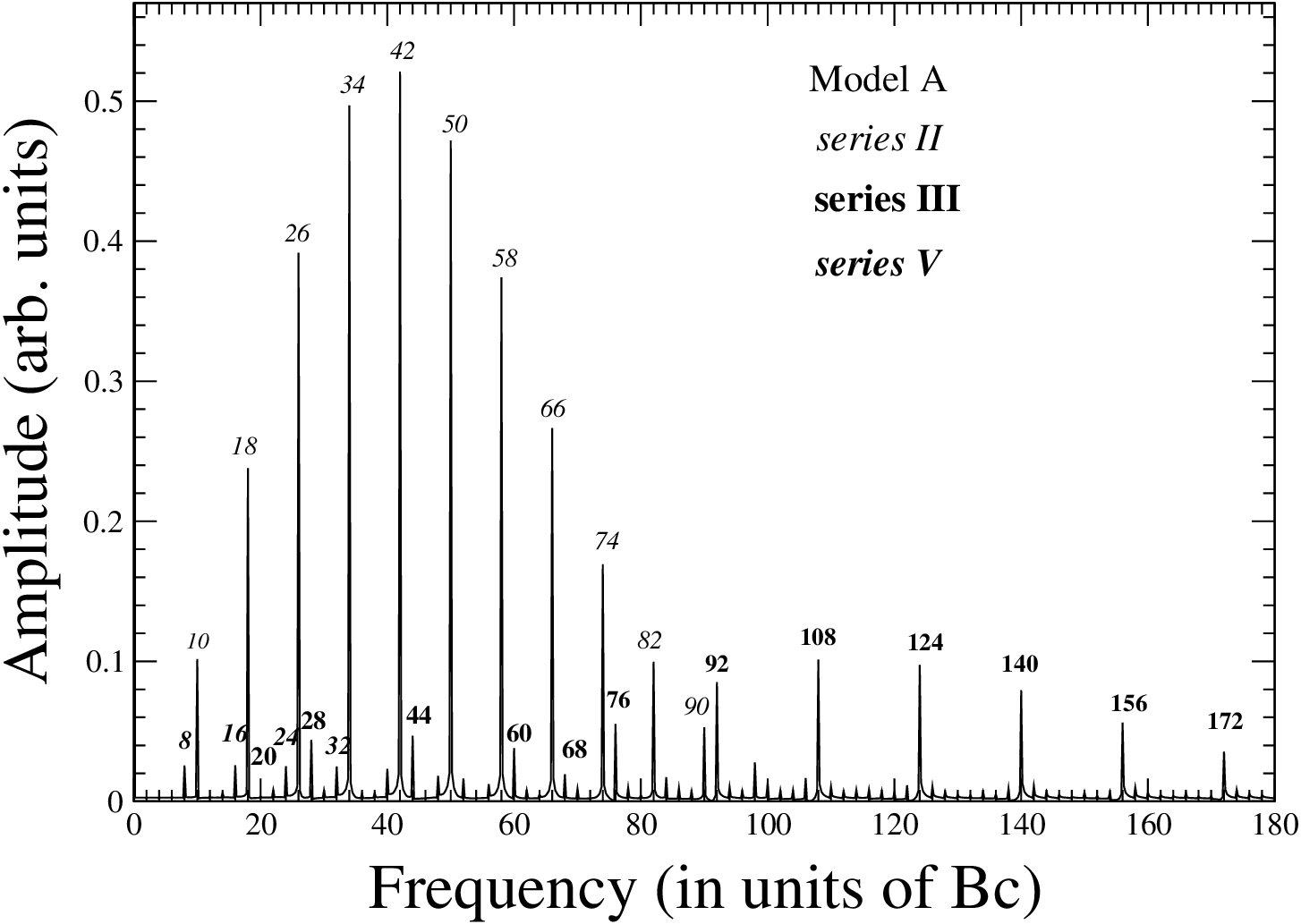} 
\par\end{centering}

\begin{centering}
$\vphantom{\frac{\frac{\frac{}{}}{}}{\frac{\frac{}{}}{}}}$ 
\par\end{centering}

\begin{centering}
\includegraphics[scale=0.28]{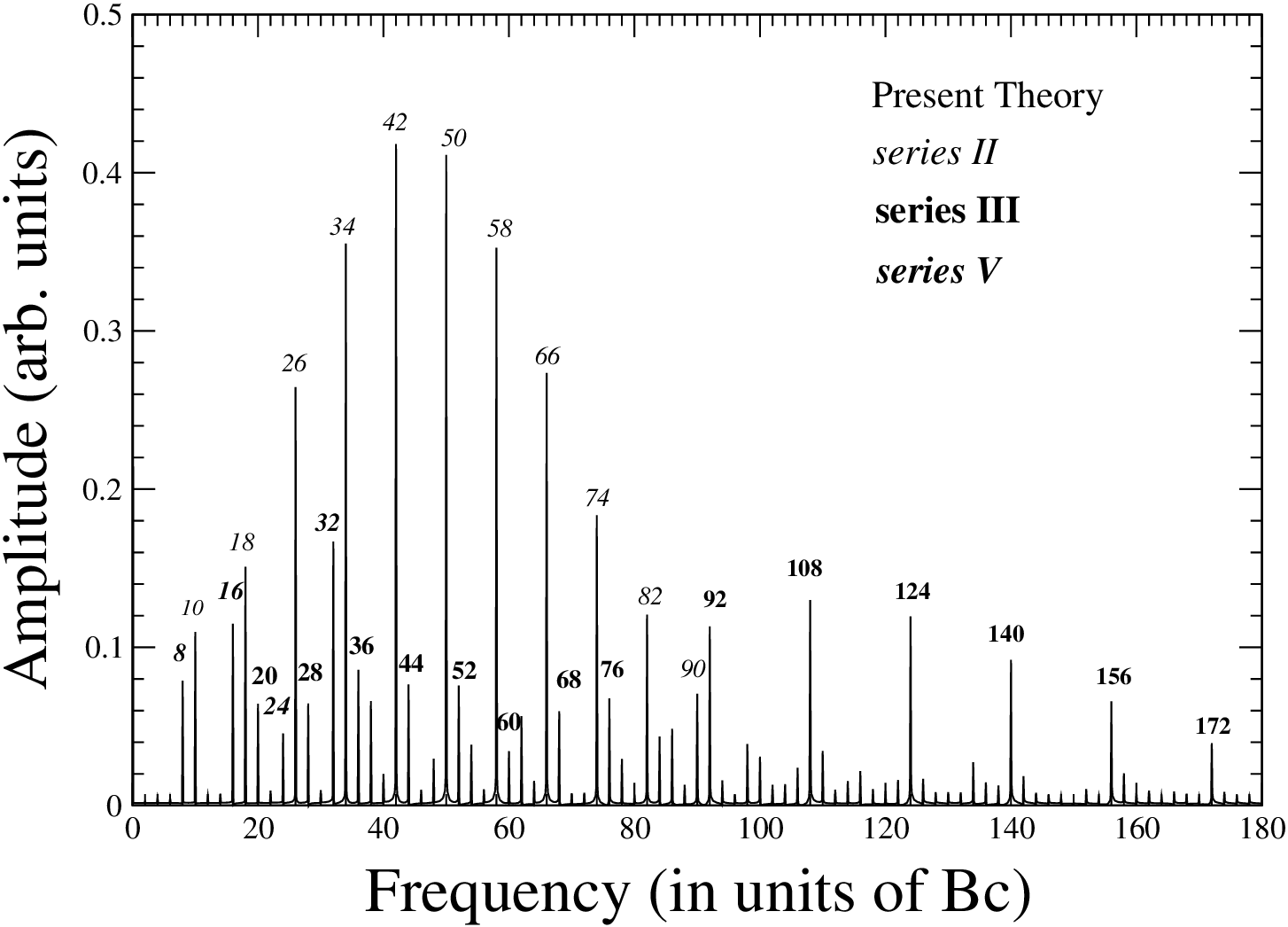} ~~~~~\includegraphics[scale=0.28]{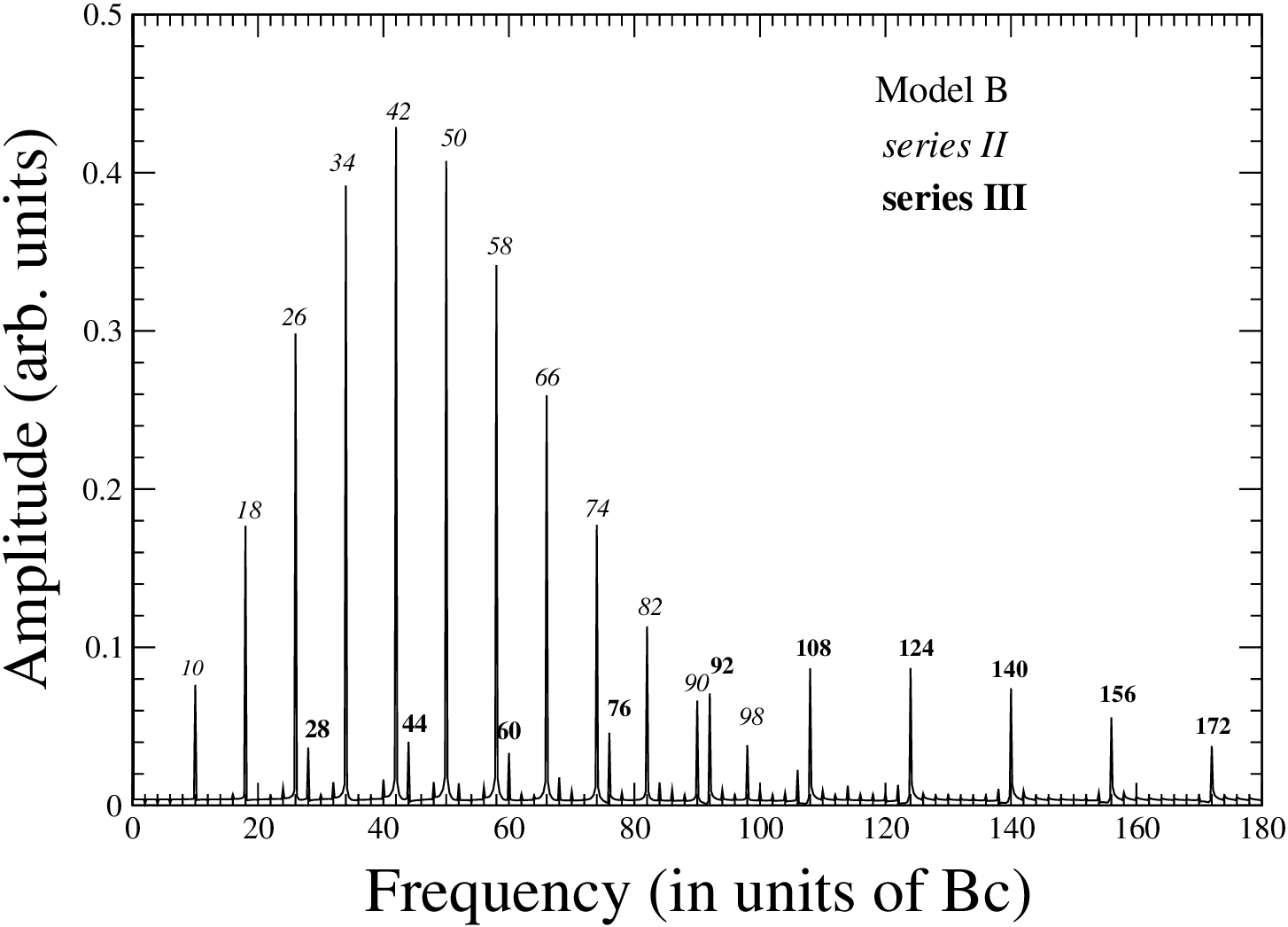} 
\par\end{centering}

\caption{\label{fig:ComSpe}Theoretical Fourier spectrum of the dynamic $19$th
harmonic signal for $\mathrm{O_{2}}$; pump intensity $I=0.5\times10^{14}\,\mathrm{W/cm^{2}}$,
probe intensity $I=1.2\times10^{14}\,\mathrm{W/cm^{2}}$W/cm$^{2}$,
duration $40\,\mathrm{fs}$, wavelength $800\,\mathrm{nm}$, and temperature
$200\,\mathrm{K}$. The calculations are done using the present theory,
Eq. (\ref{OurThe}), (left-lower panel), model A Eq.(\ref{mod:A}),
(right-upper panel), and model B, Eq. (\ref{mod:B}), (right-lower
panel). For comparison, the experimental spectrum (left-upper panel)
is also shown.}
\end{figure*}

In this sub-section we briefly discuss two alternative definitions
of HHG signals that have been employed earlier and compare them with
the definition of the HHG signal of the present theory, and with experimental
data. The present theory defines the quantum transition amplitudes
for the linearly independent reference states $|\chi_{i}(t)>,i\equiv\{ e,J_{0}M_{0}\}$,
(consisting of the product of the ground electronic and the coherent
rotational wavepacket states) to obtain the independent harmonic emission
probabilities, and in accordance with the quantum statistical theory
averages the latter to define the HHG signal (cf. e.g. Eq.(\ref{HhgSig1}):
\begin{equation}
\begin{array}{c}
S^{(n)}\left(t_{d}\right)=\hphantom{\left|\left\langle \Phi_{J_{0}M_{0}}\left(t_{d},\theta\right)\left|T_{e}^{(n)}(\theta,\alpha)\right|\Phi_{J_{0}M_{0}}\left(t_{d},\theta\right)\right\rangle \right|^{2}}\\
{\cal {C}}\sum_{J_{0}M_{0}}\rho(J_{0})\left|\left\langle \Phi_{J_{0}M_{0}}\left(t_{d},\theta\right)\left|T_{e}^{(n)}(\theta)\right|\Phi_{J_{0}M_{0}}\left(t_{d},\theta\right)\right\rangle \right|^{2}\end{array}\label{OurThe}\end{equation}
 It is worth noting that the quantum amplitude calculation in the
present theory corresponds to the {}``adiabatic nuclei'' approximation
\cite{cha-56,fai-72}, in which the matrix elements with respect to
the rotational wavepacket states are evaluated at the level of the
adiabatic amplitude-operator, $T^{(n)}(\theta)$, and {\textit{not}}
at the level of the adiabatic probability- operator, $\left|T^{(n)}(\theta)\right|^{2}$,
that occurs in the more drastic {}``frozen nuclei'' approximation.
In this theory, as in the laboratory, the operational angle is the
relative polarization angle $\alpha$, and {\textit{not}} the angle
between the polarization direction and the molecular axis, $\theta$.
In fact, the angle $\theta$ is a coordinate that is, as appropriate
for a quantum formulation, to be \textit{integrated over} to obtain
the quantum transition amplitude with respect to the rotational wavepacket
states.

In the present notation, the two other definitions of the HHG signal
that have been used earlier (to be referred to below as A and B) are,
(i) definition A (cf. Eq. (22) of \cite{mad-06} and Eq. (6) of \cite{mad-07}):
\begin{eqnarray}
 &  & S_{A}^{(n)}\left(t_{d}\right)=\nonumber \\
 &  & {\cal {C}}\left|\sum_{J_{0}M_{0}}\rho(J_{0})\left\langle \Phi_{J_{0}M_{0}}\left(t_{d},\theta\right)\left|T_{e}^{(n)}(\theta)\right|\Phi_{J_{0}M_{0}}\left(t_{d},\theta\right)\right\rangle \right|^{2}\,\label{mod:A}\end{eqnarray}
 and, (ii) definition B (cf. Eq. (12) of \cite{zho-05b} and Eq. (4)
of \cite{le-06}): \begin{eqnarray}
 &  & S_{B}^{(n)}\left(t_{d}\right)=\nonumber \\
 &  & {\cal {C}}\sum_{J_{0}M_{0}}\rho(J_{0})\left\langle \Phi_{J_{0}M_{0}}\left(t_{d},\theta\right)\left|\left|T_{e}^{(n)}(\theta,0)\right|^{2}\right|\Phi_{J_{0}M_{0}}\left(t_{d},\theta\right)\right\rangle \,\,\label{mod:B}\end{eqnarray}
 Clearly the HHG signals according to models A and B differ with each
other, and they differ from the present definition, Eq. (\ref{OurThe})
above.

We note that model A, Eq. (\ref{mod:A}), defines the statistically
averaged signal by weighting the individual amplitudes (!) first,
and then taking the absolute square of the weighted sum. This runs
counter to the quantum statistical theoretical approach of averaging
the probabilities (not amplitudes) and/or the expectation values of
Hermitian observables themselves, and not their Fourier transforms
(that are proportional to the emission amplitudes). Furthermore, the
definition of model A (Eq. (\ref{mod:A})) makes the signal to depend
on the mixed products of the statistical weights that are in principle
\textit{independent}.

The signal defined by model B, Eq. (\ref{mod:B}), is seen to depend
on the weighted sum of the diagonal matrix elements (between the rotational
wavepacket states) of the {}`` probability operator'' $\left|T^{(n)}(\theta)\right|^{2}$
-- this, of course, is not equal to the weighted sum of the absolute
squares of the diagonal matrix elements of the transition operator
$T^{(n)}(\theta)$. In model B the above circumstance is a consequence
of the more drastic {}``frozen nuclei'' approximation and an effective
inclusion of {\textit{all}} transitions, those between the same
wavepacket states ({}``elastic-like'') \textit{as well as} those
between the different wavepacket states ({}``inelastic-like'').
However, unlike in model A (Eq. (\ref{mod:A})), in model B (Eq. (\ref{mod:B}))
the weighted statistical sum is taken, in accordance with the quantum
statistical theory, at the level of the probabilities.

In Fig. \ref{fig:ComSpe} we show a comparison of the experimental
data (upper-left panel) of the $F.T.$ of the dynamic signal ($\alpha=0$)
of the 19th harmonic, for the case of O$_{2}$ with the results of
the test calculations from: the present theory theory (lower-left
panel) using Eq. (\ref{OurThe}), model A (upper-right panel) using
Eq. (\ref{mod:A}), and model B (lower-right panel) using Eq.(\ref{mod:B}),
keeping everything else the same -- the parameters are the same as
in Fig. \ref{fig:FouSpeO2}. The similarity of the calculated spectrum
from the present theory and the experimental data is seen to be very
satisfactory, that from model A is similar, except that the ratio
of the peaks of the series III to series II is generally too small
compared to the experimental data, and series V $(\boldsymbol{{\it 8,16,24,..}})\, Bc$
is rather weakly developed. In the case of model B, on the other hand,
the series V $(\boldsymbol{{\it 8,16,24,..}})\, Bc$ is simply missing.
It may be noted that the present comparison also illustrates the ability
of the experimental data at the level of the (discrete) Fourier spectrum,
to better distinguish between the various theoretical models than
may be possible at the level of the time-domain signal.

\begin{figure}
\begin{centering}
\includegraphics[scale=0.3]{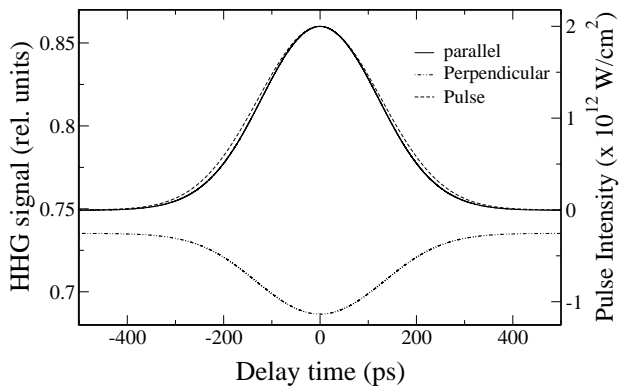} 
\par\end{centering}

\caption{\label{fig:Adi-Spe}Calculated $9$th HHG spectrum of $\mathrm{N_{2}}$
for various pump-probe polarizations angle $\alpha=0^{o}$ and $\alpha=90^{o}$;
pump intensity $I=2\times10^{12}$W/cm$^{2}$, duration $300\,\mathrm{ps}$;
probe intensity, $I=5\times10^{14}$ W/cm$^{2}$, duration $70\,\mathrm{fs}$,
and wavelength $798\,\mathrm{nm}$; Boltzmann temperature 25 K.}
\end{figure}

\subsection{Adiabatic Alignment}

Finally, we apply the present dynamic theory also to the adiabatic
case, in which we choose a long ($300\,\mathrm{ps}$) pump pulse and
a short ($70\,\mathrm{fs}$) probe pulse, as in an adiabatic alignment
experiment \cite{mar-01} used earlier. The results of our calculations
for $\mathrm{N_{2}}$, using Eqs. (\ref{NitHhgSigAlf}) for both $\alpha=0^{o}$
(solid curve), and $\alpha=90^{o}$ (dash-dot curve), are shown in
Fig. \ref{fig:Adi-Spe}. For the sake of comparison we also show the
intensity profile (dashed curve) of the pump pulse (right scale).
As can be seen immediately from the figure, in the parallel case,
the HHG signal closely follows the evolution of the long pump pulse
itself (which might be expected for an adiabatic process) and the
maximum of the signal occurs at the maximum of pulse, for $\alpha=0^{o}$.
On the other hand, a minimum is predicted for the signal at the maximum
intensity, in the perpendicular case, for $\alpha=90^{o}$. These
characteristics of the adiabatic signals for $\mathrm{N_{2}}$ are
consistent with the experimental observations made some time ago \cite{mar-01,mar-02,mar-03}.

\section{Conclusions}

To conclude, we have presented an \emph{ab initio} intense-field S-matrix
theory of dynamic alignment of linear molecules and the characteristic
HHG signals from them as detected in intense-field femtosecond pump-probe
experiments. Useful analytical expressions for the molecular alignment
and the HHG signal as a function of both the delay-time, $t_{d}$,
and the relative polarization angle, $\alpha$, between the pump and
probe pulse, are derived. Thus, we give the general HHG signal Eq.
(\ref{GenHhgSig}), the signal for $\mathrm{N_{2}}$ (generically,
active $\sigma_{g}$ orbital symmetry) Eq. (\ref{NitHhgSigAlf}),
or its leading term Eq (\ref{NitSigApp}), as well as the signal for
$\mathrm{O_{2}}$ (generically, active $\pi_{g}$ orbital symmetry)
Eq. (\ref{OxyHhgSigAlf}), or its leading term, Eq. (\ref{OxySigApp}).
They are used to make detailed analysis of the molecular alignment
and the observed experimental data for the HHG signals from coherently
rotating $\mathrm{N_{2}}$ and $\mathrm{O_{2}}$ molecules, both in
the time-domain \textit{and} in the frequency domain. The results
show a remarkable agreement between the theory and the experimental
observations. Additional predictions about the existence of critical
relative polarization angles, $\alpha_{c}$, and their relation to
the symmetry of the active orbitals and the form of the dynamic signals
are made. At a {}``magic'' angle, $\alpha_{c}\approx55^{o}$, the
dynamic HHG signals for all delay times $t_{d}$, are predicted to
approach each other closely for a linear molecule with a $\sigma_{g}$
orbital symmetry, or exhibit a {}``crossing neighborhood'', for
the $\pi_{g}$ orbital symmetry; it is also predicted to produce a
steady emission of high harmonic radiation at the magic angle from
$\mathrm{N_{2}}$. Moreover, we have investigated a number of theoretical
questions and experimental effects of general interest in connection
with the interpretation of the pump-probe HHG signals. Finally, we
have shown that the case of {}``adiabatic-alignment'', and the resulting
HHG signal, can be analyzed and understood equally well within the
present dynamical theory, using simply a long duration of the pump
pulse.\\

\begin{acknowledgments}
We thank Prof. K. Miyazaki for providing the experimental data in
digital form shown in the upper panels of Figs. \ref{fig:DynN2},
\ref{fig:DynO2}, \ref{fig:FouSpeN2}, \ref{fig:FouSpeO2}, and \ref{fig:ComSpe},
and for useful discussions. 
\end{acknowledgments}


\begin{thebibliography}{10}
\bibitem{pos-04}J.H. Posthumus, Rep. Prog. Phys. \textbf{67}, 623
(2004).

\bibitem{bec-05}A. Becker and F.H.M. Faisal, J. Phys. B \textbf{38},
R1 (2005).

\bibitem{sei-99}T. Seideman, Phys. Rev. Lett. \textbf{83}, 4971 (1999).

\bibitem{ort-99}J. Ortigoso, M. Rodriguez, M. Gupta, and B. Friedrich,
J. Chem. Phys. \textbf{110}, 3870 (1999).

\bibitem{cai-01}L. Cai, J. Marango and B. Friedrich, Phys. Rev. Lett.
\textbf{86}, 775 (2001).

\bibitem{sta-03}H. Stapelfeldt and T. Seideman, Rev. Mod. Phys. \textbf{75},
543 (2003).

\bibitem{lit-03}I.V. Litvinyuk, K.F. Lee, P.W. Dooley, D.M. Rayner,
D.M. Villaneuve, and P.B. Corkum, Phys. Rev. Lett. \textbf{90}, 233003
(2003).

\bibitem{doo-03}P.W. Dooley, I.V. Litvinyuk, K.F. Lee, D.M. Rayner,
M. Spanner, D.M. Villaneuve, and P.B. Corkum, Phys. Rev. A \textbf{68},
023406 (2003).

\bibitem{kak-04}M. Kaku, K. Masuda, and K. Miyazaki, Japan. J. Appl.
Phys. \textbf{43}, 591, (2004).

\bibitem{zei-04}D. Zeidler, J. Levesque, J. Itatani, K. Lee, P. W.
Dooley, I. Litvinyuk, D.M. Villeneuve, and P.B. Corkum in \textit{Ultrafast
Optics IV}, ed. F. Krauz \emph{et. al.} (Springer, New York, 2004)
p. 247.

\bibitem{ita-05}J. Itatani, D. Zeidler, J. Levesque, M. Spanner,
D.M. Villeneuve and P.B. Corkum, Phys. Rev. Lett. \textbf{94}, 123902
(2005).

\bibitem{kan-05}T. Kanai, S. Minemoto and H. Sakai, Nature \textbf{435},
03577 (2005) .

\bibitem{miy-05}K. Miyazaki, M. Kaku, G. Miyaji, A. Abdurrouf, and
F.H.M. Faisal, Phys. Rev. Lett \textbf{95}, 243903 (2005).

\bibitem{ita-04}J. Itatani, J. Levesque, D. Zeidler, H. Niikura,
H. Pepin, J.C. Kieffer, P.B. Corkum and D.M. Villeneuve, Nature \textbf{432},
867 (2004).

\bibitem{lev-06}J. Levesque, J. Itatani, D. Zeidler, H. Pepin, J.C.
Kieffer, P.B. Corkum, and D.M. Villeneuve, J. Mod. Opt \textbf{53},
185 (2006).

\bibitem{pat-06}S. Patchkovskii, Z. Zhao, T. Brabec, and D.M. Villeneuve,
Phys. Rev. Lett. \textbf{97}, 123003 (2006).

\bibitem{bak-06}S. Baker, J.S. Robinson, C.A. Haworth, H. Teng, R.A.
Smith, C.C. Chirila, M. Lein, J.W.G. Tisch, and J.P. Marangos, Science
\textbf{312}, 424 (2006).

\bibitem{wag-06}N.L. Wagner, A. Wuest, I.P. Christov, T. Popmintchev,
X. Zhao, M.M. Murname, and H.C. Kapteyn, PNAS, \textbf{103}, 13279
(2006).

\bibitem{fai-07}F.H.M. Faisal, A. Abdurrouf, K. Miyazaki, and G.
Miyaji, Phys. Rev. Lett. \textbf{98}, 143001 (2007).

\bibitem{fai-08}F.H.M. Faisal, and A. Abdurrouf, Phys. Rev. Lett.
\textbf{100}, 123005 (2008).

\bibitem{ros-02}F. Rosca-Pruna and M.J.J. Vrakking, J. Chem. Phys.
\textbf{116}, 6575 (2002).

\bibitem{her-50}G. Herzberg, \textit{Molecular Spectra and Molecular
Structure, I.} (Van Nostard Reinhold, New York (1950), chap. III.

\bibitem{zho-05a}X.X. Zhou, X.M. Tong, Z.X. Zhao and C.D. Lin, Phys.
Rev. A \textbf{71}, 061801(R) (2005).

\bibitem{zho-05b}X.X. Zhou, X.M. Tong, Z.X. Zhao and C.D. Lin, Phys.
Rev. A \textbf{7}2, 033412 (2005).

\bibitem{mad-06}C.B. Madsen and L.B. Madsen, Phys. Rev. A \textbf{74},
023403 (2006).

\bibitem{kak-05}M. Kaku, R. Morichi, G. Miyaji, and K. Miyazaki,
IEEE on Quantum Electronics Conference 2005, paper QWG4-3, p. 1036.

\bibitem{yos-07}K. Yoshii, G. Miyaji, K. Miyazaki, A.Abdurruf, and
F.H.M. Faisal, IEEE on CLEO - Pacific Rim 2007, p. 660 (2007).

\bibitem{fey-ed}R.P. Feynman, \emph{Quantum Electrodynamics} (Benjamin
Inc., New York, 1962).

\bibitem{sak-qm}J.J. Sakurai, \emph{Modern Quantum Mechanics} (Addison-Wesley,
New York, 1994)

\bibitem{kel-64}L.V. Keldish, Zh. Eksp. Teor. Fiz. \textbf{47}, 1945
(1964) {[}Sov. Phys. JETP \textbf{20}, 1307 (1964)].

\bibitem{fai-73}F.H.M. Faisal, J. Phys. B \textbf{6}, L89 (1973).

\bibitem{rei-80}H.R. Reiss, Phys. Rev A \textbf{22}, 1786 (1980).

\bibitem{jac-62}J.D. Jackson, \emph{Classical Electrodynamics} (J.
Wiley \& Sons, New York, 1962).

\bibitem{pre-nr}W.H. Press, B.P. Flannery, S.A. Teukolsky, and W.T.
Vetterling, \emph{Numerical Recipes: The Art of Scientific Computing}
(Cambridge University Press, 1986).

\bibitem{foot-10} We may assume as usual that the field envelopes
are slowly varying compared e.g. to the periods of the higher harmonics.

\bibitem{sal-01}P. Salières, B. Carré, L. Le Déroff, F. Grasbon,
G. G. Paulus, H. Walther, R. Kopold, W. Becker, D. B. Milosevic, A.
Sanpera, and M. Lewenstein, Science \textbf{4292}, 902 (2001).

\bibitem{foot-20} We may note that the above holds strictly for an
ideal medium with a constant velocity of propagation. In dispersive
media the frequency and the wavenumbers are related by the frequency
dependent velocity of propagation $v=c/n(\Omega)$ where $n$ is the
frequency dependent refractive index of the medium. This necessitates
a much more detailed analysis of the phase-matching condition, that
has been carried out (see e.g. \cite{sal-01}) by solving the associated
Maxwell's equations for the propagating fields, numerically in specific
cases. The results indicate rather generally an effective phase matching,
for the higher harmonics. In the presence of ionizaion, the mediumm
becomes also dissipative. In a weakly ionized medium the effect may
be taken into account by including a decay factor $e^{-\frac{1}{2}\gamma_{i}(t+t')}$
that is to be multiplied with the expression of the integrand of the
HHG amplitude (or of the dipole expectation value), where $\gamma_{i}$
is the total rate of ionization of the reference bound state $i\equiv\left|\chi_{i}(t)\right\rangle $.

\bibitem{fer-87} M. Ferray, F. Gounand, P. DOliveira, P. R. Fournier,
D. Cubaynes, J. M. Bizau, T. J. Morgan, and F. J. Wuilleumier, Phys.
Rev. Lett. \textbf{59}, 2040 (1987).

\bibitem{rho-87}A. McPherson, G. Gibson, H. Jara, U. Johann, T.S.
Luk, I.A. McIntyre, K. Boyer, and C.K. Rodes, J. Opt. Soc. Am. B \textbf{4},
595 (1987)

\bibitem{ehl-01}F. Ehlotzky, Phys. Rep. \textbf{345}, 175 (2001).

\bibitem{sei-01}T. Seideman, J. Chem. Phys. \textbf{115}, 5965 (2001).

\bibitem{lew-94}M. Lewenstein, Ph. Balcou, M. Yu. Ivanov, A. L'Huillier
and P.B. Corkum, Phys. Rev. A \textbf{49}, 2117 (1994).

\bibitem{har-65}F.E. Harris and H.H. Michels, J. Chem. Phys. \textbf{43},
S165 (1965).

\bibitem{fai-70}F.H.M. Faisal, J. Phys. B \textbf{3}, 636 (1970).

\bibitem{var-88}D. A. Varshalovich, A.M. Moskalev and V.K. Khersonskii,
\emph{Quantum Theory of Angular Momentum} (Word Scientific, Singapore,
1988).

\bibitem{zar-88}R. N. Zare, Angular Momentum: \emph{Understanding
Spatial Aspects in Chemistry and Physics} (Wiley, New York, 1988).

\bibitem{gra-65}I.S. Gradshteyn and I.M. Rhysik, \emph{Table of Integral,
Series, and Product} (Academic Press, New York, 1965).

\bibitem{ton-02}X.M. Tong, Z.X. Zhao, and C.D. Lin, Phys. Rev. A
\textbf{66}, 033402 (2002).

\bibitem{kje-05}T.K. Kjeldsen, and L.B. Madsen, Phys. Rev. A \textbf{71},
023411 (2005).

\bibitem{jor-73}W.L. Jorgensen and L. Salem, \emph{The Organic Chemist's
Book of Orbital} (Academic Press, New York, 1973)

\bibitem{jam-92}A.M. James and M.P. Lord, \emph{MacMillan's Chemical
Physical Data} (MacMillan, London, 1992).

\bibitem{hir-54}J.O. Hirschfelder, C.F. Curtis, and R.B. Bird, \emph{Molecular
Theory of Gases and Liquids} (Wiley, New York, 1954).

\bibitem{ram-07}S. Ramakrishna and T. Seideman, Phys. Rev. Lett.
\textbf{99}, 113901 (2007).

\bibitem{ram-08}S. Ramakrishna and T. Seideman, Phys. Rev. A \textbf{77},
053411 (2008).

\bibitem{miy-06}K. Miyazaki (personal communication).

\bibitem{vra-02}F. Rosca-Pruna and M.J.J. Vraking, J. Chem Phys.
\textbf{116}, 6567 (2002).

\bibitem{ave-89}I. Sh. Averbukh and N.F. Parelman, Phys. Lett. A
\textbf{139}, 449 (1989).

\bibitem{vra-96}M.J.J. Vrakking, D.M. Villeneuve, and A. Stolow,
Phys Rev. A \textbf{54}, R37 (1996).

\bibitem{blu-96}R. Bluhm, V.A. Kostelecky, and J.A. Porter, Am. J.
Phys. \textbf{64}, 994 (1996).

\bibitem{flei-06}S. Fleischer, I. Sh. Averbukh, and Y. Prior, Phys.
Rev. A \textbf{74}, 041403(R) (2006).

\bibitem{cha-56}D.M. Chase, Phys. Rev. \textbf{104}, 835 (1956).

\bibitem{fai-72}F.H.M. Faisal and A. Temkin, Phys. Rev. Lett. \textbf{28},
203 (1972).

\bibitem{tor-07}R. Torres, N. Kajumba, J. G. Underwood, J. S. Robinson,
S. Baker, J. W. G. Tisch, R. de Nalda, W. A. Bryan, R. Velotta, C.
Altucci, I. C. E. Turcu, and J. P. Marangos, Phys. Rev. Lett. \textbf{98},
203007 (2007).

\bibitem{lei-07}M. Lein, J. Phys. B \textbf{\textit{\emph{40}}},
R135 (2007).

\bibitem{mad-07}C. B. Madsen, A. S. Mouritzen, T. K. Kjeldsen, and
L. B. Madsen, Phys. Rev. A \textbf{76}, 035401 (2007).

\bibitem{le-06}A.T. Le, X.-M. Tong, and C. D. Lin Phys. Rev. A \textbf{73},
041402(R) (2006).

\bibitem{mar-01}R. Velotta, N. Hay, M.B. Mason, M. Castillejo, and
J.P. Marangos, Phys. Rev. Lett., \textbf{87}, 183901 (2001).

\bibitem{mar-02}N. Hay, R. Velotta, M.Lein, R. de Nalda, E. Hessel,
M. Castillejo, and J.P. Marangos, Phys. Rev. A \textbf{65}, 053805
(2002).

\bibitem{mar-03}N. Hay, M.Lein, R. Velotta, R. deNalda, E. Hessel,
M. Castillejo, P.L. Knight, and J.P. Marangos, J. Mod. Opt. \textbf{50},
561 (2003). 
\end{thebibliography}
\end{document}